\documentclass[aps,onecolumn,11pt,floatfix,altaffilletter,superscriptaddress,preprintnumbers,
               tightenlines,showpacs,showkeys,nofootinbib]{revtex4-2}

\makeatletter
\def\thesection{\arabic{section}}%
\def\p@section{}%
\def\thesubsection{\thesection.\arabic{subsection}}%
\def\p@subsection{}%
\def\thesubsubsection{\thesubsection.\arabic{subsubsection}}%
\def\p@subsubsection{}%
\def\appendix{%
    \par
    \setcounter{section}\z@
    \setcounter{subsection}\z@
    \setcounter{subsubsection}\z@
    \def\thesubsection{\thesection.\arabic{subsection}}%
    \def\thesubsubsection{\thesubsection.\arabic{subsubsection}}%
    \def\p@subsection{}%
    \def\p@subsubsection{}%
    \@addtoreset{equation}{section}%
    \def\theequation@prefix{\thesection}%
    \addtocontents{toc}{\protect\appendix}%
    \@ifstar{%
    \def\thesection{\unskip}%
    \def\theequation@prefix{A.}%
    }{%
    \def\thesection{\Alph{section}}%
    }%
}%
\makeatother

\usepackage[utf8]{inputenc}
\usepackage[colorlinks=true,citecolor=blue,linkcolor=blue]{hyperref}
\usepackage[normalem]{ulem}
\usepackage{amsmath,amssymb, mathrsfs, physics, aas_macros}
\usepackage{epsfig}
\usepackage{graphicx}               
\usepackage{url}
\usepackage{color}
\usepackage[dvipsnames]{xcolor}
\usepackage{soul}
\usepackage[capitalise, english]{cleveref}
\usepackage{multibib}
\usepackage{siunitx}
\usepackage{xspace}
\usepackage{enumitem}
\usepackage{fontawesome}
\usepackage{tikz}
\usepackage{fontawesome}

\newcommand\myshade{80}
\colorlet{mylinkcolor}{ForestGreen}
\colorlet{mycitecolor}{Red}
\colorlet{myurlcolor}{violet}

\hypersetup{
  linkcolor  = mylinkcolor!\myshade!black,
  citecolor  = mycitecolor!\myshade!black,
  urlcolor   = myurlcolor!\myshade!black,
  colorlinks = true
}

\DeclareSIUnit\parsec{pc}

\renewcommand{\vec}[1]{{\mathbf{#1}}}
\newcommand{\iso}[2]{{\ensuremath{{}^{#2}}\ensuremath{\rm #1}}}

\definecolor{tobycolour}{rgb}{.5,.0,.5}

\definecolor{plotorange}{RGB}{249,138,41}
\definecolor{plotgreen}{RGB}{43,151,26}
\definecolor{plotblue}{RGB}{29,47,221}
\definecolor{plotpink}{RGB}{247,37,153}
\definecolor{plotlightblue}{RGB}{104,150,234}
\definecolor{plotgrey}{RGB}{128,128,128}

\renewcommand{\iso}[2]{{\ensuremath{{}^{#2}}\ensuremath{\rm #1}}}

\newcommand{\Li}{\text{Li}}

\newcommand{\cth}[1]{\ensuremath{c_{\theta^\nu_\text{#1}}}}
\newcommand{\sth}[1]{\ensuremath{s_{\theta^\nu_\text{#1}}}}
\newcommand{\cph}[1]{\ensuremath{c_{\phi^\nu_\text{#1}}}}
\newcommand{\sph}[1]{\ensuremath{s_{\phi^\nu_\text{#1}}}}

\newcommand{\tht}[1]{\ensuremath{\theta^\nu_\text{#1}}}
\newcommand{\ph}[1]{\ensuremath{\phi^\nu_\text{#1}}}

\newcommand{\cthp}[1]{\ensuremath{c_{\theta^q_\text{#1}}}}
\newcommand{\sthp}[1]{\ensuremath{s_{\theta^q_\text{#1}}}}
\newcommand{\cphp}[1]{\ensuremath{c_{\phi^q_\text{#1}}}}
\newcommand{\sphp}[1]{\ensuremath{s_{\phi^q_\text{#1}}}}

\newcommand{\thtp}[1]{\ensuremath{\theta^q_\text{#1}}}
\newcommand{\php}[1]{\ensuremath{\phi^q_\text{#1}}}

\newcommand{\nscool}{\texttt{NScool}\xspace}

\makeatletter
\providecommand*{\diff}%
	{\@ifnextchar^{\DIfF}{\DIfF^{}}}
\def\DIfF^#1{%
	\mathop{\mathrm{\mathstrut d}}%
		\nolimits^{#1}\gobblespace}
\def\gobblespace{%
	\futurelet\diffarg\opspace}
\def\opspace{%
	\let\DiffSpace\!%
	\ifx\diffarg(%
		\let\DiffSpace\relax
	\else
		\ifx\diffarg[%
			\let\DiffSpace\relax
		\else
  			\ifx\diffarg\{%
				\let\DiffSpace\relax
			\fi\fi\fi\DiffSpace}

\begin{document}

\title{Electron and Muon Dynamics in Neutron Stars Beyond Chemical Equilibrium}

\author{Joachim Kopp}\email{jkopp@cern.ch}
\affiliation{Theoretical Physics Department, CERN,
             Esplanade des Particules, 1211 Geneva 23, Switzerland}
\affiliation{PRISMA Cluster of Excellence \& Mainz Institute for
             Theoretical Physics, \\
             Johannes Gutenberg University, Staudingerweg 7, 55099
             Mainz, Germany}

\author{Toby Opferkuch}\email{tobyopferkuch@berkeley.edu}
\affiliation{Berkeley Center for Theoretical Physics, University of California, Berkeley, CA 94720}
\affiliation{Theoretical Physics Group, Lawrence Berkeley National Laboratory, Berkeley, CA 94720}

\preprint{CERN-TH-2023-239, MITP-23-076}

\begin{abstract}
\noindent
A neutron star harbors $\mathcal{O}(10^{56})$ electrons in its core, and almost the same number of muons, with muon decay prohibited by Pauli blocking. However, as macroscopic properties of the star such as its mass, rotational velocity, or magnetic field evolve over time, the equilibrium lepton abundances (dictated by the weak interactions) change as well. Scenarios where this can happen include spin-down, accretion, magnetic field decay, and tidal deformation. We discuss the mechanisms by which a star disrupted in one of these ways re-establishes lepton chemical equilibrium. In most cases, the dominant processes are out-of-equilibrium Urca reactions, the rates of which we compute for the first time. If, however, the equilibrium muon abundance decreases, while the equilibrium electron abundance increases (or decreases less than the equilibrium muon abundance), outward diffusion of muons plays a crucial role as well. This is true in particular for stars older than about \SI{E+4}{yrs} whose core has cooled to $\lesssim \SI{20}{keV}$. The muons decay in a region where Pauli blocking is lifted, and we argue that these decays lead to a flux of $\mathcal{O}(\SI{10}{MeV})$ neutrinos. Realistically, however, this flux will remain undetectable for the foreseeable future.
\href{https://github.com/koppj/muons-in-neutron-stars}{\faGithub}
\end{abstract}

\maketitle
\tableofcontents

\section{Introduction}
\label{sec:intro}

Neutron stars are among the most fascinating objects in the Universe, and a treasure trove of information on matter under extreme conditions (see for example ref.~\cite{Oertel:2016bki} and references therein). Born in the violent collapse of a massive star, a neutron star begins its life at a very high temperature $T \gg \si{MeV}$, before rapidly cooling down, reaching a surface temperature of order keV after just one day~\cite{Yakovlev:2000jp}. Because of the large initial temperature and large chemical potentials, a neutron star's core harbors not only neutrons, but also protons, electrons, and muons \cite{Cohen:1970}. Muons are prevented from decaying because of Pauli blocking; given the large electron density, there are no unoccupied final states available that muons could decay into. The sizable population of muons can for instance be leveraged as a tool to constrain physics beyond the Standard Model, in particular related to dark matter~\cite{%
Garani:2018kkd,  
Bell:2019pyc,    
Garani:2019fpa,  
Dror:2019uea,    
Hamaguchi:2022wpz}. 

In this paper we add several new aspects to our understanding of both electrons and muons in neutron stars:
\begin{enumerate}
    \item The equilibrium lepton abundances in a neutron star are not static, but can change throughout the star's life. In the subsequent discussion, we use the term `equilibrium' to specifically denote the state of beta-equilibrium, which is maintained through the weak interactions. Notably, changes in its magnetic field or rotational velocity can affect the lepton populations.
    \item How the star responds to such changes depends on its age. In young stars ($\lesssim \SI{1E+4}{yrs}$) weak interactions (so-called Urca processes) can efficiently produce and destroy electrons and muons. In older and colder stars, however, these reactions become less efficient due to their strong temperature dependence. This implies that old neutron stars may go at least temporarily out of chemical equilibrium if their equilibrium lepton abundances change.
    \item If the equilibrium muon abundance, $n_\mu^\text{eq}$ decreases, while the corresponding quantity for electrons, $n_e^\text{eq}$ increases (or decreases less than $n_\mu^\text{eq}$), an efficient way to at least partially restore chemical equilibrium even in cold stars is muon diffusion towards the surface. Once a muon reaches a region where Pauli blocking is less severe, it will decay, and the decay electron propagates back into the core to compensate the charge imbalance left behind by the muon. We will investigate both standard free muon decay, as well as ``assisted muon decay'', where a spectator nucleon absorbs significant energy and momentum.
    \item Muon decays in the outer regions of the star can lead to a neutrino flux at energies up to tens of MeV, well above the star's core temperature. This neutrino flux could be observable above background if the equilibrium muon abundance in all neutron stars in the Milky Way were to gradually decrease over $\mathcal{O}(\si{Gyr})$ time scales. Most likely, though, the opposite happens -- stars accrete matter, which leads to an increased muon abundance.
\end{enumerate}
In the following, we will first elaborate in more detail on the conditions under which the equilibrium lepton abundances in a neutron star change (\cref{sec:ns-muons}), before discussing in detail the dynamics of muon diffusion and decay in \cref{sec:mu-diff}.  In \cref{sec:nu-fluxes} we will then investigate the resulting neutrino fluxes. We will conclude in \cref{sec:conclusions}. The appendices contain detailed discussions of lepton production and absorption via out-of-equilibrium Urca processes (\cref{sec:abs-prod}), the calculation of the muon decay width in dense electron gases (\cref{sec:mu-width}), a short summary of the different neutron star equations of state considered in this work (\cref{sec:eos}), and a discussion of backgrounds in MeV-scale neutrino experiments. Throughout the paper, we will predominantly focus on muons, as their behavior is more intricate owing to their instability in a vacuum. However, we will state at each step which of our results apply also to electrons.

The numerical codes we have developed for this paper are available from \url{https://github.com/koppj/muons-in-neutron-stars}.

\section{Electrons and Muons in Neutron Stars.}
\label{sec:ns-muons}

\begin{figure}
    \centering
    \includegraphics[width=\textwidth]{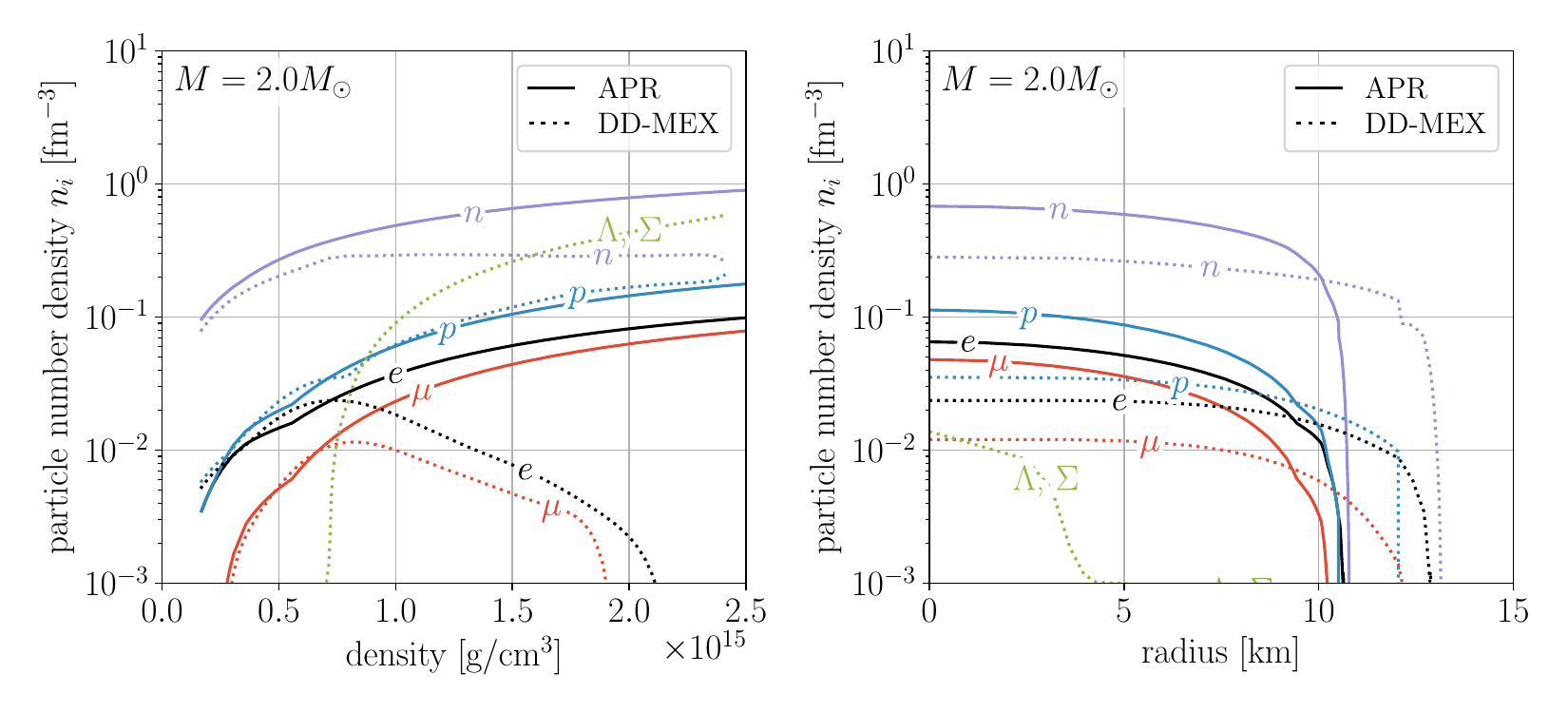}
    \caption{The abundances of electrons (black), muons (red), protons (blue), neutrons (purple), and hyperons (green) inside a $2.0 M_\odot$ neutron star for two different equations of state, namely the Akmal--Pandharipande--Ravenhall (APR) \cite{Akmal:1998cf} and DD-MEX (density-dependent relativistic mean-field) \cite{Taninah:2019cku} EOS. In the left panel, the horizontal axis shows the density, in the right panel it shows the radial coordinate. Note that the DD-MEX EOS includes hyperons, which is the reason why at very high density, the electron and muon abundances drop -- in this regime, all negative electric charge is carried by $\Sigma^-$ hyperons.}
    \label{fig:particle-abundances}
\end{figure}

In spite of the term ``neutron star'', it is well known that such stars contain significant numbers of protons, electrons, and muons (and possibly hyperons as well as more exotic matter) in addition to neutrons. This is illustrated in \cref{fig:particle-abundances}, which shows the abundances of different particle species inside a neutron star as a function of density (left) and of radius (right), for two particular equations of state.  The presence of particles other than neutrons can be understood from thermodynamic arguments: if the star contained only neutrons, their chemical potential would be so large that decay to protons and charged leptons would be energetically favorable.  Similarly, in a configuration without muons, charge neutrality would require a very large electron chemical potential, which is again thermodynamically unfavorable.  A different, but equivalent, justification for the existence of muons in neutron star cores is that they simply cannot decay because all possible quantum states for Michel electrons are occupied.  Reaching equilibrium is possible thanks to the extreme temperatures a neutron star reaches during its formation in a supernova explosion ($\gtrsim \SI{10}{MeV})$, which allow for thermal production even of heavy particles like muons. Depending on the neutron star mass and details of the equation of state (EOS), a neutron star typically harbors $\mathcal{O}(10^{56})$ electrons in its core, with the muon abundance at most a factor 2--3 smaller. 

However, the neutron star's equilibrium lepton abundances can change during the star's life. Factors that influence it include the following:
\begin{enumerate}
    \item {\bf The rotational velocity.} In a fast rotating star, the centrifugal force leads to deformation---the star is no longer spherical, but slightly oblate \cite{%
    Friedman:1986,        
    Komatsu:1989zz,       
    Cook:1993qr,          
    Stergioulas:1994ea,   
    Franzon:2016iai,      
    Silva:2020oww,        
    Konstantinou:2022,    
    Watanabe:2020vas}. This deformation comes with a slight increase in the star's volume and consequently a decrease in the core density. This in turns implies slightly reduced lepton abundances. For instance, for rotational frequencies $f \sim \SI{700}{Hz}$ (corresponding to the fastest-spinning millisecond pulsar discovered to date, PSR~J1748-2446ad, with $f = \SI{716}{Hz}$ \cite{Hessels:2006ze}) the predicted equilibrium muon abundance is several per mille lower than in a non-rotating star. For the limiting case of a star rotating at its Kepler frequency, $f_K$ (beyond which the centrifugal force for lead to immediate disintegration), ref.~\cite{Konstantinou:2022} predicts at most a $1\%$ change in density, irrespective of the underlying equation of state. We estimate that at intermediate rotation frequencies, the change in core density and therefore the relative change in muon abundance, $\Delta N_\mu / N_\mu$ scales as $f^2$, given that this is the scaling of the centrifugal force:
    \begin{align}\label{eq:dNmu-rot}
        \frac{\Delta N_\mu}{N_\mu} &\propto 0.01 \left(\frac{f}{f_K}\right)^2
                  \simeq \num{4e-4} \bigg( \frac{f}{\SI{700}{Hz}} \bigg)^2 \,.
    \end{align}
    Here, $f_K = (2/3)^{3/2} (G_N M_\star/R_\star^3)^{1/2}$, and on the right-hand side we have taken $M_\star = 2 M_\odot$ and $R_\star = \SI{12.1}{\km}$ for the mass and radius of the static non-rotating neutron star, respectively.
  
    Neutron stars radiate off rotational energy over time and spin down. The time scale over which this happens varies widely, between $\sim \SI{100}{yrs}$ and \SI{10}{Gyr} \cite{Manchester:2004bp}. Over these time scales, the equilibrium lepton abundances in the star are therefore expected to experience a slight increase.

    \item {\bf Accretion.} A neutron star in a tight binary systems can accrete substantial amounts of matter from its companion star \cite{%
      Tauris:2011ck,      
      Liu:2011ss,         
      Tauris:2012jp,      
      Li:2021zfx}.        
    In this process, which typically occurs over giga-year timescales, two competing effects influence the electron and muon abundances: on the one hand, the increase in the star's mass leads to an increase in the equilibrium abundances. Simultaneously, accretion can spin up the star, which favors a decrease in the lepton abundances. As discussed above, the latter effect is only a per mille level perturbation, while accretion can easily be at the 10\% level. This leads us to the conclusion that also during accretion-powered spin-up, the equilibrium lepton abundances \emph{increase}. The additional electrons originate simply from the accretion flow, but as the accreted material does not contain muons due to its lower density, the star must produce them internally in significant numbers to maintain equilibrium. We will discuss below the mechanisms through which this can happen, and the limitations to the muon production efficiency.
  
    \item {\bf The magnetic field.} Similar to rotation, also a strong magnetic field leads to a deformation of the star, a decrease of the core density, and a reduction in the electron and muon abundances compared a non-magnetized star \cite{%
    Bandyopadhyay:1997kh, 
    Chakrabarty:1997ef,   
    Broderick:2000pe,     
    Suh:2000ni,           
    Mao:2001cv,           
    Broderick:2001qw,     
    Wei:2005aga,          
    Noronha:2007wg,       
    Rabhi:2009ih,         
    Ferrer:2010wz,        
    Strickland:2012vu,    
    Sinha:2013dfa,        
    Casali:2013jka,       
    Lopes:2014vva,        
    Franzon:2016iai,      
    Dexheimer:2016yqu,    
    Coelho:2016lcf,       
    Gomes:2017zkc,        
    Negreiros:2018cjk,    
    Gomes:2019paw,        
    Thapa:2020ohp,
    Watanabe:2020vas,
    Rather:2021azv,       
    Chatterjee:2021wsr,   
    Rather:2022bmm}.      
    The physics behind the reduced core density is magnetic pressure: the magnetic field resists the increase in energy density that compression would cause. In the extreme magnetic field of a magnetar (up to \SI{e15}{Gauss} at the surface, possibly up to \SI{e18}{Gauss} in the core), this can be a per cent level effect.
  
    We are not aware of any mechanism through which a neutron star's magnetic field can significantly \emph{increase} over time. (It has been noted in ref.~\cite{Price:2006fi} that the magnetic field could rapidly increase to gigantic values -- even larger than in magnetars -- during a binary neutron star merger. We will not consider this scenario here as the merging stars will not have time to equilibrate before collapsing into a black hole. (Binary neutron star mergers might avoid collapse only if the initial masses of the stars are very small and the equation of state is rather stiff, allowing for neutron star masses significantly above $2 M_\odot$.)
  
    Neutron stars born with strong magnetic fields tend to slowly expel them from their core \cite{%
      Goldreich:1992,    
      Thapa:2020ohp,     
      Watanabe:2020vas}. 
    The exact processes responsible for this are not understood in detail, but mechanisms under discussion include Ohmic dissipation (energy loss caused by the currents carrying the magnetic field experiencing Ohmic resistance) and ambipolar diffusion (rearrangement of magnetic field lines that occurs when the magnetic field moves electrons and protons under the constraint of maintaining charge neutrality), possibly aided by Hall drift (electron motion due to the Hall effect, which does not directly dissipate energy, but can amplify Ohmic losses) \cite{Goldreich:1992}. Also the timescale over which magnetic fields decay is a topic of active study
    \cite{%
      Jones:2005km,      
      Elfritz:2015vom,   
      Bransgrove:2017jzs,
      Igoshev:2021ewx}.  
    What is clear is that the details of the process depend heavily on the magnetic field strength. Weaker magnetic fields ($\lesssim \SI{e13}{G}$ in the core) appear to be expelled relatively fast ($< \si{Myr}$) \cite{Bransgrove:2017jzs}, while fields above this threshold may only be expelled over Gyr timescales, if at all \cite{Elfritz:2015vom}. On the other hand, only the strongest magnetic fields (core field strength $\sim \SI{e18}{G}$) lead to appreciable changes in the star's composition \cite{Thapa:2020ohp, Watanabe:2020vas}.

    \item {\bf Tidal deformation.} When a neutron star comes close to another compact star, tidal forces can deform the star in much the same way as the centrifugal force can in case of rapid rotation. Such close encounters can happen for neutron stars in dense clusters of stars, or for stars in tight binary systems.  Taking the analogy between tidal and centrifugal forces further, we can use \cref{eq:dNmu-rot} to estimate the change in the muon abundance due to tidal deformation by calculating, for a given tidal force, the rotation frequency that leads to an equivalent centrifugal force. This implies that the muon abundance in the tidally deformed star differs by
    \begin{align}
        \frac{\Delta N_\mu}{N_\mu} &\simeq \num{1.7e-6} \times \bigg( \frac{M_2}{M_\odot} \bigg)
                                                               \bigg( \frac{d}{\SI{1000}{km}} \bigg)^{-2} \,,
                                             \qquad\quad\,\,\, \text{(general tidal deformation)}
                                                                \label{eq:dNmu-tidal}\\
                                   &\simeq \num{1.7e-6} \times \bigg( \frac{M}{M_\odot} \bigg)^{-1/2}
                                                               \bigg( \frac{\tau}{\SI{10000}{sec}} \bigg)^{-1/2}
                                           \quad \text{(deformation during binary inspirals)} \nonumber
    \end{align}
    from the abundance in an isolated star of the same mass. Once again, the relative change in the electron abundance can be expected to be similar to the relative change in the muon abundance. In \cref{eq:dNmu-tidal}, $M_2$ is the mass of the passing object, and $d$ is the separation of the two stars. In the second line, which is valid for tidal deformation during binary inspirals, we have expressed the distance between the stars in terms of the time to coalescence, $\tau$ \cite{Maggiore:2007ulw}. In doing so, we have assumed both binary partners to have equal mass, $M$. (The full dependence on the individual masses, $M_1$ and $M_2$, is $\Delta N_\mu/N_\mu \propto [M_1/M_2 \times (M_1+M_2)]^{-1/2}$.) \Cref{eq:dNmu-tidal} is based on a Newtonian treatment of gravity. It neglects general relativistic corrections, which reach a few per cent at $d \sim \SI{100}{km}$. Such small separations are reached only during binary inspirals, where they correspond to a time to coalescence of $\tau \lesssim \SI{0.4}{sec}$.
  
    We see that tidal deformation will only appreciably affect the equilibrium lepton abundances in neutron stars that are very close to another compact object, that is, very shortly before a binary coalescence event. The timescales over which the disruption happens in this case are so short that the star has no time to react and re-equilibrate.
\end{enumerate}

\section{Lepton Production, Absorption, Diffusion and Decay.}
\label{sec:mu-diff}

\subsection{General Arguments and Back-of-the-Envelope Estimates}
\label{sec:back-of-envelope}

When the equilibrium lepton abundances, which correspond to the thermodynamically most advantageous configuration, change, the star will try to respond. In young neutron stars ($\lesssim \SI{e6}{yrs}$, with core temperatures $T \sim \SI{100}{keV}$), this will happen dominantly through Standard Model weak interactions, notably through ``modified Urca processes'' of the form $n + n + \ell \leftrightarrow p + n + \nu_\ell$.\footnote{The term ``Urca process'' was allegedly coined by Gamow and Schoenberg in 1939 to emphasize the fact that these processes result ``in a rapid disappearance of thermal energy from the interior of the star, similar to the rapid disappearance of money from the pockets of gamblers at the Casino da Urca [in Rio de Janeiro]'' \cite{Gamow:1970}. Indeed, by absorbing and emitting muons and electrons, Urca processes produce an abundant thermal neutrino flux that constitutes the main energy loss mechanism for a neutron star during the first $\sim \SI{e6}{yrs}$ of its life \cite{Chiu:1964zza, Shapiro:1983du, Haensel:1995,Yakovlev:1995, Yakovlev:2000jp}.} In thermal equilibrium, the rate of these processes scales $\propto G_F^2 m_n^{*3} m_p^* p_{Fp} / (m_\pi^4 \mu_\ell^2) \cdot T^6$, where $G_F$ is the Fermi constant, $m_n^*$ and $m_p^*$ are the effective nucleon masses in the dense neutron star medium, $m_\pi$ is the neutral pion mass, $p_{Fp}$ is the proton's Fermi momentum, and $\mu_\ell$ is the lepton chemical potential ($\ell = e, \mu$). $p_{Fp}$ and $\mu_\ell$ are both of $\mathcal{O}(\SI{100}{MeV})$.  The detailed derivation of the Urca rates, both in thermal equilibrium and out of equilibrium, is given in \cref{sec:abs-prod}, for the equilibrium case see also Refs.~\cite{Yakovlev:1995, Yakovlev:2000jp, Shapiro:1983du}.

The strong temperature scaling of the Urca rate implies that lepton absorption and production through weak interactions become inefficient after several thousand years as the star cools. If the equilibrium abundances of a cold neutron star change by one of the mechanisms outlined in \cref{sec:ns-muons} above, the star will therefore react only very slowly, and it will spend some time out of chemical equilibrium. One exception is the case where the equilibrium muon abundance decreases, but the equilibrium electron abundance increases, or decreases less than the equilibrium muon abundance. In this case, the star may react quickly, but not only via Urca absorption. Rather, two additional processes become relevant: muon diffusion and decay.

In fact, with absorption via Urca processes suppressed, muons are able to diffuse over macroscopic distances in old stars, allowing them to eventually leave the core if doing so is thermodynamically favorable (see below for a discussion of the conditions under which this is the case). Once outside the core, they become unstable and decay. More precisely, muon decay becomes allowed once the electron Fermi momentum, $p_{Fe}$, drops below $m_\mu/2$. (A Michel electron energy of $m_\mu/2$ corresponds to the extremal kinematic configuration where the electron recoils against two collinear neutrinos.) These decays produce a flux of neutrinos which will be the main focus of \cref{sec:nu-fluxes} below.  The Michel electron will then diffuse back into the core to restore charge neutrality.

We can estimate the mean free path of muons and electrons inside a neutron star by starting from their thermal conductivity $\kappa_\ell$, which in turn is an output of established neutron star cooling calculations. We use in particular the public code \nscool \cite{Page:2006ud, Page:2005fq, 1998nspt.conf..183P}.  The thermal conductivity is related to the lepton relaxation time $\tau_\ell$ by~\cite{Shternin:2007ee}
\begin{align}
    \kappa_\ell = \frac{\pi^2 T n_\ell \tau_\ell}{3 p_{F \ell}} \,.
    \label{eq:kappa-mu}
\end{align}
where $T$ is the local temperature, $n_\ell$ the lepton number density, and $p_{F\ell}$ the lepton Fermi momentum.  From the relaxation time, we obtain the mean free path
\begin{align}
    \lambda_\ell &= \tau_\ell v_{F\ell} \,,
    \label{eq:lambda-mu}
\end{align}
with $v_{F\ell}$ the leptons' Fermi velocity. (For $\ell = e$, this will be essentially the speed of light.) We use $v_{F\ell}$ as an estimate for the average velocity, taking into account that only leptons close to their respective Fermi surfaces are sufficiently mobile to contribute to the thermal conductivity. As $\lambda_\ell$ is roughly inversely proportional to $T$, it increases from around \SI{e-8}{cm} hours after the supernova explosion to \SI{e-6}{cm} after \SI{e5}{yrs}. For a lepton undergoing a random walk the average net travel distance after a time interval $\Delta t$ is then
\begin{align}
    \ev{\Delta x} = \sqrt{\Delta t \, v_{F\ell} \lambda_\ell} \,,
    \label{eq:delta-x}
\end{align}
For $v_{F\ell} \sim \mathcal{O}(1)$, we see that diffusion over $\mathcal{O}(\si{km})$ distances may be possible over timescales of order years.

Let us now discuss in more detail electrostatic effects. When a muon or electron diffuses away from the core of a neutron star, it leaves behind a proton or $\Sigma^+$ hyperon. Hadrons are more tightly bound in the neutron star lattice and therefore cannot follow the muon---they can only be destroyed in Urca reactions.  If the charge imbalance becomes too large, it will prevent further lepton diffusion. Therefore, every electron or muon diffusing from the core of the star to the outer layers needs to be replaced by another lepton diffusing the opposite way. For electrons, this implies that there cannot be any net outward diffusion. For muons diffusing away from the core, however, charge neutrality may be restored if the Michel electrons produced in muon decay travel back inside. This will indeed happen if the star is in an out-of-equilibrium state where the abundance of muons in the core is larger than the equilibrium abundance, and the abundance of electrons is smaller. We are, unfortunately, not aware of an astrophysical scenario where this is the case. What can occur, though, is a situation where the number of excess muons is larger than the number of excess electrons. In fact, in typical configurations where the equilibrium Fermi energies of both lepton species, $E_{F\mu}^\text{eq}$ and $ E_{Fe}^\text{eq}$, are the same, a small decrease in $E_{F\mu}^\text{eq} = E_{Fe}^\text{eq}$ translates into a larger reduction in the muon number density than in the electron number density. Then, muon outward diffusion and decay, combined with inward diffusion of Michel electrons, can bring an out-of-equilibrium star into a thermodynamically more favorable state, though it cannot fully restore chemical equilibrium. The latter has to be achieved through Urca reactions, which, as we will show quantitatively in \cref{sec:mu-diff}, are typically much slower.

Regarding muon decays, note that the requirement $p_{Fe} < m_\mu/2$ may be avoided in presence of to a spectator nucleon which absorbs excess momentum,
\begin{align}
    p + \mu^- \to p + e^- + \nu_\mu + \bar\nu_e \,.
    \label{eq:assisted-mu-decay}
\end{align}
This process, which we call ``assisted muon decay'', may allow muons to decay further inside the star than without a spectator. We consider only protons as spectators here as the coupling between the muon and the spectator is via photon exchange. A na\"ive back-of-the-envelope estimate for the rate of assisted muon decay is
\begin{align}
    \Gamma_\text{amd} &\sim \frac{\alpha^2 G_F^2 T^8 k_{Fe}^2 k_{Fp}^2}{m_\mu^7} \,, \nonumber\\
    &\simeq \SI{5.7e-27}{sec^{-1}} \times 
    \bigg( \frac{k_{Fp}}{\SI{250}{\MeV}} \bigg)^2
    \bigg( \frac{k_{Fe}}{\SI{200}{\MeV}} \bigg)^2
    \bigg( \frac{T}{\SI{e8}{K}} \bigg)^8 \,.
    \label{eq:Gamma-mu-ass-estimate}
\end{align}
The factor $\alpha^2 G_F^2$ in this estimate describes the electromagnetic coupling of the muon or electron to the spectator nucleon as well as the weak interaction destroying the muon. The remaining factors originate from considering the 12-dimensional phase space; a factor $T^6$ arises from the neutrino momenta, which are of order $T$ in regions with severe Pauli blocking ($p_{Fe} \sim p_{F\mu} \gtrsim m_\mu$). For the electron and proton we acquire factors $k_{Fe}^2 T$ and $k_{Fp}^2 T$,  respectively from scattering around their Fermi surfaces. The remaining factor $m_\mu^7$ restores the correct dimensions for the rate. Assisted muon decay will be discussed in more detail in \cref{sec:noneq-absorption} below, and a rigorous derivation of its rate will be given in \cref{sec:abs-prod}.\Cref{eq:Gamma-mu-ass-estimate} shows that assisted muon decay is severely impeded compared to free muon decay. Nevertheless, it may still be relevant in regions where the latter process is Pauli-blocked. 

Let us estimate the neutrino flux from a neutron star losing a significant fraction of the muons in its core via diffusion and decay:
\begin{align}
  \phi_\nu
    &\simeq \frac{\Delta N_\mu}{4 \pi d^2 \tau_\text{NS}} \notag\\
    &\sim \SI{1.3e-10}{cm^{-2}\,sec^{-1}}
          \bigg( \frac{\Delta N_\mu / N_\mu}{0.001} \bigg)
          \bigg( \frac{d}{\SI{10}{kpc}} \bigg)^{-2}
          \bigg( \frac{\SI{1}{Gyr}}{\tau_\text{NS}} \bigg) \,.
    \label{eq:phi-nu}
\end{align}
Here, $\Delta N_\mu$ is the number of muons lost, $\tau_\text{NS}$ is the time scale over which this loss happens, and $d$ is the distance to the neutron star. In the second line, we have replaced $\Delta N_\mu$ by the relative number of expelled muons, $\Delta N_\mu / N_\mu$ assuming a total muon abundance of \num{5e55}. (The latter number varies by an $\mathcal{O}(1)$ factor depending on the mass of the neutron star and the equation of state.) Clearly, this neutrino flux is too small to be observable for the foreseeable future. At $\sim \mathcal{O}(\SI{10}{MeV})$ energies, experiments are just now reaching flux sensitivities $\sim \SI{1}{cm^{-2}\,sec^{-1}}$, corresponding to the expected flux of diffuse supernova neutrinos \cite{Beacom:2010kk}. But at sub-MeV energies, no experiment with even comparable sensitivity is on the horizon -- and even if it was, it would have to contend with the large solar neutrino background.

Detection prospects would be significantly better if \emph{all} \num{e8}--\SI{e9} neutron stars in the Milky Way \cite{Sartore:2010} were to suffer substantial muon loss simultaneously. This would be the case if spin-down led to a \emph{decrease} rather than an increase in the equilibrium muon abundance. In fact, the results from ref.~\cite{Watanabe:2020vas} suggest that this is the case for certain equations of state (notably the NL3$\omega\rho$ equation of state~\cite{Horowitz:2000xj} in regular neutron stars, and the DD-ME2 equation of state \cite{Lalazissis:2005} in magnetars). However, we believe this conclusion to be spurious as we are not aware of any physical mechanism that would explain why the volume of a star should increase as it spins down.

\subsection{Out-of-Equilibrium Lepton Production}
\label{sec:noneq-production}

We have argued in \cref{sec:ns-muons} above that the equilibrium electron and muon abundances in a neutron star can change due to changes in its rotational velocity or magnetic field, or through accretion or tidal deformation. We have conjectured that in many of these dynamic scenarios, the star will not be able to respond immediately and will therefore remain in an out-of-equilibrium configuration for an extended period of time. Let us now put this statement on a more quantitative foundation by studying the rates of out-of-equilibrium lepton production, absorption, and decay processes.  Detailed derivations of these rates are given in \cref{sec:abs-prod}; here we only quote the results and discuss their physical implications.

For lepton production, it is convenient to parameterize the deviation from equilibrium through the quantity
\begin{align}
    B \equiv \beta (\mu_p - \mu_n + \mu_\ell) \,,
    \label{eq:B}
\end{align}
with $\beta = 1/T$ and $\ell = e, \mu$, using the fact that, in beta-equilibrium the muon, electron, neutron, and proton chemical potentials satisfy $\mu_\mu = \mu_e = \mu_n - \mu_p$. The lepton production rate per unit volume via the direct Urca process $n \to p + \ell^- + \bar\nu_\ell$ is then found to be (see \cref{eq:Gamma-DU-prod-final,eq:I-DU-prime})\footnote{By assuming interactions with individual nucleons, we neglect the possible impact of nucleon superfluidity in the core of the neutron star. We expect superfluidity to suppress Urca rates, but the out-of-equilibrium corrections we are interested in here would remain unchanged. In other words, the ratio of the out-of-equilibrium and equilibrium rates is expected to be largely independent of these corrections.}
\begin{align}
    \Gamma_\text{DU}^\text{prod}
        &= \frac{m_n^{\star} m_p^\star \mu_\ell \Theta_\text{np$\ell$}}{4\pi^5 \beta^5} G_F^2 \cos^2 \theta_C 
           (1 + 3 g_A^2) I^\prime_\text{DU} \,,
    \label{eq:Gamma-DU-prod-maintext}
    \intertext{with}
    I^\prime_\text{DU}
        &= -\frac{B^2\pi^2}{2} \left( B + \log[1+e^{-B}] - \log[1+e^B] \right)
                                   - (B^2 + \pi^2) \Li_3(z) - 6 B \Li_4(z) 
                                   - 12 \Li_5(z) \,.
    \label{eq:I-DU-prime-maintext}
\end{align}
In \cref{eq:Gamma-DU-prod-maintext}, $m_n^\star$ and $m_p^\star$ are the neutron and proton effective masses, respectively, which take into account the effects of many-body interactions on the particles' dispersion relations, $G_F$ is the Fermi constant, $\theta_C$ the Cabibbo angle, and $g_A \simeq 1.26$ the axial coupling of the nucleon. The Heaviside function $\Theta_\text{np$\ell$} \equiv \Theta(|\vec{p}_2| - \left| |\vec{p}_3| - |\vec{p}_1|\right|)\, \Theta(|\vec{p}_3| + |\vec{p}_1| - |\vec{p}_2|)$ enforces the \emph{triangle inequality} $p_{Fn} \leq p_{Fp} + p_{F\ell}$, which is a precondition for direct Urca production to occur and is typically only satisfied in the cores of the heaviest neutron stars ($\gtrsim 2 M_\odot$).  The functions $\Li_n(z)$ appearing in $I^\prime_\text{DU}$ are polylogarithms, while $z = - \text{exp}(-B)$.

As detailed in \cref{sec:abs-prod}, the strategy for arriving at \cref{eq:Gamma-DU-prod-maintext} is to first neglect the neutrino momentum compared to the nucleon and charged lepton momenta, to neglect any anisotropy in the matrix elements, and then to first evaluate the angular part of the phase space integral. The energy integrals are evaluated afterwards, making use of the residue theorem multiple times. Throughout the derivation, it is assumed that the nucleon and charged lepton energies are close to the respective Fermi surfaces of these particle, that is, that the star is not too hot. 

Following this strategy, we can also compute the rate of out-of-equilibrium modified Urca processes of the form $n + n \to n + p + \ell^- + \bar\nu_\ell$. Here, the spectator nucleon can absorb momentum, thus rendering this process viable also in light neutron stars.  Our result is
\begin{align}
    \Gamma_\text{MU}^\text{prod}
        &= \frac{G_F^2 g_A^2 m_n^{\star3} m_p^\star \,p_{Fp}}{\pi^9\beta^7}
           \frac{p_{F\ell}}{\mu_\ell}
           \alpha_n \beta_n \bigg( \frac{g_{\pi N N}}{m_\pi} \bigg)^4
           I^\prime_\text{MU} \,,
    \label{eq:Gamma-MU-prod-maintext}
    \intertext{with}
    I^\prime_\text{MU}
        &= -\frac{3}{8} B^2 \pi^4 \left(B + \log[1+e^{-B}] - \log[1+e^B]\right)
           - \frac{1}{12}(B^2 + \pi^2)(B^2 + 9\pi^2) \Li_3(z) \notag\\
        &\qquad - B (B^2 + 5\pi^2) \Li_4(z)
                - 2 (3 B^2 + 5\pi^2) \Li_5(z)
                - 20 B \Li_6(z)
                - 30 \Li_7(z) \,.
    \label{eq:I-MU-prime-maintext}
\end{align}
The notation used here is the same as in \cref{eq:Gamma-MU-prod-maintext} above, with the additional appearance of the proton and lepton Fermi momenta $p_{Fp}$ and $p_{F\ell}$, respectively, with the neutral pion mass $m_\pi$, with the factor $g_{\pi N N} \simeq 1$ parameterizing the pion--nucleon vertex, and with
\begin{align}
    \alpha_n \simeq 1.76 - 0.634 (n_0/n_n)^{2/3} \,,
    \qquad\qquad
    \beta_n \simeq 0.68  
    \label{eq:alpha-beta}
\end{align}
(see \cref{sec:matrix-elements} and ref.~\cite{Friman:1979ecl, Yakovlev:1995} for details). In the expression for $\alpha_n$, $n_n$ is the neutron number density, and $n_0 \simeq \SI{0.16}{fm^{-3}}$ is the canonical nuclear density. The prime on $I^\prime_\text{MU}$ indicates that this integral is closely related to, but different from, the integral $I$ introduced in refs.~\cite{Friman:1979ecl, Yakovlev:1995} in the calculation of the Urca \emph{luminosity}.

\begin{figure}
    \centering
    \begin{tabular}{cc}
        \includegraphics[width=0.48\textwidth]{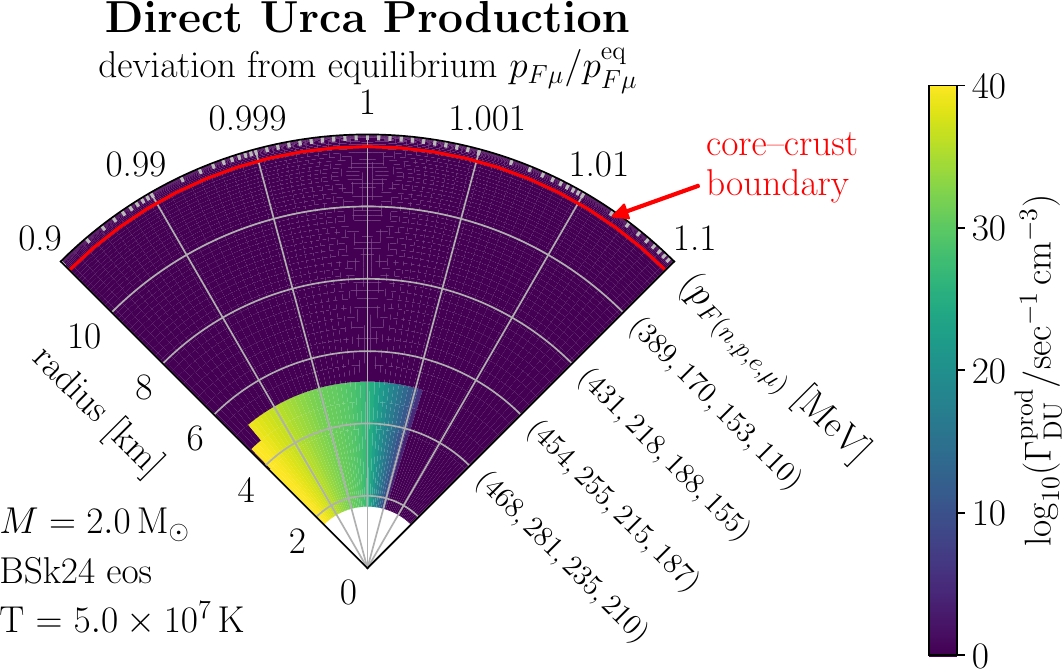} \quad &
        \includegraphics[width=0.48\textwidth]{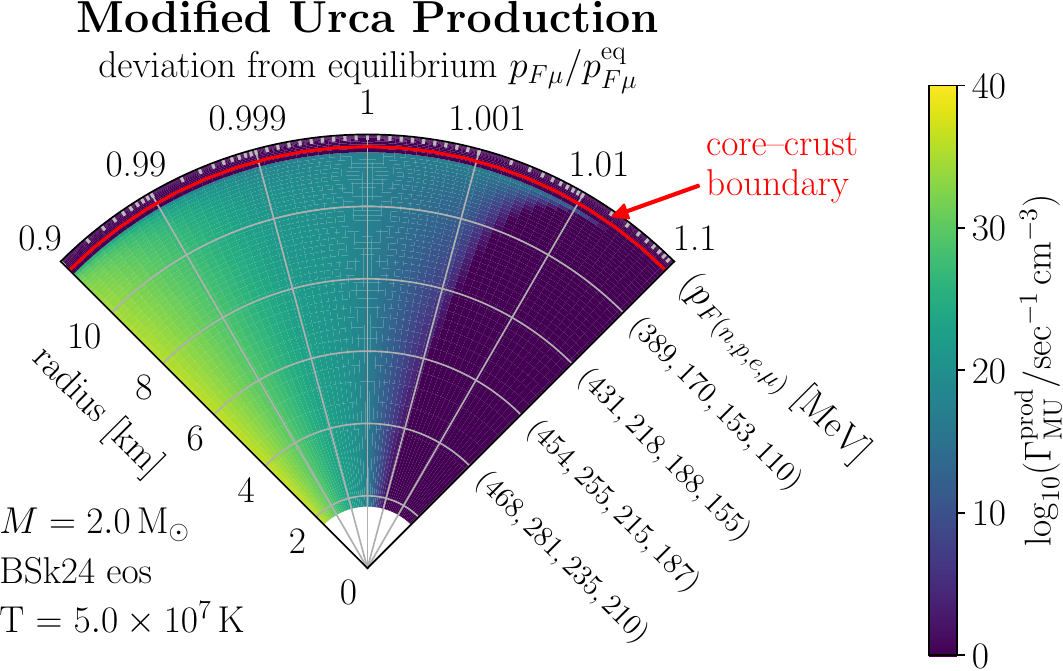}
    \end{tabular}
    \caption{Direct \emph{(left)} and modified \emph{(right)} Urca muon production rate as a function of the radial location in the star (radial axis) and of the departure from equilibrium (azimuthal axis). We have assumed a \SI{2.0}{M_\odot} neutron star, heavy enough to sustain direct Urca production in its core. For this plot, we have made the (in general inaccurate) assumption of a constant temperature throughout the star. For electrons (not shown here), the direct Urca production rate is identical to the one for muons, with the obvious replacement $p_{F\mu}, p_{F\mu}^\text{eq} \to p_{Fe}, p_{Fe}^\text{eq}$, since $\mu_e = \mu_\mu$; the modified electron Urca production rate differs by a factor $p_{Fe}^\text{eq} / p_{F\mu}^\text{eq} \sim \mathcal{O}(1)$ from the one shown here, see \cref{eq:Gamma-MU-prod-maintext}.}
    \label{fig:Urca-prod-pizza-plot}
\end{figure}

In \cref{fig:Urca-prod-pizza-plot}, we show the direct and modified muon production rates as a function of radial location for an exemplary neutron star at a core temperature of \SI{5e7}{Kelvin}, which is reached after about 100~years. Results for electrons are similar. Through the polylogarithms in \cref{eq:I-DU-prime-maintext,eq:I-MU-prime-maintext}, the muon production rate depends very strongly on the departure from beta-equilibrium. Already a sub-per-mille-level deviation of $B$ from unity leads to muon production rates that are increased ($B < 1$) or decreased ($B > 1$) by several orders of magnitude. Note also the peculiar, but not unexpected, behavior of the direct Urca rate. First, it is non-zero only in the very inner region of the star. Moreover, for a particular range of radii, direct Urca production is allowed close to beta equilibrium, but becomes impossible further away.

Note also that \cref{eq:Gamma-DU-prod-maintext,eq:Gamma-MU-prod-maintext} exhibit a very strong temperature dependence. In beta equilibrium, this dependence is due to the high negative power of $\beta$ appearing in the prefactor. Away from equilibrium, also the factor $\beta$ in the definition of $B$ is relevant, which implies that at low temperature, the Urca rates are sensitive to smaller deviations of $\mu_\ell$ from its equilibrium value than at high temperatures. On the other hand, the overall rates at low $T$ are lower because of the prefactor.

\subsection{Out-of-Equilibrium Lepton Absorption and Decay}
\label{sec:noneq-absorption}

\begin{figure}
    \centering
    \begin{tabular}{cc}
        \includegraphics[width=0.48\textwidth]{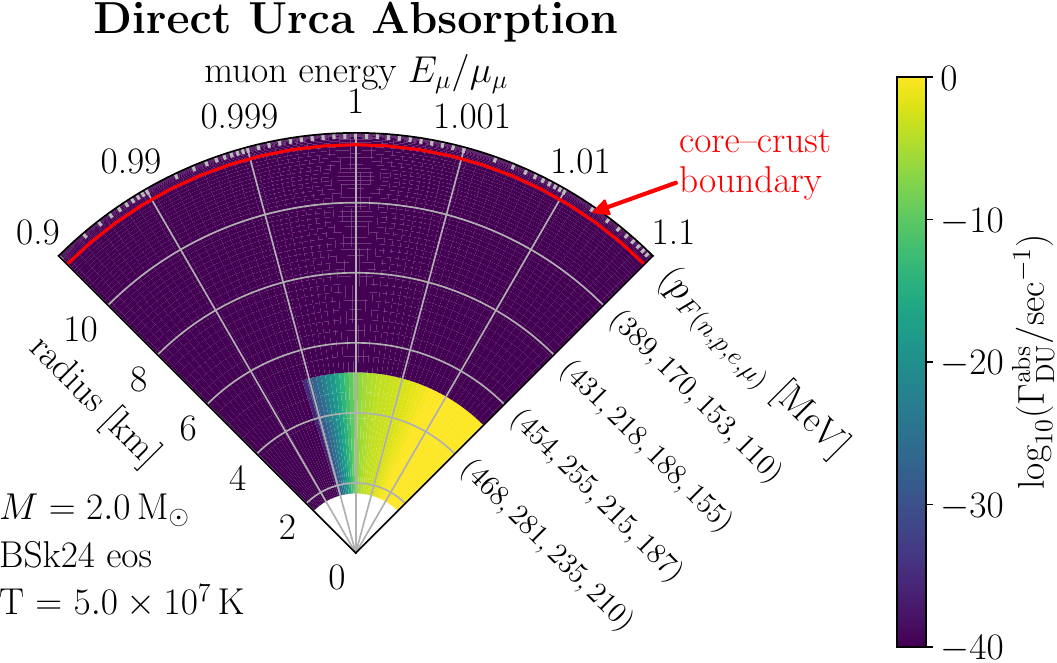} \quad &
        \includegraphics[width=0.48\textwidth]{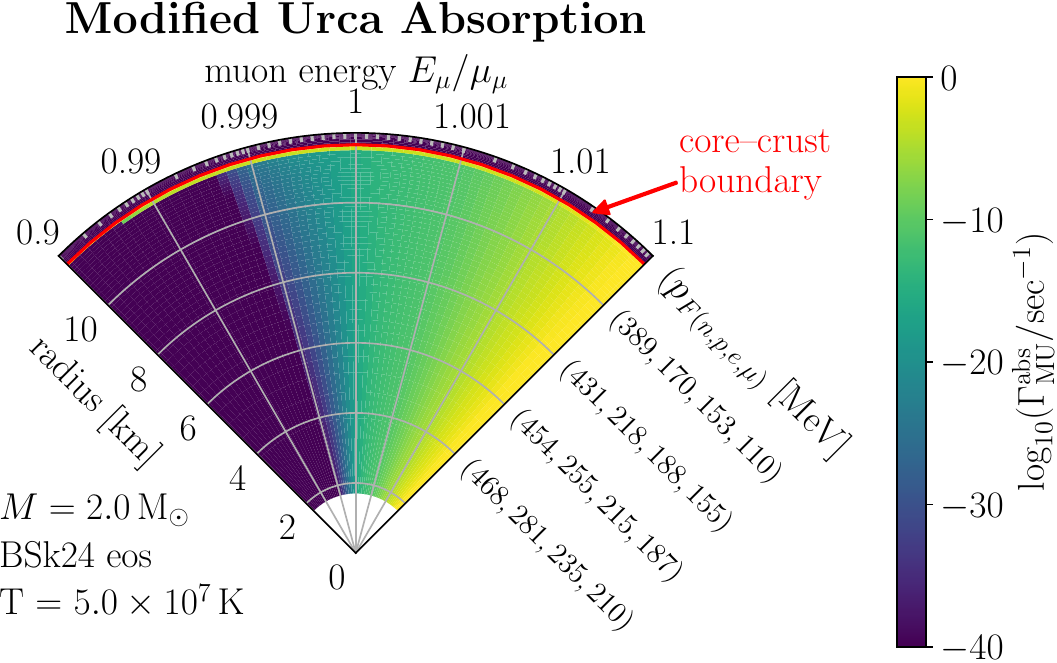} \\[0.5cm]
        \multicolumn{2}{c}{\includegraphics[width=0.48\textwidth]{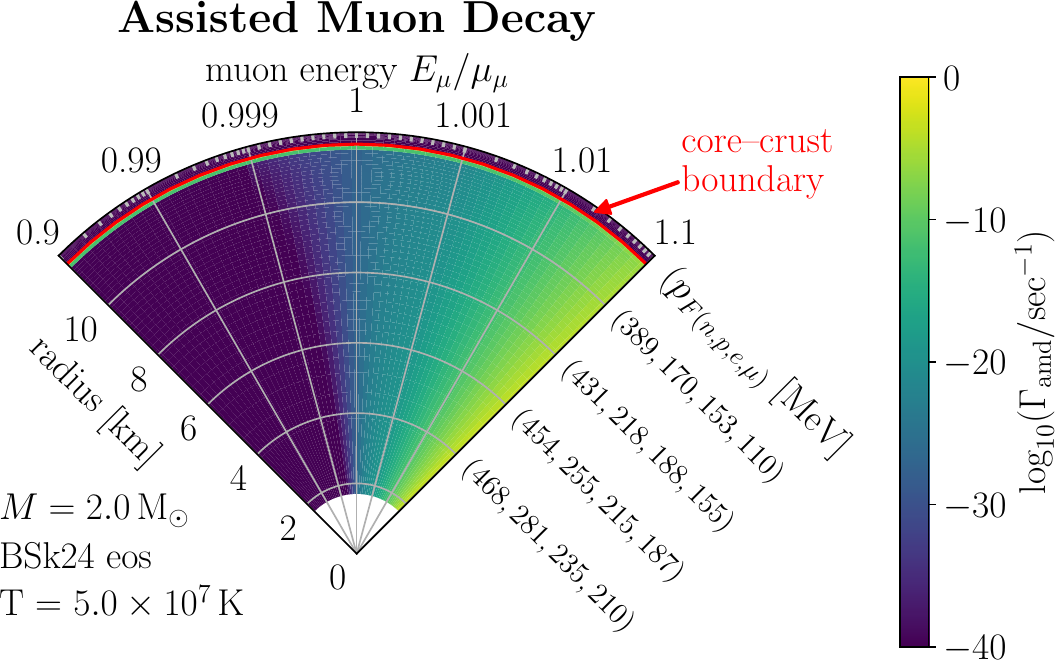}}
    \end{tabular}
    \caption{Direct \emph{(top left)} and modified \emph{(top right)} Urca muon absorption rates, as well as assisted muon decay rate \emph{(bottom)} as a function of the radial location in the star (radial axis) and of the muon energy relative to the Fermi energy (azimuthal axis). We have assumed a \SI{2.0}{M_\odot} neutron star, heavy enough to sustain direct Urca processes in its core. For this plot, we have made the (in general inaccurate) assumption of a constant temperature throughout the star.  For electrons (not shown here), the direct Urca absorption rate is identical to the one for muons, with the obvious replacement $E_\mu, \mu_\mu \to E_e, \mu_e$; the modified electron Urca absorption rate differs by a factor $p_{Fe} / p_{F\mu} \sim \mathcal{O}(1)$ from the one shown here, see \cref{eq:Gamma-MU-abs-maintext}. Assisted decay exists only for muons.}
    \label{fig:Urca-abs-pizza-plot}
\end{figure}

In full analogy to lepton production, we can also calculate the lepton absorption rate away from beta equilibrium. In this case, for a lepton of energy $E_\ell$, the deviation from beta equilibrium is parameterized as
\begin{align}
    A \equiv \beta (\mu_n - \mu_p - E_\ell) \,.
    \label{eq:A}
\end{align}
The direct Urca absorption rate is then
\begin{align}
    \Gamma_\text{DU}^\text{abs}
        &= \frac{m_n^\star m_p^\star\Theta_\text{np$\ell$}}{4\pi^3 \beta^4  p_{F\mu}} G_F^2 \cos^2 \theta_C (1 + 3 g_A^2)    I_\text{DU}^{\prime\prime} \,,
    \label{eq:Gamma-DU-abs-maintext}
    \intertext{with}
     I_\text{DU}^{\prime\prime} &= 2 \left[A \, \Li_3(e^{-A}) + 3 \, \Li_4(e^{-A}) \right] \,,
     \label{eq:I-DU-pprime-maintext}
\end{align}
and with the same notation as in \cref{sec:noneq-production}. Similarly, the modified Urca absorption rate is
\begin{align}
    \Gamma_\text{MU}^\text{abs}
        &= \frac{G_F^2 \cos^2\theta_C g_A^2 m_n^{\star3} m_p^\star \,p_{Fp}}{\pi^7 \beta^6 E_\mu^2} \alpha_n \beta_n
           \bigg( \frac{g_{\pi N N}}{m_\pi} \bigg)^4 I_\text{MU}^{\prime\prime} \,,
    \label{eq:Gamma-MU-abs-maintext}
\intertext{with}
    I_\text{MU}^{\prime\prime} &= \frac{1}{3}
        \Big[ (A^3 + 4\pi^2 A) \Li_3(e^{-A})
            + 3 (3 A^2 + 4\pi^2) \Li_4(e^{-A})
            + 36 A \Li_5(e^{-A})
            + 60 \Li_6(e^{-A}) \Big] \,.
    \label{eq:I-MU-maintext}
\end{align}
Finally, for assisted muon decay, we find
\begin{align}
    \Gamma_\text{amd}
        &= \frac{16 (m_p^*)^2 \mu_e E_\mu^2 \alpha^2 G_F^2}
                {\pi^5 \beta^8 m_\mu^8} I_\text{amd} \,,
    \label{eq:Gamma-amd-maintext}
\intertext{with the definitions}
    I_\text{amd}
        &= - 2 \, (C^2 + \pi^2) \, \Li_6(-e^{-C})
           - 24 \, A \, \Li_7(-e^{-C})
           - 84 \, \Li_8(-e^{-C}) \,.
    \label{eq:I-amd-maintext}
\intertext{and}
    C &= \beta (\mu_e - E_\mu) \,.
\end{align}
We show these rates in the ``pizza-slice'' plots of \cref{fig:Urca-abs-pizza-plot} for $\ell = \mu$. Once again, results for the electron Urca rates will be very similar to the ones for muons; of course, assisted decay is possible only for muons. As for Urca production (\cref{fig:Urca-prod-pizza-plot}), we observe a very strong dependence of the rates on the deviation from equilibrium -- note that the color scale in \cref{fig:Urca-abs-pizza-plot} covers 40 orders of magnitude. And once again, direct Urca processes are only allowed in the very core of the star.

\subsection{Interplay of Absorption, Diffusion, and Decay}
\label{sec:interplay}

\begin{figure}
    \centering
    \includegraphics[width=\textwidth]{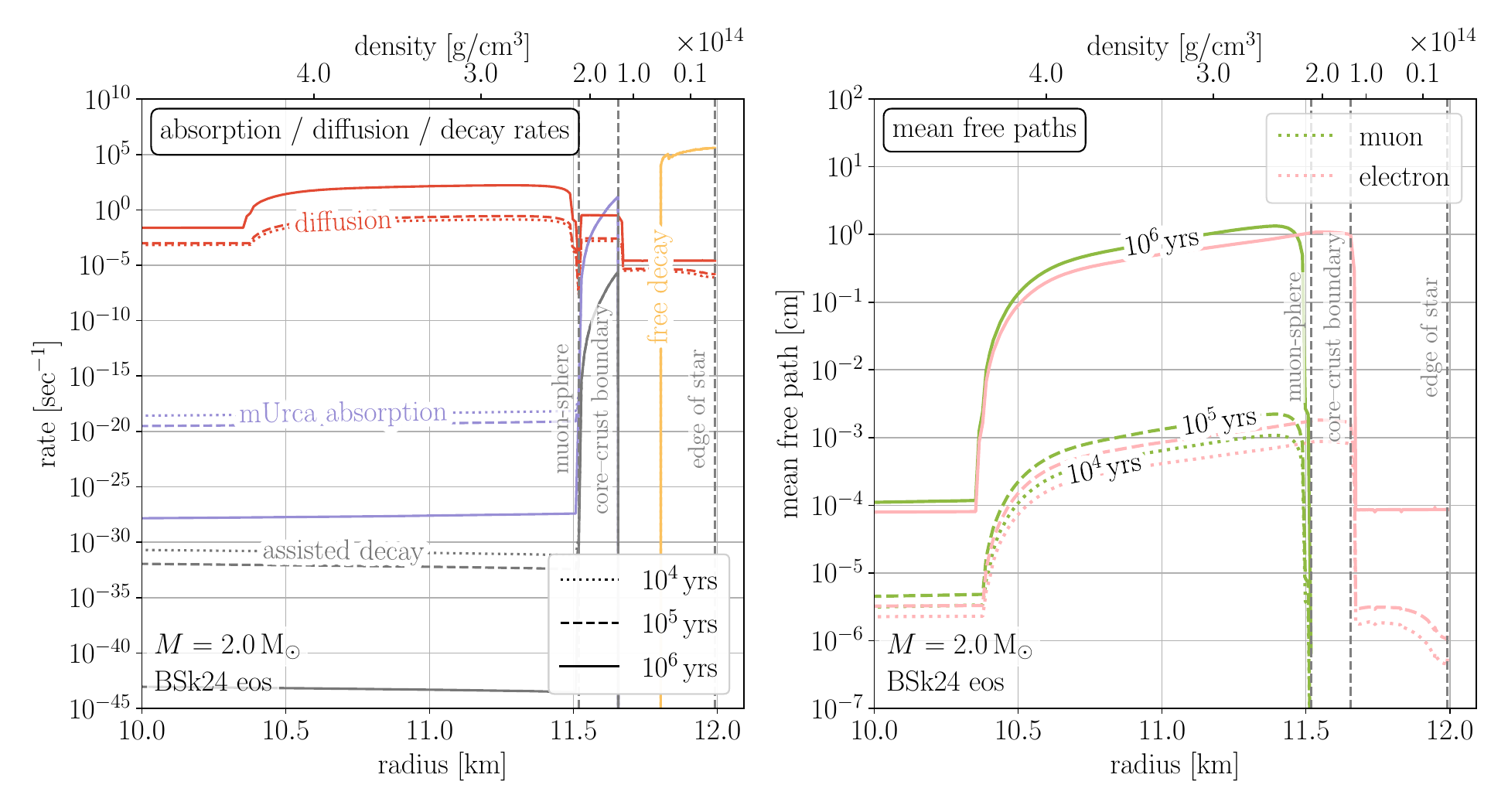}
    \caption{\emph{Left:} muon absorption, diffusion, and decay rates in the outer layers of a neutron star. \emph{Right:} electron and muon mean free paths. We assume a $2 M_\odot$ neutron star with the BSk24 equation of state, and we show results for three different ages, namely $\SI{e4}{yrs}$ (dotted), $\SI{e5}{yrs}$ (dashed), and $\SI{e6}{yrs}$ (solid). We also indicate the edge of the muon-sphere, the core--crust boundary, and the edge of the star. \emph{Beyond the muon-sphere, diffusing muons pass through a region of strong absorption, before reaching layers in which their decay is allowed.}}
    \label{fig:rates-mfps}
\end{figure}

For muons, the interplay between different loss processes, namely absorption, diffusion, and decay is particularly complex in the outer layers of a neutron star, where the macroscopic properties of the background matter change rapidly.  We therefore investigate this region in more detail in \cref{fig:rates-mfps}. The left panel in this figure shows the muon absorption, diffusion and decay rates around the muon-sphere (the radius beyond which the equilibrium muon abundance vanishes), the core--crust boundary (defined as in \nscool\ as the radius at which the density drops below a certain threshold), and the edge of the star (which we define here for simplicity as the radius beyond which there are no more neutrons). The right panel shows the electron and muon mean free paths, which are important in determining the muon diffusion efficiency.

We clearly observe qualitatively different behavior in the different regions delineated by these boundaries:
\begin{itemize}
    \item {\bf Deep inside the star}, muons can diffuse fairly quickly, and at the late times considered here unimpeded by absorption or decay. Note that the diffusion rate in \cref{fig:rates-mfps} is somewhat arbitrarily defined as the time it takes a muon to travel on average \SI{100}{m} according to \cref{eq:delta-x}. The increase in the diffusion rate at $r \simeq \SI{10.3}{km}$ is related to an increase in the muon thermal conductivity at that radius, as determined by \nscool.

    \item {\bf Beyond the muon-sphere}, the Urca absorption and assisted decay rates shoot up because any muon finding itself in this region that has no background muons is by definition out-of-equilibrium. And we have seen above that the out-of-equilibrium absorption and assisted decay rates can be enormous. The diffusion rate changes slightly as well, but this is to some extent related to the fact that we have to change the way we calculate it. Without the notion of a muon thermal conductivity $\kappa_\mu$ in this region, \cref{eq:kappa-mu} can no longer be used to determine the muon mean free path, so instead we use the mean free path of electrons. This is conservative as it tends to underestimate the true muon mean free path.

    \item {\bf The core--crust boundary} marks the radius beyond which absorption becomes irrelevant due to the steep drop in the density of spectator nucleons (see, however, ref.~\cite{Gusakov:2004mj}). The density of protons even goes to zero, and even though there is a residual possibility of muons being absorbed by interacting with protons bound in nuclei, we neglect this possibility due to the relatively low density. Instead, we set the Urca absorption rate and the assisted muon decay rate to zero where $p_{Fp} = 0$. This implies that beyond the core--crust boundary, there is a region where muon diffusion is largely unimpeded, before decaying via free decay once they reach a point where the electron Fermi momentum drops sufficiently low.
\end{itemize}
We conclude that a muon wishing to escape the confines of the neutron star encounters an absorption barrier between the muon-sphere and the core--crust boundary, which it will only be able to overcome in old and cold stars, where the diffusion rate is largest. Once beyond the core--crust boundary, however, its decay is unhindered.

\subsection{Integrated Rates}
\label{sec:integ-rates}

\begin{figure}
    \centering
    \includegraphics[width=0.6\textwidth]{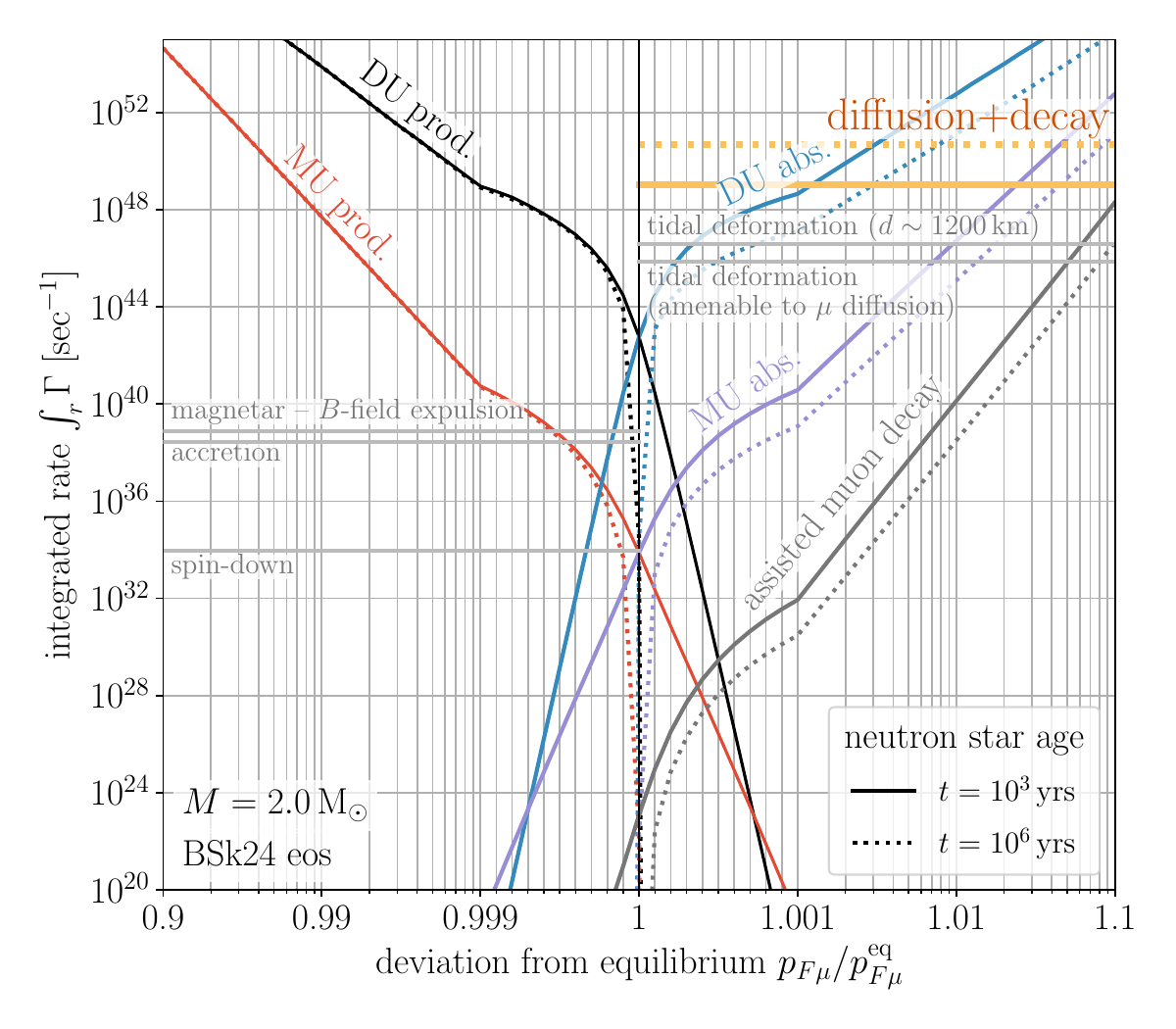}
    \caption{The integrated muon production, absorption, and decay rates from \cref{eq:Gamma-prod-integ,eq:Gamma-abs-integ,eq:Gamma-amd-integ} as a function of the deviation from equilibrium. We also show the rate at which muons can diffuse out of the neutron star. As expected, direct Urca (DU) rates are much larger than modified Urca (MU) rates where they are kinematically allowed. Assisted muon decay is always subdominant. In equilibrium ($p_{F\mu} = p_{F\mu}^\text{eq}$, indicated by the vertical gray line), the production and absorption rates are identical as expected.  For an older and colder neutron star (dotted lines), the equilibrium rates are much smaller than for a young star (solid lines). Horizontal gray lines indicate the production rate needed to maintain the star in equilibrium for several processes driving it away from equilibrium. While this plot is for muons, Urca rates for electrons are almost identical, differing at most by $\mathcal{O}(1)$ factors. Assisted decay is of course impossible for electrons.}
    \label{fig:Gamma-integ}
\end{figure}

To get a more global view on the processes affecting the electron and muon populations inside a neutron star, it is instructive to consider the radially integrated production, absorption, and decay rates,
\begin{align}
    \Gamma_\text{DU,MU}^\text{prod,integ} &\equiv \int\! {\diff r} \, 4 \pi r^2 \, \Gamma_\text{DU,MU}^\text{prod}
        &\qquad\text{electrons + muons}
        \label{eq:Gamma-prod-integ} \\
    \Gamma_\text{DU,MU}^\text{abs,integ}  &\equiv \int\! {\diff r} \, 4 \pi r^2
                                             \int\! \frac{2 \times 4\pi E_\ell p_\ell {\diff E_\ell}}
                                                         {(2\pi)^3 [e^{\beta(E_\ell - \mu_\ell)} + 1]} \,
                                             \Gamma_\text{DU,MU}^\text{abs}
        &\qquad\text{electrons + muons}
        \label{eq:Gamma-abs-integ} \\
    \Gamma_\text{amd}^\text{integ}   &\equiv \int\! {\diff r} \, 4 \pi r^2
                                             \int\! \frac{2 \times 4\pi E_\mu p_\mu {\diff E_\mu}}
                                                         {(2\pi)^3 [e^{\beta(E_\mu - \mu_\mu)} + 1]} \,
                                             \Gamma_\text{amd} \,.
        &\qquad\text{muons only}
        \label{eq:Gamma-amd-integ}
\end{align}
For the production rates, we simply integrate over the volume of the star (as $\Gamma_\text{DU,MU}^\text{prod}$ is already integrated over lepton energies), while for the absorption and decay rates, we integrate over both spatial coordinates and lepton energies, properly weighted with the leptons' Fermi--Dirac distribution function. The extra factor of two in \cref{eq:Gamma-abs-integ,eq:Gamma-amd-integ} accounts for the two lepton spin polarizations.

The integrated muon production, absorption, and diffusion/decay rates are plotted in \cref{fig:Gamma-integ} as a function of $p_{F\mu} / p_{F\mu}^\text{eq}$, which parameterizes the departure from equilibrium. As before, results for electrons are very similar, except for the absence of assisted decay. As in \cref{fig:Urca-prod-pizza-plot,fig:Urca-abs-pizza-plot}, we find again that the rates very sensitively depend on the deviations from equilibrium, which agrees with our expectations. This dependence is even more pronounced for cold neutron stars (dotted lines) than for hotter stars (solid lines): in a cold star, production (absorption/decay) shuts off completely when the muon Fermi momentum is only slightly above (below) its equilibrium value. At higher temperatures, the Fermi surface is more smeared out, softening this behavior. Note that in equilibrium, production and absorption balance each other, so that the net muon abundance remains constant. This observation serves as a useful cross-check of our results.

We also note that assisted decay is always subdominant compared to Urca processes and compared to diffusion followed by free decay (except at extreme deviations from equilibrium). This is due to the extremely strong temperature dependence of the assisted decay rate, and due to the $\alpha^2$ suppression -- in assisted decay, the coupling to the spectator is via a photon, whereas in modified Urca processes it is mediated by a pion.

We also show in \cref{fig:Gamma-integ} an estimate for the rate at which muons diffuse out of the neutron star. We obtain this estimate by computing the time $\Delta t$ it takes a muon to travel from the center of the star to the surface using \cref{eq:delta-x}, assuming the muon mean free path $\lambda_\mu$ to be constant at its median value throughout the star. We then plot $N_\mu / \Delta t$, where $N_\mu$ is the total number of muons in the star. We see that diffusion and decay is a more efficient muon loss mechanism than Urca processes, except at very large deviations from equilibrium.

\Cref{fig:Gamma-integ} allows us to gauge the deviations from equilibrium that can be expected in realistic situations. In particular, the horizontal gray lines indicate, for a few benchmark cases, the rate at which the equilibrium muon abundance changes while the star is being tidally deformed, expels its magnetic field, accretes matter from a companion star, or spins down. To compute the rates corresponding to these benchmark scenarios, we proceed as follows:
\begin{itemize}
    \item For {\bf spin-down}, we assume a typical spin-down rate of $\dot{P} = \SI{e-15}{sec/sec}$ and a typical rotation period of $P = \SI{0.5}{sec}$ \cite{Manchester:2004bp}. Based on ref.~\cite{Konstantinou:2022}, we further assume that the core density differs by $\mathcal{O}(1\%)$ between a non-rotating star and a star spinning at its Kepler frequency (the extremal frequency at which the gravitational and centrifugal forces at the equator are equal). In between these two extremes, we assume that the relative change in core density, $\delta\rho/\rho$ scales quadratically with $1/P$, based on the fact that this is how the centrifugal force scales with $P$. To compute the change in the total number of muons, $dN_\mu$, we then rescale the muon density everywhere in the star by $\delta\rho/\rho$. We also take into account the decrease in overall volume as the star becomes more spherical during spin-down. The rate plotted in \cref{fig:Gamma-integ} is given by $dN_\mu / (P/\dot{P})$, where $P/\dot{P}$ is the spin-down timescale.

    \item To model {\bf accretion}, we assume a 10\% increase in the star's mass over a time scale of \SI{1}{Gyr}. We assume that during this time, the muon density everywhere in the star grows linearly.

    \item A {\bf magnetar expelling its $B$-field} becomes more spherical (less oblate) in the same way a star spinning down does. To the best of our knowledge the modifications to the BSk equations of state in presence of a magnetic field have never been computed. Therefore, we estimate the change in muon abundance based on the calculations from ref.~\cite{Rather:2021azv} for the DD-MEX equation of state \cite{Taninah:2019cku}. In particular, we assume that the field drops linearly from \SI{2.28e17}{G} in the core / \SI{8.98e16}{G} at the surface to zero over a time scale of \SI{1}{Gyr}.

    \item {\bf Tidal deformation} can be modeled in complete analogy to spin-down, by simply replacing the centrifugal force by the tidal force. For the chosen separation, $d = \SI{1200}{km}$, of two equal-mass neutron stars, we compute the gravitational pull of the companion star and determine the frequency at which the star would need to rotate for the centrifugal force to be equal to this tidal force. We then compute the impact on the muon abundance in the same way as we did for spin-down, using for the time scale of tidal deformations $d / (\diff R / \diff t)$, where $\diff R / \diff t$ is the rate at which the orbital separation shrinks due to gravitational wave emission \cite{Maggiore:2007ulw}.
\end{itemize}
While any of these processes is ongoing, the star will settle in an out-of-equilibrium state where the rate at which the equilibrium abundance increases or decreases (given by the corresponding horizontal gray line) equals the muon production, absorption, or decay rate (colored lines). For instance, we can read off from \cref{fig:Gamma-integ} that a \SI{2.0}{M_\odot} neutron star accreting from a companion will settle down at $1 - p_{F\mu} / p_{F\mu}^\text{eq} \lesssim 10^{-4}$ thanks to strong direct Urca production. (If only modified Urca processes were active, the deviation from equilibrium would be $1 - p_{F\mu} / p_{F\mu}^\text{eq} \simeq \num{5e-4}$.) Magnetars pushing out their magnetic field are expected to depart from equilibrium by similar amounts, while stars undergoing merely conventional spin-down are even closer to equilibrium. Tidal deformations prior to a merger can in principle drive a star much further from equilibrium; however, the phase where this happens lasts only a very short time. In the example given in \cref{fig:Gamma-integ} -- a binary system of two equal-mass neutron stars with a separation of \SI{1200}{km} -- the merger is only 46 minutes away.

\section{Neutrino Fluxes from Muon Decay}
\label{sec:nu-fluxes}

While we are currently not aware of any mechanism by which a neutron star's equilibrium muon abundance may \emph{decrease} significantly over extended periods of time, it may well be that such mechanisms exist. As we have discussed in \cref{sec:back-of-envelope} above, we then expect muon diffusion and decay to play a role, and to lead to the emission of $\mathcal{O}(\SI{10}{MeV})$ neutrinos. In the following, we will estimate the flux and spectrum of this hypothetical neutrino flux. In doing so we assume deviations from equilibrium are small and that direct Urca absorption is kinematically forbidden.

\subsection{Monte Carlo Simulation}
\label{sec:mc}

We have written a simple Monte Carlo simulation which tracks individual muons inside a neutron star \cite{github}. We use the \nscool package \cite{Page:2006ud, Page:2005fq, 1998nspt.conf..183P} to solve the Tolman--Oppenheimer--Volkoff (TOV) equations \cite{Tolman:1939jz, Oppenheimer:1939ne, Glendenning:1997wn, Zdunik:2016vza} to construct a neutron star based on an equation of state, and then to simulate the thermal evolution of the star over the first $\sim \SI{e6}{yrs}$, until its effective surface temperature has dropped to \SI{1}{keV}. We then use snapshots of the star at fixed times as a constant background in which we evolve ensembles of muons. In doing so, we proceed as follows:
\begin{description}[before={\renewcommand\makelabel[1]{\bfseries ##1.}}, wide=0\parindent]
    \item[{\scshape 1}] {\scshape Generation of a muon ensemble.}
    Each ensemble of muons consists of \num{1000} particles whose radial distribution is given by the solution to the TOV equations. We consider only muons with momenta close to the Fermi surface as only these muons are mobile enough to diffuse. In particular we determine the number density of ``mobile'' muons at a given radius as
    \begin{align}
        \int\diff E_\mu \, \frac{8\pi \, p_{F\mu}^2}{(2\pi)^3} \, w(E_\mu,\mu_\mu) \, f(E_\mu,\mu_\mu) \,,
        \label{eq:mobile-mu}
    \end{align}
    with the Fermi--Dirac distribution
    \begin{align}
        f(E_\mu,\mu_\mu) \equiv \frac{1}{1 + \exp[(E_\mu - \mu_\mu) / T]} \,.
        \label{eq:FD-distribuion}
    \end{align}
    The weight factor
    \begin{align}
        w(E_\mu,\mu_\mu) = f(E_\mu,\mu_\mu) \, [1 - f(E_\mu,\mu_\mu)] \,,
    \end{align}
    which selects a small momentum interval around the Fermi surface, is identical to the factor appearing in the calculation of the thermal conductivity due to muons \cite{Gnedin:1995lgf}.
  
    \item[{\scshape 2}] {\scshape Muon diffusion.}
    Each muon's location is tracked in time steps $\Delta t$ taken to be significantly longer than the average collision time (to keep the code's running time manageable), but much smaller than the timescale over which the neutron star's properties change significantly (to keep the computation accurate). The mean distance $\Delta x$ a muon travels during a time step $\Delta t$ is given by \cref{eq:delta-x}. In the simulation the diffusion distance in each time step is drawn from a normal distribution with width $\Delta x$, while the direction of diffusion is drawn from a uniform angular distribution. Once a muon has propagated out of the region of high ambient muon density towards larger radii, where only electrons are abundant, we cannot use \cref{eq:kappa-mu,eq:lambda-mu} to estimate its mean free path. Instead we use the electron mean free path as a proxy in this region.
    
    To account for gravity, we add in each time step an extra radial shift
    \begin{align}
        \Delta r_\text{grav} = \frac{1}{2} \frac{F_g}{m_\mu} \Delta t \, \frac{\lambda_\mu}{v_{F\mu}} \,,
    \end{align}
    corresponding to the displacement towards the center of the star between collisions, times the number of collisions in a time interval $\Delta t$. The gravitational force at a specific radius $r$ is estimated as
    \begin{align}
        F_g = - \frac{G_N r \, m_\mu M_\text{NS}}{R_\text{NS}^3}
              + \frac{G_N^2 r (r^2 + 3 R_\text{NS}^2) m_\mu M_\text{NS}^2}{2 R_\text{NS}^6} \,,
    \end{align}
    including the leading general relativistic corrections in an interior Schwarzschild geometry.  Here $G_N$ is Newton's constant while $M_\text{NS}$ and $R_\text{NS}$ are the mass and radius of the neutron star, respectively.
    
    The time step $\Delta t$ is adapted dynamically for each muon to ensure that the average radial displacement $|\Delta x_\text{diff}| + |\Delta r_\text{grav}|$ stays between \SI{50}{m} and \SI{500}{m} in the core of the neutron star, and between \SI{15}{m} and \SI{40}{m} at radii within \SI{0.5}{km} or the core--crust boundary.
    
    \item[{\scshape 3}] {\scshape Muon absorption.}
    In each time step, we check whether the muon is absorbed.  The rates for the relevant Urca reactions are computed as discussed in \cref{sec:noneq-absorption,sec:mu-absorption} using methods from refs.~\cite{Shapiro:1983du, Friman:1979ecl}. The energy of each muon in each time step is chosen as $\mu_\mu + x \cdot T$, where $x$ is a random number drawn from the a distribution $\propto \exp(x) / [1 + \exp(x)]$.  This is again motivated by the factors appearing under the phase space integrals in the calculation of the thermal conductivity \cite{Gnedin:1995lgf}.
  
    \item[{\scshape 4}] {\scshape Muon decay.}
    We also check in each time step whether the muon decays, accounting for the reduction of the decay width by Pauli blocking due to the large electron chemical potential, $\mu_e$, see \cref{sec:mu-width}. In particular, we include a factor $1 - f(E_e, \mu_e)$ in the phase space integral, where $f(E_e, \mu_e)$ is the Fermi--Dirac distribution for an electron with energy $E_e$ and chemical potential $\mu_e$.  Besides regular free muon decay, we also include assisted muon decay (see \cref{sec:back-of-envelope,sec:assisted-mu-decay}), even though the latter is subdominant.
    
    \item[{\scshape 5}] {\scshape Neutrino Spectrum.}
    Each muon decay yields two neutrinos whose 4-momenta we randomly draw based on the kinematics of three-body decay with the appropriate squared matrix element. Once again, Pauli blocking is taken into account as explained in \cref{sec:mu-width}.

    We take the neutrino emission to be isotropic, neglecting possible $\mathcal{O}(1)$ violations of this assumption which could be caused by strong magnetic fields polarizing the decaying muons. (Magnetic fields may also lead to anisotropic diffusion; but as long as the decays are isotropic, this would not affect the angular distribution of neutrinos.)
   
    \item[{\scshape 6}] {\scshape Neutrino evolution outside the neutron star:}
    We redshift each neutrino by a factor $\sqrt{1 - 2 G_N M_\text{NS} / r}$ where $r$ is the radius at which the neutrino is produced. We neglect here the departure of the gravitational potential from strict $1/r$ scaling inside the neutron star because muon decays happen only near the crust, which is thin and much less dense than the core.
   
    We also include neutrino oscillations, assuming non-adiabatic flavor transitions. In other words, we use oscillation probabilities in vacuum, in particular
    \begin{align}
        \begin{split}
          P(\nu_\mu   \to \nu_e)     &\simeq \frac{1}{4} \sin^2 2 \theta_{12} \,, \\
          P(\bar\nu_e \to \bar\nu_e) &\simeq 1 - \frac{1}{2} \sin^2 2\theta_{12} \,.
        \end{split}
        \label{eq:P-osc}
    \end{align}
    As there are no $\mu^+$ in a neutron star, $\nu_\mu$ and $\bar\nu_e$ are the only neutrino flavors produced. In \cref{eq:P-osc}, we have neglected the mixing angle $\theta_{13}$ and have set $\theta_{23} = \pi/4$. For $\theta_{12}$, we use a value of $33.82^\circ$~\cite{Esteban:2018azc}.
\end{description}

\subsection{Neutrino Spectra}
\label{sec:nu-spectra}

\begin{figure}
    \centering
    \includegraphics[width=0.6\textwidth]{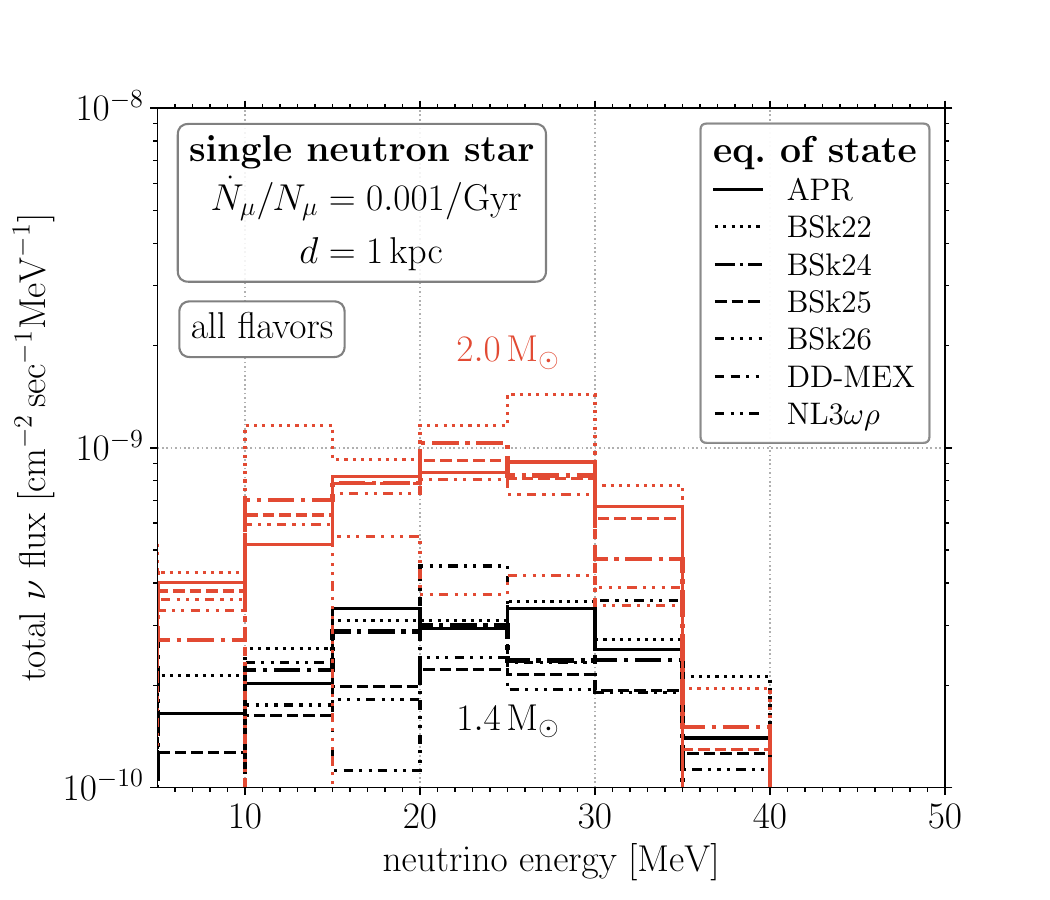}
    \caption{Neutrino flux at Earth from muon diffusion and decay in a single neutron star at $d = \SI{1}{kpc}$ losing 0.1\% of its muons over a time span of \SI{1000}{yrs}. Different colors indicate different neutron star masses, while different line styles correspond to different core equations of state. We include in particular the APR \cite{Akmal:1998cf}, BSk (Skyrme-type density functional) \cite{Pearson:2018tkr}, DD-MEX (density-dependent relativistic mean-field) \cite{Taninah:2019cku}, and NL3$\omega\rho$ \cite{Horowitz:2000xj} equations of state. Overall, we find that the flux from a single neutron star will be undetectably small even for the large muon loss rate assumed here (for which we are not aware of any possible astrophysical origin).}
    \label{fig:single-ns}
\end{figure}

In \cref{fig:single-ns}, we show the expected all-flavor neutrino flux at Earth from a single neutron star \SI{1}{kpc} away losing 0.1\% of its muons over a time span of \SI{1}{Gyr}. (We re-emphasize that we are not aware of a mechanism that actually leads to such loss of muons -- but this does not mean that no such mechanism can exist.) We show results for different neutron star masses (different colors) and equations of state (different line styles). Not unexpectedly, we observe a significant dependence on the neutron star mass: at the same muon loss rate $\dot{N}_\mu / N_\mu$, a $\SI{2}{M_\odot}$ star (light red) expels significantly more muons than a $\SI{1.4}{M_\odot}$ star (dark red), simply because it contains more muons. We also indicate in \cref{fig:single-ns} the dependence of the neutrino flux on the equation of state (different line styles), concluding that different choices could change the neutrino flux by almost an order of magnitude. This can be traced back to the fact that different equations of state predict different muon abundances.

From the overall scale in \cref{fig:single-ns}, it is clear that the neutrino flux from a single neutron star, even a nearby one, will be unobservably small for the foreseeable future. Current experimental sensitivities are around $\SI{.1}{cm^{-2} sec^{-1} MeV^{-1}}$, whereas we predict fluxes that are 8--9 orders of magnitude smaller, even for the sizeable muon loss rate assumed here.

The picture would look somewhat more promising if \emph{all} the neutron stars in Milky Way (of which there are between $10^8$ and $10^9$ \cite{Sartore:2010}) were losing muons simultaneously. To study this---still fairly hypothetical---scenario in more detail, we have generated \num{1000} pseudo-galaxies with random populations of neutron stars, picking the exact number of stars in each population from the above range. We also choose a random equation of state (APR, BSk22, BSk24, BSk25, BSk26, DD-MEX, or NL3$\omega\rho$) for each pseudo-galaxy, see \cref{sec:eos}.\footnote{We have used a fixed model for neutron and proton superfluidity across all equations of state, see \cref{sec:matrix-elements} for details. However, we do not expect superfluidity to play a significant role in this rate. For muons escaping the muon-sphere the absorption rate is dominated by the out-of-equilibrium enhancement, see \cref{fig:rates-mfps}.} Keeping the equation of state fixed would reduce the uncertainty in our results by a few tens of per cent.  For the radial locations of neutron stars in the galaxy, we choose an exponential distribution with a scale length of \SI{2.48}{kpc} \cite{Xiang:2018}, and we assume the thickness of the Milky Way's disk to be \SI{1}{kpc}. Neutron star masses are assumed to follow the ``preferred'' mass function from ref.~\cite{Alsing:2017bbc}.

\begin{figure*}
    \hspace*{-0.3cm}
    \begin{tabular}{cc}
        \includegraphics[width=0.54\textwidth]{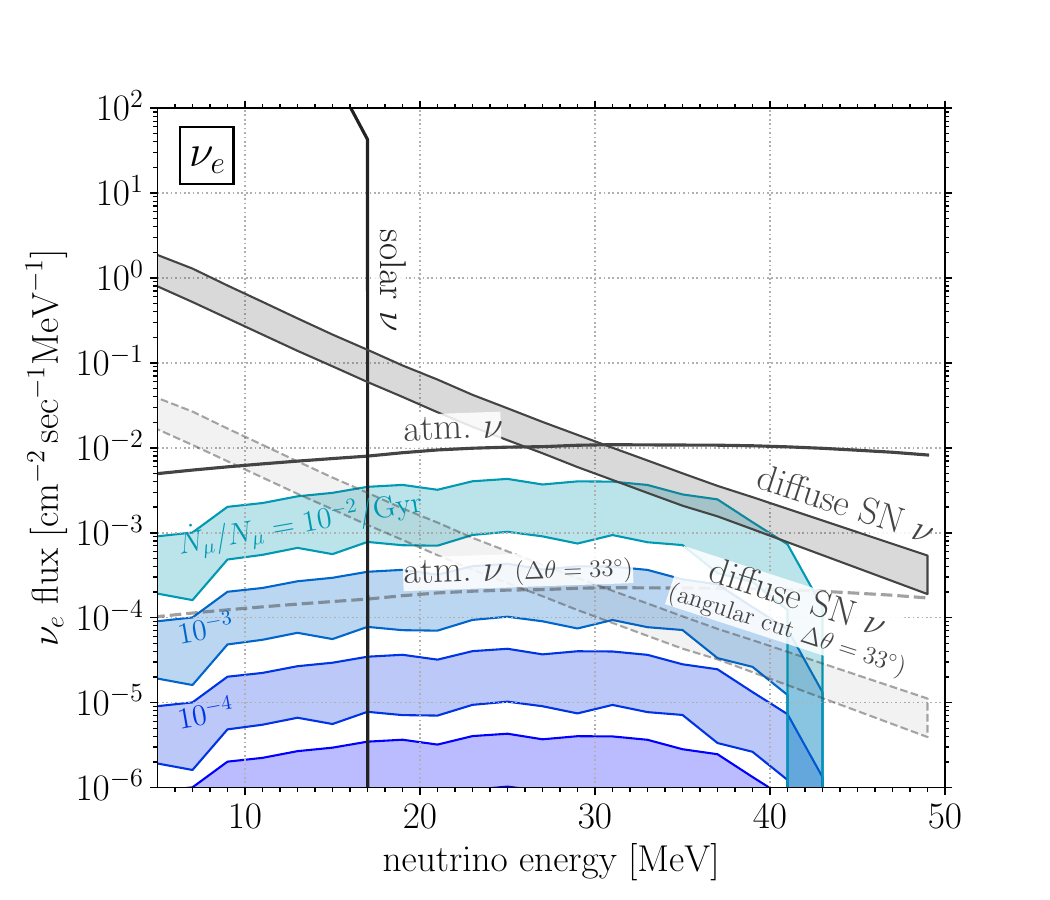} \hspace{-0.6cm} &
        \includegraphics[width=0.54\textwidth]{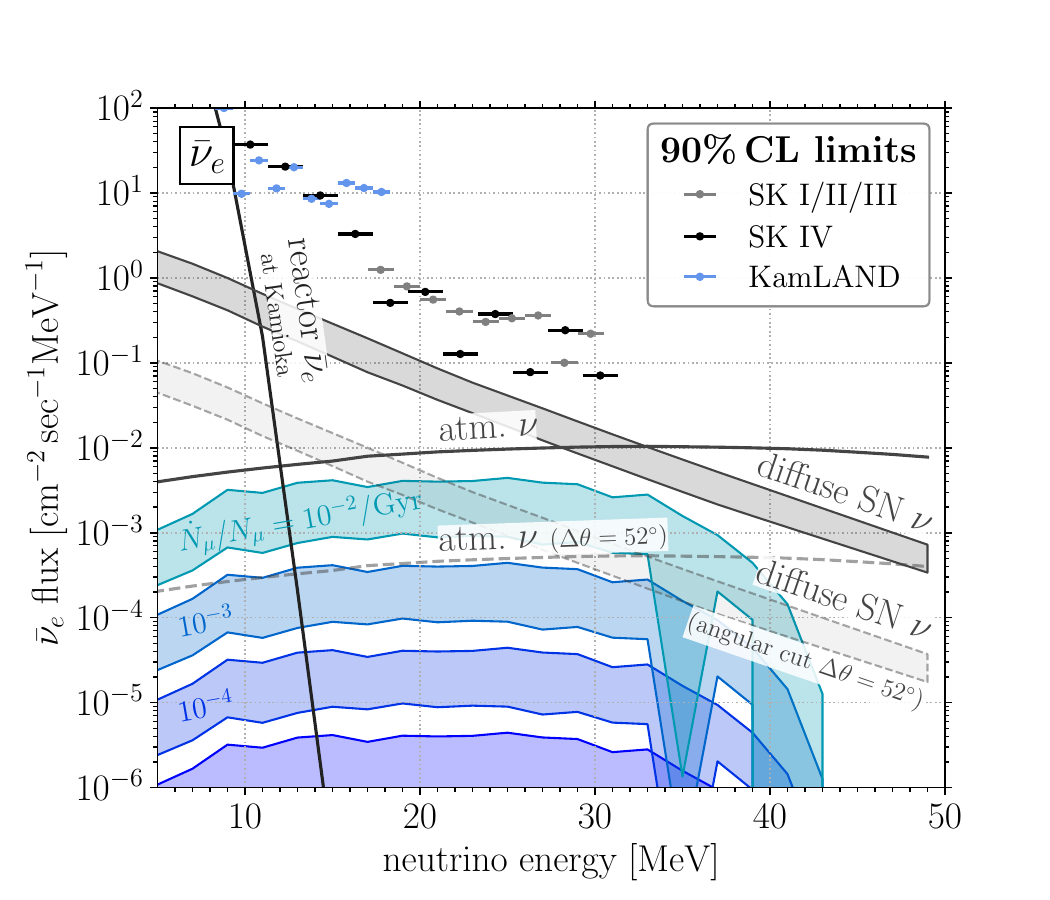}
    \end{tabular}
    \caption{Spectra of electron neutrinos (left) and electron anti-neutrinos (right) at Earth from muon decay in neutron stars, assuming a mechanism exists that incentivizes neutron stars to slowly expel muons. Each of the colored bands corresponds to a muon loss rate $\dot{N}_\mu / N_\mu$, and the width of the bands shows the uncertainty due to the number of neutron stars in the Milky Way, their masses, and their spatial distribution.  We compare to backgrounds due to solar neutrinos \cite{Bahcall:2004pz}, atmospheric neutrinos \cite{Battistoni:2005pd}, diffuse supernova neutrinos \cite{Moller:2018kpn}, and reactor neutrinos at Kamioka \cite{Beacom:2003nk}, see \cref{sec:backgrounds} for details. Finally, for $\bar\nu_e$, we show the 90\% confidence level exclusion limits from KamLAND ({\color{plotlightblue}light-blue}) and SuperKamiokande ({\color{plotgrey}gray} \& black) \cite{Super-Kamiokande:2021jaq}.  While an observation above background in detection channels without angular resolution (notably inverse beta decay) would be possible only for extremely large $\dot{N}_\mu / N_\mu$, prospects would be better in a detector exploiting neutrino--electron scattering, which can use angular resolution to suppress backgrounds. We have here assumed an angular resolution corresponding to the $1\sigma$ deviation between the incoming neutrino and outgoing electron direction in neutrino--electron scattering. As this deviation varies slightly between different neutrino flavors, the angular resolutions in the two panels are not identical.}
    \label{fig:spectrum-comparison}
\end{figure*}

The resulting predictions for the integrated neutrino flux are shown as colored bands in \cref{fig:spectrum-comparison}, with the width of the bands corresponding to the 68\% confidence level spread among the \num{1\,000} pseudo-galaxies. We only show fluxes for $\nu_e$ and $\bar\nu_e$ as other neutrino flavors are far more difficult to detect at $\mathcal{O}(\SI{10}{\MeV})$ energies. Comparing to the relevant backgrounds \cite{Bahcall:2004pz, Battistoni:2005pd, Ferrari:2005zk, Boehlen:2014, Cocco:2004ac, Beacom:2010kk, Moller:2018kpn, Esteban:2018azc}, we find that in detectors without directional sensitivity only a hypothetical signal from neutron stars losing $\sim 1\%$ of their muons over a Gyr time scale may be observable in the face of atmospheric and diffuse supernova neutrino backgrounds. This is the situation that most current and future neutrino detectors will be facing: neither DUNE observing $\nu_e + \iso{Ar}{40}$ interactions \cite{DUNE:2020ypp, Capozzi:2018dat, Zhu:2018rwc} nor HyperKamiokande \cite{Hyper-Kamiokande:2018ofw} and JUNO \cite{JUNO:2021vlw, JUNO:2023vyz} observing inverse beta decay will have angular resolution.  For a detection channel with decent angular resolution (for instance DUNE or HyperKamiokande observing neutrino--electron scattering), also signals that are 1--2 orders of magnitude smaller could be observable. This is possible because most of the signal will come from the Galactic Center region, which covers only a $\sim 10^\circ \times 10^\circ$~degree region in the Sky.  Assuming a detector resolution of $\sim 40^\circ$, based on the typical angular distance between the direction of an incoming $\sim \SI{25}{MeV}$ neutrino and the outgoing electron in neutrino--electron scattering, the atmospheric and diffuse supernova backgrounds can be suppressed to the level of the dashed lines in \cref{fig:spectrum-comparison}, while the solar and reactor backgrounds can be eliminated completely. However, due to the smaller cross section for neutrino--electron scattering, significantly larger detector masses and/or exposure times are required.

\section{Summary and Conclusions.}
\label{sec:conclusions}

We have discussed the dynamics of electrons and muons in neutron stars, considering in particular situations where the star departs from equilibrium. This can happen for instance due to changes in its rotational velocity or magnetic field, due to accretion, or due to tidal interactions with another compact object. Under such conditions, the star's core density, and therefore the equilibrium particle abundances in its core, change. In response to this change the processes that create and destroy muons---direct and modified Urca reactions---go out of equilibrium. We have for the first time calculated these out-of-equilibrium Urca rates and have used our results to estimate the time scales over which the star adjusts to the new equilibrium conditions. 

Moreover, we have for the first time studied a process we dubbed assisted muon decay (that is, muon decay in presence of a spectator nucleon), which opens an additional muon depletion channel in regions where regular muon decay is Pauli-blocked. We have, however, found assisted muon decay to be subdominant compared to Urca absorption.

We have finally pointed out that in neutron stars older than about \SI{10000}{yrs}, an efficient way to deplete muons is their diffusion out of the core into a region where their decay is no longer Pauli-blocked. This could be relevant in case of tidal interactions, which can deform or spin up a neutron star, leading to a decrease in its equilibrium core muon abundance. It is quite possible that there are astrophysical mechanisms that we are not aware of leading to a similar outcome. (The other mechanisms discussed in this paper lead to an increase rather than decrease of the muon abundance.)

Developing this argument further, we have then shown that muon depletion via diffusion and decay implies a flux of neutrinos with energies up to about \SI{40}{MeV} which, to the best of our knowledge, has never been studied before. This flux could be observable above the atmospheric and diffuse supernova neutrino backgrounds if all neutron stars in the Milky Way were to expel 0.01--1\% of their muons over Gyr-timescales.

A possible direction of further study is \emph{lateral} diffusion of leptons: a neutron star in a tight binary system experiences tidal forces that cause it to bulge in the direction of the companion star as well as diametrically to it. As the neutron star spins and orbits its companion, the bulge moves around it in complete analogy to the tides on the Earth's oceans. This leads to moving density gradients in the neutron star that should cause electrons and muons to drift laterally. The relevant time scale is the spin period, which can be long in a very old neutron star.

\section*{Acknowledgments.}

This work has benefited tremendously from discussions with John Beacom, Omar Benhar, Kfir Blum, Christopher Dessert, Jordy de Vries, Veronica Dexheimer, Farrukh Fattoyev, Tassos Fragos, Hans-Thomas Janka, Simon Knapen, Kei Kotake, Nadav Outmezguine, Dany Page, Ishfaq Rather, Sanjay Reddy, Ben Safdi, Albert Stebbins, and Urs Wiedemann. We are particularly grateful to Veronica Dexheimer and Ishfaq Rather for help with the equations of state discussed in ref.~\cite{Rather:2021azv}. The authors' work has been partially supported by the European Research Council (ERC) under the European Union's Horizon 2020 research and innovation program (grant agreement No.\ 637506, ``$\nu$Directions''). TO is supported by the DOE Early Career Grant
DESC0019225.

\appendix
\section{Electron and Muon Production, Absorption, and Decay}
\label{sec:abs-prod}

In this appendix, we show in detail the calculation of the charged lepton production and absorption rates in out-of-equilibrium neutron stars. To simplify the discussion and the notation, we show only results for muons here. The Urca rates for electrons can be obtained from the ones for muons with the obvious replacements $m_\mu \to m_e$, $\mu_\mu \to \mu_e$, $p_{F\mu} \to p_{Fe}$, etc. Assisted muon decay has no electron analogue.

The three classes of processes we will discuss in the following are:
\begin{align*}
    &\text{Direct-Urca (DU) Processes:}   &\qquad n           &\to p + \mu + \bar\nu_\mu \,, \\
    &                                     &\qquad \mu + p     &\to n + \nu_\mu \,, \\[0.2cm]
    &\text{Modified-Urca (MU) Processes:} &\qquad n + n       &\to n + p + \mu + \bar\nu_\mu \,, \\
    &                                     &\qquad n + p + \mu &\to n + n + \nu_\mu \,, \\[0.2cm]
    &\text{Assisted Muon Decay:}          &\qquad p + \mu     &\to p + e + \nu_\mu + \bar\nu_e \,. \\
\end{align*}
Direct and indirect Urca processes in equilibrium have been widely studied in the literature, while to the best of our knowledge out-of-equilibrium Urca reactions and assisted muon decay have not been considered before.

Additional branches of modified Urca processes exist, including a proton instead of a neutron as the spectator nucleon. However, for the equations of state and neutron star masses considered this branch is subdominant. Importantly, direct Urca reactions are only allowed if the Fermi momenta of protons ($p_{Fp}$), neutrons ($p_{Fn}$), and muons ($p_{F\mu}$) satisfy the inequality $p_{Fn} < p_{Fp} + p_{F\mu}$. This can be understood from momentum conservation, together with the fact that the neutron star is in beta equilibrium ($\mu_n = \mu_p + \mu_\mu$), which implies that neutrons, protons, and muons involved in the Urca reactions have energies within $\sim \pm T$ of their respective Fermi energies, while the energy of the neutrinos is of order $T$ and therefore negligible.

Our determination of the muon production, absorption, and decay rates is based largely on methods from Refs.~\cite{Yakovlev:1995, Friman:1979ecl} for modified Urca processes, and on Ref.~\cite{Lattimer:1991ib} for direct Urca reactions. The relevant techniques are also described in Ref.~\cite{Shapiro:1983du}. In these references, the main focus lies on the neutrino luminosity which determines the neutron star cooling rate. Here we are instead interested in the underlying rates.

\subsection{Matrix Elements}
\label{sec:matrix-elements}

The matrix element for the direct Urca reactions follows from completely standard techniques (see Refs.~\cite{Yakovlev:2000jp, Lattimer:1991ib, Yakovlev:1995}):
\begin{align}
    \sum_\text{spins} |\mathcal{M}_\text{DU}|^2
        &= 8 G_F^2 E_\mu E_\nu \cos^2\theta_C (1 + 3 g_A^2)
         \equiv E_\mu E_\nu \sum_\text{spins} | \widetilde{\mathcal{M}}_\text{DU}|^2 \,.
    \label{eq:M-DU}
\end{align}
Here, $\theta_C$ is the Cabbibo angle, $G_F$ is the Fermi constant, and $g_A \sim 1.26$ is the nucleon axial-vector coupling. We have averaged over the outgoing neutrino direction, simplifying the energy dependencies. In the last equality of \cref{eq:M-DU}, we have separated the energy dependent factors that will be integrated over from the constant piece of the matrix element. Note that the nucleon wave functions are normalized to unity, while the leptons have Lorentz-invariant normalization.

For the matrix element of modified Urca processes, a number of assumptions are made in determining the interaction of the nucleon system:
\begin{enumerate}
    \item The nucleon--nucleon interactions are factorized into long-distance and short-distance components. For the long-distance interactions the one-pion-exchange (OPE) component is calculated, which given the typical nucleon separation is expected to dominate. $\rho$-meson exchange is neglected. While for the short-distance interactions an effective description using Landau theory is adopted.
    \item The nucleon system is assumed to be non-relativistic.
    \item The matrix element is determined to leading order in an expansion of the nucleon propagator in the inverse nucleon mass.
    \item Diagrams with the exchange of the initial/final state nucleon legs are neglected.   
\end{enumerate}
Based on these assumptions the matrix element for modified Urca processes is (see Refs.~\cite{Friman:1979ecl, Yakovlev:1995})
\begin{align}
    \sum_\text{spins} \left| \mathcal{M}_\text{MU}\right|^2
        &=  \frac{256 G_F^2  \cos^2\theta_C g_A^2 E_\nu E_\mu}{(E_\nu+E_\mu)^2}
            \bigg[ 2 \left(\frac{g_{\pi N N}}{m_\pi}\right)^4
                      \left(\frac{k^2}{k^2 + m_\pi^2}\right)^2
                   \label{eq:murca-matrixelement-full}\\
        &\qquad  + 2 \left[\left(g^\prime - g\right) + \left(f^\prime - g^\prime\right)\right] 
                     \left(\frac{g_{\pi N N}}{m_\pi}\right)^2 \frac{k^2}{k^2 + m_\pi^2} + 3 \left[\left(g^\prime - g\right)^2 + \left(f^\prime - g^\prime \right)^2 \right]
            \bigg] \,,  \notag \\
        &= \frac{ E_\nu E_\mu}{(E_\nu+E_\mu)^2}
           \sum_\text{spins} \left| \widetilde{\mathcal{M}}_\text{MU} \right|^2 \,,
    \label{eq:murca-matrixelement}
\end{align}
where $g_{\pi N N} \simeq 1$ is the coupling constant in the $\pi N N$ interaction vertex, and $m_\pi$ is the neutral pion mass. $g^\prime$, $g$ and $f^\prime$ are the parameters of the effective Landau theory, while $k$ is the momentum exchange between the nucleons. Following Refs.~\cite{Friman:1979ecl, Yakovlev:1995}, $\alpha_n$ is used to encode the density dependence of the parameters entering in the square brackets of \cref{eq:murca-matrixelement-full}. While a fudge factor $\beta_n$ is applied to correct for the effect of short distance interactions neglected in the OPE potential used above. In what follows for comparison purposes we will utilize the matrix element in the form \cite{Yakovlev:1995}
\begin{align}
    \sum_\text{spins} \left| \mathcal{M}_\text{MU}\right|^2
        \simeq 256 G_F^2 \cos^2\theta_C g_A^2 \frac{E_\nu E_\mu}{(E_\nu+E_\mu)^2} \alpha_n \beta_n\,.
    \label{eq:murca-approx-form}
\end{align}
We take $\alpha_n = 1.76 - 0.634 (n_0/n_n)^{2/3}$ (where $n_n$ is the neutron density and $n_0 = \SI{0.16}{\femto\meter^{-3}}$) and $\beta_n = 0.68$ as implemented in \nscool. In our Monte Carlo simulations from \cref{sec:mc}, neutron and proton superfluidity effects are included through the extraction of the neutrino luminosity from \nscool, which amounts to modifications of the $\alpha_n$ and $\beta_n$ coefficients. Here we again use default \nscool settings. This corresponds to refs.~\cite{Schwenk:2002fq} and \cite{Page:2009fu} (model `a') for the ${}^1\text{S}_0$ and ${}^3\text{P}_2$ neutron pair gaps, respectively. While for protons ref.~\cite{proton-superfluidity-ref} is used for the ${}^1\text{S}_0$ gap.

To calculate the matrix element for assisted muon decay, we use the Mathematica packages {\tt FeynArts~3.11} \cite{Kublbeck:1990xc, Hahn:2000kx} and {\tt FeynCalc~9.3.1} \cite{Mertig:1990an, Shtabovenko:2016sxi, Shtabovenko:2020gxv} to handle the algebra. We make the following approximations:
\begin{enumerate}
    \item We assume the nucleon to be at rest before \emph{and} after the interaction. This is equivalent to treating its electromagnetic field as a static, classical, Coulomb field.
    \item We treat the electron as ultra-relativistic.
    \item We neglect neutrino momenta (of order $T$) compared to the muon and electron momenta. This is justified by the large chemical potentials of the charged leptons.
\end{enumerate}
The result for the spin-averaged squared matrix element is
\begin{align}
    \sum_\text{spins} \left| \mathcal{M}_\text{amd}\right|^2
        &= \frac{4096 \pi^2 \alpha^2 G_F^2
                 E_\mu^3  E_e  E_{\bar\nu_e} E_{\nu_\mu}}{m_\mu^8} 
         \equiv E_\mu^3  E_e E_{\bar\nu_e} E_{\nu_\mu}
                \sum_\text{spins} \left| \widetilde{\mathcal{M}}_\text{amd} \right|^2 \,,
    \label{eq:Msq-assisted-mu-decay}
\end{align}
with the obvious notation for the various particle energies. We have once again assumed the nucleon wave functions to be normalized to unity, while for the leptons we use the usual Lorentz-invariant normalization.

\subsection{Muon Production}
\label{sec:mu-production}

We write the muon production rate per unit volume per unit time as the sum of the direct and modified Urca rates,
\begin{align}
    \Gamma^\text{prod} &= \Gamma_\text{DU}^\text{prod} 
                        + \Gamma_\text{MU}^\text{prod} \,.
\end{align}

\subsubsection{Muon Production via the Direct Urca Process}
\label{sec:DU-prod}

We begin with the computation of $\Gamma_\text{DU}^\text{prod}$ before turning to the more involved modified Urca case.  The rate of interest is obtained by integrating over the phase space for the process
\begin{align}
    n(p_1) \to p(p_2) + \mu (p_3\text{ or } p_\mu) + \overline{\nu}_\mu(p_\nu)\,, 
\end{align}
yielding
\begin{align}
    \Gamma_\text{DU}^\text{prod} &=
        \int\!\bigg[ \prod_{j=1}^3 \frac{\diff^3 p_j}{(2\pi)^3} \bigg]
        \frac{\diff^3 p_\nu}{2E_\nu(2\pi)^3} \frac{(2\pi)^4}{2 E_\mu} \delta^{(4)}(p_f - p_i)
        \mathscr{L}_\text{blocking} \sum_\text{spins} \left| \mathcal{M}_\text{DU}\right|^2\,.
    \label{eq:phasespace-start}
\end{align}
In the above expressions, $j = 1, 2$ labels the nucleons, while $p_3 \equiv p_\mu$ and $p_\nu$ are the muon and neutrino 4-momenta, respectively. Note that the integration measures for nucleons and leptons are different because of the different wave function normalizations assumed in the matrix element. In the 4-momentum-conserving $\delta$-function, $p_i = p_1$ and $p_f = p_2 + p_\mu + p_\nu$ are the sums of the initial- and final-state 4-momenta, respectively. Lastly \cref{eq:phasespace-start} contains the phase space factor 
\begin{align}
    \mathscr{L}_\text{blocking} = f_1 (1-f_2) (1-f_3) \,,
    \label{eq:L-blocking-DU-prod}
\end{align}
which describes the distribution of initial state neutrons as well as Pauli blocking for the proton and the muon in the final state. $f_k$ denotes the Fermi--Dirac phase space distribution for the $k$-th particle species, that is,
\begin{align}
    f_k \equiv f_k(E_k, \mu_k) &= \frac{1}{1 + \exp[(E_k - \mu_k)/T]}\,.
\end{align}
Here, $E_k$ is the particle energy, $\mu_k$ is the chemical potential (or Fermi energy), and $T$ is the local temperature of the neutron star. In order to analytically evaluate the phase space integrals in \cref{eq:phasespace-start}, we exploit the hierarchy between the Fermi energies and the temperature of the system. Consequently the integrals are dominated by the region within $\mathcal{O}(T)$ about the equilibrium Fermi momenta of the nucleons and muons.  We begin by introducing spherical coordinates in momentum space to separate the angular integrals from the integrals over the absolute value of the momenta:
\begin{align}
    \Gamma_\text{DU}^\text{prod} &= \frac{1}{4(2\pi)^{8}}
        \int\!\bigg[ \prod_{j=1}^3 |\vec{p}_j|^2 \diff |\vec{p}_j| \diff \Omega_j \bigg]
        \frac{\diff |\vec{p}_\nu| \diff \Omega_\nu E_\nu}{E_\mu} \,
        \delta^{(3)}(\vec{p}_f - \vec{p}_i) \delta\left(E_f - E_i\right)
        \mathscr{L}_\text{blocking} \sum_\text{spins} \left| \mathcal{M}_\text{DU}\right|^2\,.
    \label{eq:Gamma-DU-prod-1}
\end{align}
Neglecting the neutrino momentum of order $T$ compared to the much larger nucleon and muon momenta in the $\delta$-function, the angular integrals can be performed
\begin{align}
    \int\!\bigg[\prod_{j=1}^3 \diff\Omega_j \bigg] \diff\Omega_\nu\,\delta^{(3)}(\vec{p}_f - \vec{p}_i)
        &= 4\pi \int\!\bigg[ \prod_{j=1}^3 \diff\Omega_j \bigg]
           \delta^{(3)}(\vec{p}_2 + \vec{p}_3 - \vec{p}_1) \notag\\
        &= \frac{4\pi}{|\vec{p}_2|^2} \int\!\bigg[ \prod_{j=1}^3 \diff\Omega_j \bigg]
           \delta(|\vec{p}_2| - |\vec{p}_1 - \vec{p}_3|) \, \delta^{(2)}(\Omega_2 - \Omega_{1-3}) \notag\\
        &= \frac{4\pi}{|\vec{p}_2|^2} \int\! \diff\Omega_1 \diff\Omega_3 \,
           \delta(|\vec{p}_2| - \sqrt{\vec{p}_1^2 + \vec{p}_3^2 - 2 |\vec{p}_1| |\vec{p}_3| \cos\theta_1}) \,
           \Theta_\text{np$\mu$}
                                                                                                  \notag\\
        &= \frac{2\pi (4\pi)^2}{|\vec{p}_1| |\vec{p}_2| |\vec{p}_3|}\, \Theta_\text{np$\mu$}\,.
\end{align}
Here, $\Theta_\text{np$\mu$} = \Theta(|\vec{p}_2| - \left| |\vec{p}_3| - |\vec{p}_1|\right|)\, \Theta(|\vec{p}_3| + |\vec{p}_1| - |\vec{p}_2|)$ is a Heaviside step-function ensuring this process is present only when the \textit{triangle inequality} of the sum of three momenta is satisfied, $p_{Fn} \leq p_{Fp} + p_{F\mu}$. This is necessary to ensure that the argument of the delta function has real roots in $|\vec{p}_2|$, and it is the deeper mathematical reason for why direct Urca processes are forbidden in most neutron stars. Next, we express the momentum integrals in terms of energy integrals using $\diff |\vec{p}_\nu| = \diff E_\nu$ for the neutrinos and $|\vec{p}| \diff |\vec{p}| \simeq m^\star \diff E$ for the massive particles. Here $m^\star$ is the effective particle mass which, for nucleons, includes the effects of many-body interactions inside the neutron star.  For muons, it is simply given by the muon Fermi energy, $m_\mu^\star = \mu_\mu$, as can be inferred from the relativistic dispersion relation, expanded around $p_{F\mu}$. The rate now becomes
\begin{align}
     \Gamma_\text{DU}^\text{prod} &= \frac{m_n^{\star} m_p^\star \mu_\mu \Theta_\text{np$\mu$}}{(2\pi)^{5}}
         \int\!\bigg[ \prod_{j=1}^3\diff E_j \bigg] \diff E_\nu \, \frac{E_\nu}{E_\mu} \,
         \delta(E_f - E_i) \,
         \mathscr{L}_\text{blocking} \sum_\text{spins} \left| \mathcal{M}_\text{DU}\right|^2\,.
    \label{eq:Gamma-DU-prod-2}
\end{align}
Turning to dimensionless integration variables, we introduce $x_j \equiv \pm \beta (E_j - \mu_j)$, where the plus sign is for incoming particles and the minus sign for outgoing particles, $\beta \equiv 1/T$, and $x_\nu \equiv \beta E_\nu$. Firstly with these definitions the Pauli-blocking factor simplifies to
\begin{align}
    \mathscr{L}_\text{blocking} &= \prod_{j=1}^3 \frac{1}{1+e^{x_j}} \,,
\end{align}
while the remaining delta function becomes
\begin{align}
    \delta(E_f - E_i) &= \delta(E_2 + E_\mu + E_\nu - E_1) \,,\notag \\
        &= \delta\Big( \beta^{-1} \Big[x_\nu -  \sum_{j=1}^3 x_j \Big] + \mu_p + \mu_\mu - \mu_n \Big)
    \equiv \beta \, \delta\left(\sum_{j=1}^3 x_j-\overline{x}_\nu\right)\,,
\end{align}
with the definitions
\begin{align}
    \overline{x}_\nu \equiv x_\nu + B
    \qquad\text{and}\qquad
    B \equiv \beta(\mu_p - \mu_n + \mu_\mu)
\end{align}
The latter quantity is a measure of how far the neutron star is from equilibrium. Combining these results gives
\begin{align}
    \Gamma_\text{DU}^\text{prod} &= \frac{m_n^\star m_p^\star \mu_\mu\Theta_\text{np$\mu$}}{(2\pi)^{5} \beta^5}
        \sum_\text{spins} \left| \widetilde{\mathcal{M}}_\text{DU}\right|^2
        \int\!\bigg[\prod_{j=1}^3 \diff x_j \frac{1}{1+e^{x_j}} \bigg] \diff x_\nu \,
        x_\nu^2 \, \delta\Big( \sum_{j=1}^3 x_j-\overline{x}_\nu \Big)  \,,
    \label{eq:Gamma-DU-prod-3}
\end{align}
where we have pulled out the energy dependence of the spin-summed matrix element, see \cref{eq:M-DU}. The final set of integrals to evaluate is thus:
\begin{align}
    I_\text{DU}^\prime \equiv \int_0^\infty \! \diff x_\nu x_\nu^2 \, J_\text{DU} \,,
    \qquad\text{with}\qquad
    J_\text{DU} = \int_{-\infty}^\infty \! \Big[\prod_{j=1}^3 \diff x_j \frac{1}{1+e^{x_j}} \Big]
        \delta\Big(\sum_{j=1}^3 x_j-\overline{x}_\nu \Big)
    \label{eq:J-DU-1}
\end{align}
Here, we have adjusted the lower integration limits in $J_\text{DU}$ from $-\beta \mu_j$ to $-\infty$ to simplify the integration. This modification introduces only a negligible error, as the region with $x_j < -\beta \mu_j$ corresponds to fermions deep within the Fermi sea. Due to Pauli blocking, the contributions from fermions in these lower lying states are exponentially suppressed. To evaluate $J_\text{DU}$, we use standard complex integration techniques as described for instance in Appendix~F of Ref.~\cite{Shapiro:1983du}.  More precisely, we introduce an auxiliary variable $z$ and write
\begin{align}
    J_\text{DU}
        &= \int_{-\infty}^\infty \! \frac{dz}{2\pi} \, e^{i z \Big(\sum_{j=1}^3 x_j-\overline{x}_\nu \Big)}
           \bigg[ \int \! \diff x \frac{1}{1 + e^x} \bigg]^3         \notag\\
        &= \int_{-\infty}^\infty \! \frac{dz}{2\pi} \, e^{-i z \overline{x}_\nu}
           \bigg[ \int \! \diff x \frac{e^{i z x}}{1 + e^x} \bigg]^3
            \label{eq:first-int}
\end{align}
\begin{figure}
    \centering
    \includegraphics[width = 0.495\textwidth]{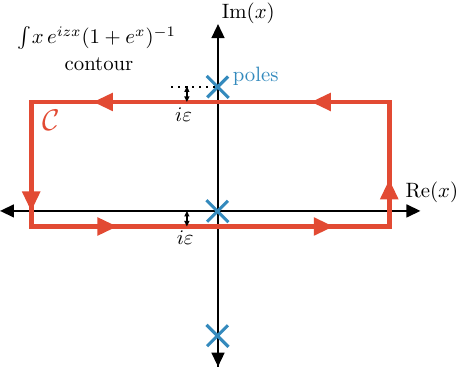}
    \includegraphics[width = 0.495\textwidth]{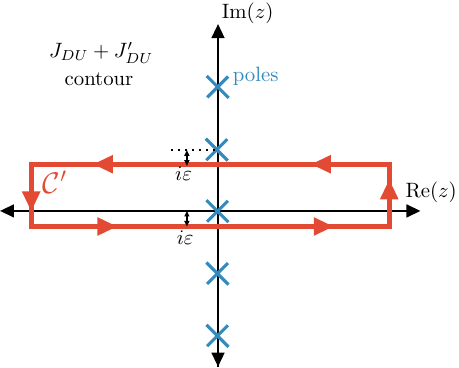}
    \caption{(\textbf{Left}) Complex contour $\mathcal{C}$ for evaluation of the integral in \cref{eq:first-int}.  (\textbf{Right}) Complex contour $\mathcal{C}^\prime$ that results from summing the contributions $J_\text{DU}$ and $J_\text{DU}^\prime$ for evaluation of the integral in \cref{eq:J-DU-complex-contour}.}
    \label{fig:complex-contour}
\end{figure}

\noindent
The integral in square brackets, $f(z) \equiv \int \! \diff x \, e^{i z x} (1 + e^x)^{-1}$, has an integrand that has poles at $x = i (2n+1) \pi$, with $n \in \mathbb{Z}$.  It can be evaluated by closing the integration contour as shown in the left-hand panel of \cref{fig:complex-contour}. The segment along the real axis is just $f(z)$, the added segment at $\Im(x) = 2\pi$ gives $-e^{-2 \pi z} f(z)$, and the two vertical segments vanish in the limit $\Re(x) \to \pm\infty$. Putting everything together and using the residual theorem, we obtain $f(z) - e^{-2\pi z} f(z) = -2 \pi i \, e^{-\pi z}$, or, equivalently, $f(z) = -i \pi / \sinh(\pi z)$ \,. The remaining integral over $z$,
\begin{align} \label{eq:second-int}
    J_\text{DU} = \int_{-\infty-i\epsilon}^{\infty-i\epsilon} \!
                  \frac{dz}{2\pi} \, e^{-i z \overline{x}_\nu}
                  \bigg( \frac{-i \pi}{\sinh(\pi z)} \bigg)^3 \,,
\end{align}
converges only if $z$ has an infinitesimal negative imaginary part $\epsilon$, hence the modification of the integration boundaries compared to \cref{eq:J-DU-1}.  We can construct a closed integration contour as shown in the right-hand panel of \cref{fig:complex-contour}, where once again the vertical segments do not contribute in the limit $\Re(z) \to \pm \infty$.  The upper horizontal segment yields
\begin{align}
    J_\text{DU}' &= \int_{\infty+i-i\epsilon}^{-\infty+i-i\epsilon} \!
                    \frac{dz}{2\pi} \, e^{-i z \overline{x}_\nu}
                    \bigg( \frac{-i \pi}{\sinh(\pi z)} \bigg)^3 \,, \notag\\
                 &= \int_{\infty-i\epsilon}^{-\infty-i\epsilon} \!
                    \frac{dz'}{2\pi} \, e^{-i z' \overline{x}_\nu - \overline{x}_\nu}
                    \bigg( \frac{i \pi}{\sinh(\pi z')} \bigg)^3 \,, \notag\\
                 &= e^{-\overline{x}_\nu} J_\text{DU} \,.
\end{align}
Combining the two contour segments and using the residue theorem, it then follows that
\begin{align}
     J_\text{DU} + J_\text{DU}^\prime 
        = (1 + e^{-\overline{x}_\nu}) J_\text{DU}
        = \frac{1}{2} (\pi^2 + x^2) \,,
        \label{eq:J-DU-complex-contour}
\end{align}
or, equivalently,
\begin{align}
    J_\text{DU} &= \frac{\pi^2 + \overline{x}_\nu^2}{2(1 + e^{\overline{x}_\nu})} \,,
    \label{eq:J-DU-2}
\end{align}
The final integration over $x_\nu$ in \cref{eq:J-DU-1} is now readily performed:
\begin{align}
    I^\prime_\text{DU}
        &= \int_{B}^\infty \! \diff \overline{x}_\nu
           \frac{(\overline{x}_\nu-B)^2 (\pi^2+\overline{x}_\nu^2)}
                {2 (1 + e^{\overline{x}_\nu})} \,, \\ 
        &= -\frac{B^2\pi^2}{2} \left( B + \log[1+e^{-B}] - \log[1+e^B] \right)
                                   - (B^2 + \pi^2) \Li_3(z) - 6 B \Li_4(z) 
                                   - 12 \Li_5(z) \,,
    \label{eq:I-DU-prime}
\end{align}
where $z = -\exp(-B)$ and $\Li_n(z)$ is the polylogarithm. In beta equilibrium, where $B=0$ and $z=-1$, \cref{eq:I-DU-prime} simplifies to
\begin{align}
    \lim_{B\to 0} I^\prime_\text{DU}
        &= \frac{3}{4} \left(\pi^2\zeta(3) + 15 \zeta(5)\right)
        \simeq 20.6\,.
\end{align}
Putting everything together, our final expression for the direct Urca muon production rate now reads
\begin{align}
    \Gamma_\text{DU}^\text{prod}
        &= \frac{m_n^{\star} m_p^\star \mu_\mu \Theta_\text{np$\mu$}}{4\pi^5 \beta^5} G_F^2 \cos^2 \theta_C 
           (1 + 3 g_A^2) I^\prime_\text{DU} \,,
    \label{eq:Gamma-DU-prod-final}
\end{align}
with $I^\prime_\text{DU}$ from \cref{eq:I-DU-prime}. \\

\noindent
For practical purposes, it is useful to relate $\Gamma_\text{DU}^\text{prod}$ to the neutrino luminosity (energy loss in neutrinos per unit volume per unit time assuming beta equilibrium) from direct muon Urca processes, $Q_\text{DU}^\nu$, which is tracked, for instance, by \nscool. $Q_\text{DU}^\nu$ is obtained by inserting an extra factor $E_\nu = x_\nu / \beta$ under the integral in \cref{eq:Gamma-DU-prod-1}, which propagates into the integrand in \cref{eq:Gamma-DU-prod-3}.  $Q_\text{DU}^\nu$ is therefore given by an expression analogous to \cref{eq:Gamma-DU-prod-final}, but with $I^\prime_\text{DU}$ replaced by
\begin{align}
    I_\text{DU} = \beta^{-1} \int_0^\infty \diff x_\nu x_\nu^3 J_\text{DU}
                \quad\xrightarrow{B \to 0}\quad
                  \beta^{-1} \! \int_{0}^\infty \! \diff x_\nu 
                  \frac{\pi^2 + x_\nu^2}{2 (1 + e^{x_\nu})}
                = \frac{457}{5040} \frac{\pi^6}{\beta} \,.
    \label{eq:I-DU}
\end{align}
In other words,
\begin{align}
    Q^\nu_\text{DU} &= \Gamma_\text{DU}^\text{prod} \frac{2 I_\text{DU}}{I_\text{DU}^\prime}
                     = \frac{457}{10080}
                       \frac{\pi m_n^{\star} m_p^\star \mu_\mu \Theta_\text{np$\mu$}}
                            {\beta^6} G_F^2 \cos^2 \theta_C (1+3g_A^2) \,,
    \label{eq:DUneutrinoemission}
\end{align}
in agreement with eq.~(120) of ref.~\cite{Yakovlev:2000jp}. The factor $2$ multiplying the ratio $I_\text{DU}/I^\prime_\text{DU}$ arises because neutrinos are emitted both when muons are produced and when they are absorbed. (The production and absorption rates are equal in beta equilibrium.)

\subsubsection{Muon Production via the Modified Urca Process}
\label{sec:MU-prod}

The rate of muon production from the modified Urca process, 
\begin{align}
    n(p_1) + n(p_2) \to n(p_3) + p(p_4) + \mu (p_5\text{ or } p_\mu) + \overline{\nu}_\mu(p_\nu)\,,
\end{align}
follows from similar arguments as the direct Urca rate. The starting point is again a phase space integral which now includes also the momenta of the spectator nucleons:
\begin{align}
    \Gamma_\text{MU}^\text{prod} &=
        \int\!\bigg[ \prod_{j=1}^5 \frac{\diff^3 p_j}{(2\pi)^3} \bigg]
        \frac{\diff^3 p_\nu}{2 E_\nu(2\pi)^3} \frac{(2\pi)^4}{2 E_\mu s}
        \delta^{(4)}(p_f - p_i) \mathscr{L}_\text{blocking}
        \sum_\text{spins} \left| \mathcal{M}_\text{MU}\right|^2\,.
    \label{eq:Gamma-MU-prod-0}
\end{align}
Here, $s=2$ is a symmetry factor, $j = 1, \dotsc, 4$ labels the nucleons in the system $1 + 2 \to 3 + 4$, while $p_\mu = p_5$ and $p_\nu$ are the muon and neutrino four-momenta, respectively. In the 4-momentum-conserving delta function $p_i$ and $p_f$ are the sums of the initial- and final-state four-momenta. The phase space factor that also accounts for Pauli blocking is
\begin{align}
    \mathscr{L}_\text{blocking} = f_1 f_2 (1-f_3) (1-f_4) (1-f_5) \,,
\end{align}
with the same conventions as in \cref{eq:L-blocking-DU-prod}. We begin as before by separating the angular and absolute momenta integrations:
\begin{align}
    \Gamma_\text{MU}^\text{prod} &=
        \frac{1}{8 (2\pi)^{14}} \int\!\bigg[ \prod_{j=1}^5 |\mathbf{p}_j|^2 \diff |\mathbf{p}_j| \diff \Omega_j \bigg] 
        \frac{\diff |\mathbf{p}_\nu| \diff \Omega_\nu E_\nu}{E_\mu} \,
        \delta^{(3)}(\vec{p}_f - \vec{p}_i) \delta(E_f - E_i)
        \mathscr{L}_\text{blocking} \sum_\text{spins} \left| \mathcal{M}_\text{MU}\right|^2\,.
    \label{eq:Gamma-MU-prod-1}
\end{align}
To proceed, we make two important assumptions:
\begin{enumerate}
    \item As in \cref{sec:DU-prod}, we assume a hierarchy between the Fermi energies and the temperature of the system, so that the integrals are dominated by the region within $\mathcal{O}(T)$ about the equilibrium Fermi momenta of the nucleons and electrons. 
    \item The angular integration is simplified in scenarios where the neutron Fermi momentum is much larger than the Fermi momenta of protons and muons, $p_{Fn} \gg p_{Fp}, p_{Fe}, p_{F\mu}$.
\end{enumerate}
Neglecting the neutrino momentum of order $T$ and assuming $|\vec{p}_2| \gg |\vec{p}_4 + \vec{p}_5|$ (which amounts to the second assumption above) the angular integrals can be performed in analogy to \cref{sec:DU-prod}:
\begin{align}
    \int \bigg[ \prod_{j=1}^5 \diff \Omega_j \bigg] \diff \Omega_\nu \delta^{(3)}(\vec{p}_f - \vec{p}_i)
        = 4 \pi \int \bigg[ \prod_{j=1}^5 \diff\Omega_j \bigg]
          \delta^{(3)}(\vec{p}_3 + \vec{p}_4 + \vec{p}_5 - \vec{p}_1 - \vec{p}_2)
        \simeq \frac{2\pi (4\pi)^4}{|\vec{p}_1| |\vec{p}_2| |\vec{p}_3|} \,.
\end{align}
Expressing the momentum integrals in terms of energy integrals by using $\diff |\vec{p}_\nu| \simeq \diff E_\nu$ for the neutrino and $|\vec{p}| \diff |\vec{p}| \simeq m^\star \diff E$ for massive particles, the rate becomes
\begin{align}
     \Gamma_\text{MU}^\text{prod} &=
         \frac{2m_n^{\star3} m_p^\star}{(2\pi)^{9}}
         \int \bigg[\prod_{j=1}^5\diff E_j \bigg] \diff E_\nu |\vec{p}_4| |\vec{p}_\mu| E_\nu  \,
         \delta(E_f - E_i)  \mathscr{L}_\text{blocking}
         \sum_\text{spins} \left| \mathcal{M}_\text{MU}\right|^2\,.
    \label{eq:Gamma-MU-prod-2}
\end{align}
As in the discussion following \cref{eq:Gamma-DU-prod-2}, we now turn to dimensionless integration variables $x_j = \pm \beta (E_j - \mu_j)$, where the $\pm$ refers to incoming/outgoing states. Firstly with these definitions the Pauli-blocking factor simplifies to
\begin{align}
    \mathscr{L}_\text{blocking} &= \prod_{j=1}^5 \frac{1}{1+e^{x_j}} \,,
\end{align}
while the remaining delta function becomes 
\begin{align}
    \delta(E_f - E_i) &= \delta(E_3 + E_4 + E_5 + E_\nu - E_1 - E_2) \,,\notag\\
        &= \delta\Big( \beta^{-1} \Big[x_\nu - \sum_{j=1}^5 x_j \Big]
                     + \mu_p + \mu_\mu - \mu_n \Big)
         = \beta \, \delta\Big(\sum_{j=1}^5 x_j - \overline{x}_\nu\Big) \,.
\end{align}
Combining these results gives 
\begin{align}
    \Gamma_\text{MU}^\text{prod} &=
        \frac{2m_n^{\star3} m_p^\star \, p_{Fp}}{(2\pi)^{9} \beta^7}
        \frac{p_{F\mu}}{\mu_\mu}
        \sum_\text{spins} \left| \widetilde{\mathcal{M}}_\text{MU}\right|^2
        \int \! \bigg[ \prod_{j=1}^5\diff x_j \frac{1}{1+e^{x_j}} \bigg]
        \diff x_\nu \, x_\nu^2 \, \delta\Big(\sum_{j=1}^5 x_j-\overline{x}_\nu\Big)  \,,
    \label{eq:Gamma-MU-prod-3}
\end{align}
where we have pulled out the energy dependence of the spin-summed matrix element, see \cref{eq:murca-matrixelement}, and set the momentum of the proton to its Fermi momentum. The last step is the evaluation of the final integrals:
\begin{align}
    I^\prime_\text{MU} &= \int_0^\infty \diff x_\nu \, x_\nu^2 J_\text{MU} \,,
\end{align}
with
\begin{align}
    J_\text{MU} &= \int\!\bigg[\prod_{j=1}^5\diff x_j \frac{1}{1+e^{x_j}} \bigg] 
        \delta\Big(\sum_{j=1}^5 x_j-\overline{x}_\nu\Big) \,.
\end{align}
Following standard complex integration techniques as described above in \cref{sec:DU-prod} or in Appendix~F of Ref.~\cite{Shapiro:1983du}, $J_\text{MU}$ evaluates to
\begin{align}
    J_\text{MU} &= \frac{1}{1 + e^{\overline{x}_\nu}} 
        \left( \frac{3\pi^4}{8} + \frac{5\pi^2}{12} \overline{x}_\nu^2
             + \frac{1}{24} \overline{x}_\nu^4 \right) \,.
\end{align}
The final integration over $x_\nu$ is then
\begin{align}
    I^\prime_\text{MU}
        &= \int_{B}^\infty \! \diff \overline{x}_\nu
           \frac{(\overline{x}_\nu - B)^2}{1 + e^{\overline{x}_\nu}}
           \left( \frac{3\pi^4}{8} + \frac{5\pi^2}{12} \overline{x}_\nu^2
                + \frac{1}{24} \overline{x}_\nu^4 \right) \,, \\ 
        &= -\frac{3}{8} B^2 \pi^4 \left(B + \log[1+e^{-B}] - \log[1+e^B]\right)
           - \frac{1}{12}(B^2 + \pi^2)(B^2 + 9\pi^2) \Li_3(z) \notag\\
        &\qquad - B (B^2 + 5\pi^2) \Li_4(z)
                - 2 (3 B^2 + 5\pi^2) \Li_5(z)
                - 20 B \Li_6(z)
                - 30 \Li_7(z) \,,
    \label{eq:I-MU-prime}
\end{align}
where again $z = -\exp(-B)$. In beta equilibrium ($B = 0$), \cref{eq:I-MU-prime} simplifies to
\begin{align}
    \lim_{B\to 0} I^\prime_\text{MU}
        &= \frac{3}{32} \left(6\pi^4 \zeta(3) + 100 \pi^2 \zeta(5) + 315\zeta(7) \right)
        \simeq 191.6 \,.
\end{align}
Our final expression for the muon production rate via modified Urca reactions is
\begin{align}
    \Gamma_\text{MU}^\text{prod}
        &= \frac{G_F^2 \cos^2\theta_C g_A^2 m_n^{\star3} m_p^\star \,p_{Fp}}{\pi^9\beta^7}
           \frac{p_{F\mu}}{\mu_\mu}
           \alpha_n \beta_n \bigg( \frac{g_{\pi N N}}{m_\pi} \bigg)^4
           I^\prime_\text{MU} \,,
    \label{eq:Gamma-MU-prod-4}
\end{align}
with $I^\prime_\text{MU}$ from \cref{eq:I-MU-prime}. Note that the factor $p_{F\mu}/\mu_\mu$ arises for muons but not for electrons because the rest-mass of the muon is not negligible compared to the Fermi momenta in the problem. \\

\noindent
As for the direct Urca case, it is once again useful to relate $\Gamma_\text{MU}^\text{prod}$ to the neutrino luminosity due to modified Urca processes involving electrons, $Q^{\nu,e}_\text{MU}$, which is commonly found in the literature \cite{Yakovlev:2000jp}:
\begin{align}
    Q^{\nu,e}_\text{MU}
        &= \Gamma_\text{MU}^\text{prod} \frac{\mu_\mu}{p_{F\mu}}\frac{2 I_\text{MU}}{I^\prime_\text{MU}}
         = \frac{11513}{30240} \frac{G_F^2 \cos^2\theta_C g_A^2 m_n^{\star3} m_p^\star \, p_{Fp}}
                                    {2\pi \beta^8} \alpha_n \beta_n
           \bigg( \frac{g_{\pi N N}}{m_\pi} \bigg)^4 \,,
    \label{eq:MUneutrinoemission}
\end{align}
with
\begin{align}
    \lim_{B \to 0} I_\text{MU}
        &\equiv \lim_{B \to 0} \beta^{-1} \int_0^\infty \diff x_\nu \, x_\nu^3
                                                       J_\text{MU} \notag\\
        &= \beta^{-1} \int_{0}^\infty \diff x_\nu
           \frac{x_\nu^3}{e^{x_\nu}+1} \left( \frac{3\pi^4}{8}
                                             +\frac{5\pi^2}{12} x_\nu^2
                                             + \frac{1}{24} x_\nu^4 \right)
         = \frac{11513}{120960} \frac{\pi^8}{\beta} \,.
    \label{eq:I-MU}
\end{align}
The factor $2$ multiplying the ratio $I_\text{MU}/I^\prime_\text{MU}$ in \cref{eq:MUneutrinoemission} arises as the neutrino luminosity includes the process in both the forward and inverse direction.  The result in \cref{eq:MUneutrinoemission} agrees with that of Ref.~\cite{Yakovlev:2000jp}, but is in contradiction with earlier results and with the routines implemented in \nscool, where the final result depends on $p_{Fe}$ rather than $p_{Fp}$. We have corrected this in the version of \nscool used to produce the numerical results shown in this paper. \\

\noindent
To summarize this section, we put together both the direct Urca and modified Urca contributions to the muon production rate and obtain
\begin{align}
    \Gamma^\text{prod} &= \Gamma_\text{DU}^\text{prod}
        + \Gamma_\text{MU}^\text{prod} \,, \\[0.2cm]
    &= Q^\nu_\text{DU} f_\text{DU}
         + Q^{\nu,e}_\text{MU} f_\text{MU}\,, 
\end{align}
where 
\begin{align}
    f_\text{DU} &= \frac{I_\text{DU}^\prime}{2 I_\text{DU}}
                 = \frac{2520}{457} \frac{\beta}{\pi^6} I_\text{DU}^\prime \,, \\
    f_\text{MU} &= \frac{p_{F\mu}}{\mu_\mu} \frac{I_\text{MU}^\prime}{2 I_\text{MU}}
                 = \frac{60480}{11513} \frac{\beta}{\pi^8} \frac{p_{F\mu}}{\mu_\mu} I_\text{MU}^\prime \,,
\end{align}
and with $I_\text{DU}^\prime$ from \cref{eq:I-DU-prime}, $I_\text{MU}^\prime$ from \cref{eq:I-MU-prime}.

\subsection{Muon Absorption}
\label{sec:mu-absorption}

We now turn to the calculation of the absorption rate of muons (per unit time), which as before contains two contributions 
\begin{align}
    \Gamma^\text{abs} &= \Gamma_\text{DU}^\text{abs}  + \Gamma_\text{MU}^\text{abs} \,.
    \label{eq:Gamma-abs}
\end{align}

\subsubsection{Muon Absorption via the Direct Urca Process}
\label{sec:DU-abs}

The rate in \cref{eq:Gamma-abs} is dominated by the direct Urca process,
\begin{align}
    p(p_1) + \mu (p_3\text{ or } p_\mu) \to n(p_2)  + \overline{\nu}_\mu(p_\nu) \,,
\end{align}
where kinematically accessible, with the rate given by
\begin{align}
    \Gamma_\text{DU}^\text{abs} &=
        \int \bigg[ \prod_{j=1}^2 \frac{\diff^3 p_j}{(2\pi)^3} \bigg] \frac{\diff^3 p_\nu}{2E_\nu(2\pi)^3}
        \frac{(2\pi)^4}{2 E_\mu} \delta^{(4)}\left(p_f - p_i\right) \mathscr{L}_\text{blocking} \, \frac{1}{2}\sum_\text{spins} \left| \mathcal{M}_\text{DU}\right|^2\,.
    \label{eq:Gamma-DU-abs-1}
\end{align}
The factor $1/2$ in front of the matrix element squared takes care of the spin averaging of the additional fermion in the initial state compared to the muon production case. Specifying the form of the Pauli-blocking factor 
\begin{align}
    \mathscr{L}_\text{blocking} &= f_1 (1-f_2) \,,
\end{align}
as well as performing the angular integrals results in 
\begin{align}
    \Gamma_\text{DU}^\text{abs} &= \frac{m_n^\star m_p^\star \Theta_\text{np$\mu$}}{4(2\pi)^3}
        \int \bigg[ \prod_{j=1}^2 \diff E_j \bigg] \diff E_\nu
        \frac{E_\nu^2}{|\vec{p}_\mu|} \, \delta(E_f-E_i) \, f_1 (1-f_2)
        \sum_\text{spins} \left| \widetilde{\mathcal{M}}_\text{DU}\right|^2 \,.
\end{align}
Using the same arguments as in \cref{sec:mu-production} and introducing again the dimensionless variables $x_j = \pm \beta (E_j - \mu_j)$, the delta function becomes 
\begin{align}
    \delta(E_f-E_i)
        &= \delta(E_\nu+ E_2 - E_1 - E_\mu) \,,\\
        &= \delta\Big( \beta^{-1} \Big[ x_\nu - \sum_{j=1}^2 x_j \Big] + \mu_n - \mu_p - E_\mu \Big)
         = \delta\Big( \beta^{-1} \Big[ \widetilde{x}_\nu-\sum_{j=1}^2 x_j \Big] \Big) \,,
\end{align}
where we define
\begin{align}
    \widetilde{x}_\nu \equiv x_\nu + A
    \qquad\text{and}\qquad
    A\equiv \beta (\mu_n - \mu_p - E_\mu) \,.
    \label{eq:x-nu-tilde}
\end{align}
$A$ is a measure of the muons' distance to the Fermi surface. In dimensionless integration variables, the absorption rate can now be written as
\begin{align}
    \Gamma_\text{DU}^\text{abs} &= \frac{m_n^\star m_p^\star\Theta_\text{np$\mu$}}{4(2\pi)^3 \beta^4 p_{F\mu}}
        \int \bigg[ \prod_{j=1}^2 \diff x_j(1+e^{x_j})^{-1} \bigg] \diff \widetilde{x}_\nu
        (\widetilde{x}_\nu - A)^2 \delta\Big( \widetilde{x}_\nu - \sum_{j=1}^2 x_j \Big)
        \left|\widetilde{\mathcal{M}}_\text{DU}\right|^2 \,.
    \label{eq:Gamma-DU-abs-2}
\end{align}
The remaining integrals are
\begin{align}
    I_\text{DU}^{\prime\prime}
        &= \int_A^\infty \! \diff \widetilde{x}_\nu \left(\widetilde{x}_\nu-A\right)^2
           J^\prime_\text{DU} \,.
\end{align}
With
\begin{align}
    J_\text{DU}^\prime
        &= \int \bigg[\prod_{j=1}^2 \diff x_j\frac{1}{1+e^{x_j}} \bigg]
           \delta\Big(\widetilde{x}_\nu - \sum_{j=1}^2 x_j \Big)
         = \frac{\widetilde{x}_\nu}{e^{\widetilde{x}_\nu} - 1} \,,
\end{align}
they evaluate to
\begin{align}
     I_\text{DU}^{\prime\prime} &= 2 \left[A \, \Li_3(e^{-A}) + 3 \, \Li_4(e^{-A}) \right]\,.
     \label{eq:I-DU-pp}
\end{align}
Putting everything together we get the expression for the direct Urca muon absorption rate,
\begin{align}
    \Gamma_\text{DU}^\text{abs}
        &= \frac{m_n^\star m_p^\star\Theta_\text{np$\mu$}}{4\pi^3 \beta^4 p_{F\mu}} G_F^2 \cos^2 \theta_C (1 + 3 g_A^2) I_\text{DU}^{\prime\prime} 
                                  \notag\\
        &= Q^\nu_\text{DU} \frac{\beta \pi^2}{4 \mu_\mu p_{F\mu}}\frac{I_\text{DU}^{\prime\prime}}{I_\text{DU}} \,,
\end{align}
with $I_\text{DU}$ from \cref{eq:I-DU} and $I_\text{DU}^{\prime\prime}$ from \cref{eq:I-DU-pp}.

\subsubsection{Muon Absorption via the Modified Urca Process}

We now turn to muon absorption via the modified Urca process,
\begin{align}
    \mu(p_\mu) + n(p_1) + p(p_2) \to n(p_3) + n(p_4) + \nu_\mu(p_\nu) \,,
\end{align}
where the quantities in parenthesis are again the four-momenta of the particles in question. In order to determine the rate we take the matrix element of \cref{eq:murca-matrixelement} and perform the phase space integration as before (but keeping the muon 4-momentum fixed). That is, 
\begin{align}
    \Gamma_\text{MU}^\text{abs}
        &= \int \! \bigg[ \prod_{j=1}^4 \frac{\diff^3 p_j}{(2\pi)^3} \bigg]
           \frac{\diff^3 p_\nu}{2E_\nu(2\pi)^3} \frac{(2\pi)^4}{2E_\mu s} \delta^{(4)}(p_f - p_i) \,
           \mathscr{L}_\text{blocking} \frac{1}{2} \sum_\text{spins} \left| \mathcal{M}_\text{MU}\right|^2\,,
    \label{eq:Gamma-MU-abs-0}
\end{align}
where $s=2$ as before. The two differences compared to \cref{eq:Gamma-MU-prod-0} are the omission of the integration over muon momenta and the Pauli-blocking factor, which changes to 
\begin{align}
    \mathscr{L}_\text{blocking} = f_1 f_2 (1-f_3) (1-f_4) \,.
\end{align}
As before we begin by performing the angular integrals
\begin{align}
    \int\!\bigg[\prod_{j=1}^4 \diff \Omega_j \bigg] \diff \Omega_\nu \, \delta^{(3)}(\vec{p}_f - \vec{p}_i)
        &= 4\pi \int\!\bigg[\prod_{j=1}^4 \diff \Omega_j \bigg]
           \delta^{(3)}(\vec{p}_3 + \vec{p}_4 - \vec{p}_1 - \vec{p}_2 - \vec{p}_\mu) \notag\\
        &= \frac{2\pi \, (4\pi)}{|\vec{p}_3|}  \int\!\diff \Omega_2 \diff \Omega_4
           \frac{1}{|\vec{p}_1||\vec{p}_s - \vec{p}_4|}
        \simeq  \frac{2\pi (4\pi)^3}{|\vec{p}_1| |\vec{p}_3| |\vec{p}_4|} \,,
\end{align}
where $\vec{p}_s =\vec{p}_2 + \vec{p}_\mu - \vec{p}_\nu$, and we have used the approximation $p_{Fn} \gg p_{Fp}$ which implies $|\vec{p}_s - \vec{p}_4| \simeq |\vec{p}_4|$. Inserting this result into \cref{eq:Gamma-MU-abs-0}, pulling the energy-dependence out of the matrix elements as in \cref{eq:murca-matrixelement}, and replacing momentum integrals by energy integrals, we obtain for the rate
\begin{align}
    \Gamma_\text{MU}^\text{abs}
        &= \frac{m_n^{\star3} m_p^\star \, p_{Fp}}{2(2\pi)^7}
           \int\!\bigg[\prod_{j=1}^4\diff E_j \bigg] \diff E_\nu \frac{E_\nu^2}{(E_\nu + E_\mu)^2} \,
           \delta(E_f - E_i)  \mathscr{L}_\text{blocking}
           \sum_\text{spins} \left| \widetilde{\mathcal{M}}_\text{MU}\right|^2 \,.
    \label{eq:Gamma-MU-abs-1}
\end{align}
Pulling $|\vec{p}_2| \simeq p_{Fp}$ out of the integral is justified by the fact that only nucleons whose energy falls within $\sim \pm T$ from the Fermi surface contribute significantly to the rate, combined with $T \ll p_{Fp}$. Introducing dimensionless variables $x_j = \pm \beta (E_j - \mu_j)$, the energy conserving delta function becomes
\begin{align}
    \delta(E_f - E_i)
        &= \delta(E_3 + E_4 + E_\nu - E_1 - E_2 - E_\mu) \,,\notag \\
        &= \delta\Big( \beta^{-1} \Big[x_\nu -  \sum_{j=1}^4 x_j \Big] + \mu_n - \mu_p - E_\mu \Big)
          = \beta \, \delta\Big(\widetilde{x}_\nu - \sum_{j=1}^4 x_j\Big) \,,
\end{align}
with $\widetilde{x}_\nu \equiv x_\nu + A$ and $A \equiv \beta (\mu_n - \mu_p - E_\mu)$ as in \cref{eq:x-nu-tilde}. The absorption rate is now
\begin{align}
    \Gamma_\text{MU}^\text{abs}
        &= \frac{m_n^{\star3} m_p^\star \, p_{Fp}}{2(2\pi)^7 \beta^6 E_\mu^2}
           \int\!\bigg[\prod_{j=1}^4\diff x_j \frac{1}{1+e^{x_j}}\bigg] \diff \widetilde{x}_\nu
           \left(\widetilde{x}_\nu - A\right)^2
           \delta\Big(\widetilde{x}_\nu - \sum_{j=1}^4 x_j \Big)
           \sum_\text{spins} \left| \widetilde{\mathcal{M}}_\text{MU}\right|^2 \,.
    \label{eq:Gamma-MU-abs-2}
\end{align}
In this expression we have neglected the neutrino energy in comparison to the muon energy in the denominator. We perform the remaining integrals in the complex plane using the same techniques as above (see also Ref.~\cite{Shapiro:1983du}). The result is
\begin{align}
    I_\text{MU}^{\prime\prime} &= \int_A^\infty \! \diff \widetilde{x}_\nu
                                      \left(\widetilde{x}_\nu - A\right)^2
                                      J_\text{MU}^\prime \,.
\end{align}
with
\begin{align}
    J_\text{MU}^\prime &= \int \! \bigg[\prod_{j=1}^4\diff x_j \frac{1}{1+e^{x_j}}\bigg]
                          \delta\Big(\widetilde{x}_\nu - \sum_{j=1}^4 x_j\Big)
                        = \frac{4\pi^2 \widetilde{x}_\nu + \widetilde{x}_\nu^3}{6 (e^{\widetilde{x}_\nu} - 1)} \,,
\end{align}
and therefore
\begin{align}
    I_\text{MU}^{\prime\prime} = \frac{1}{3}
        \Big[ (A^3 + 4\pi^2 A) \Li_3(e^{-A})
            + 3 (3 A^2 + 4\pi^2) \Li_4(e^{-A})
            + 36 A \Li_5(e^{-A})
            + 60 \Li_6(e^{-A}) \Big] \,.
    \label{eq:I-MU-pp}
\end{align}
The modified Urca absorption rate is thus
\begin{align}
    \Gamma_\text{MU}^\text{abs}
        &= \frac{m_n^{\star3} m_p^\star \, p_{Fp}}{2(2\pi)^7 \beta^6 E_\mu^2}
           \sum_\text{spins} \left| \widetilde{\mathcal{M}}_\text{MU} \right|^2 I_\text{MU}^{\prime\prime} \,, \\
        &= \frac{G_F^2 \cos^2\theta_C g_A^2 m_n^{\star3} m_p^\star \,p_{Fp}}{\pi^7 \beta^6E_\mu^2}\alpha_n \beta_n
           \bigg( \frac{g_{\pi N N}}{m_\pi} \bigg)^4 I_\text{MU}^{\prime\prime}
         = \frac{\beta\pi^2}{2 E_\mu^2} Q_\text{MU}^\nu \frac{I_\text{MU}^{\prime\prime}}{I_\text{MU}} \,,
    \label{eq:Gamma-MU-abs-3}
\end{align}
with $I_\text{MU}^{\prime\prime}$ from \cref{eq:I-MU-pp} and $I_\text{MU}$ from \cref{eq:I-MU}.\\

\noindent
In summary, the total (direct + modified Urca) muon absorption rate is
\begin{align}
    \Gamma^\text{abs} 
        &= \Gamma_\text{DU}^\text{abs} + \Gamma_\text{MU}^\text{abs} \,,\\
        &= Q^\nu_\text{DU} h_\text{DU} + Q^\nu_\text{MU} h_\text{MU}\,, 
\end{align}
where 
\begin{align}
    h_\text{DU} &= \frac{\beta \pi^2}{4 \mu_\mu p_{F\mu}}\frac{I_\text{DU}^{\prime\prime}}{I_\text{DU}}
                 = \frac{1260\beta^2}{457\pi^4\mu_\mu p_{F\mu}} I_\text{DU}^{\prime\prime}\,, \\
    h_\text{MU} &= \frac{\beta\pi^2}{2 E_\mu^2} \frac{I_\text{MU}^{\prime\prime}}{I_\text{MU}}
                 = \frac{60480\beta^2  }{11513\pi^6 E_\mu^2} I_\text{MU}^{\prime\prime} \,,
\end{align}
and where $I_\text{DU}^{\prime\prime}$ and $I_\text{MU}^{\prime\prime}$ can be read off from \cref{eq:I-DU-pp} and \cref{eq:I-MU-pp}, respectively.

\subsection{Assisted Muon Decay}
\label{sec:assisted-mu-decay}

The calculation of the rate of assisted muon decay,
\begin{align}
    \mu(p_\mu) + p(p_1) \to p(p_2) + e(p_e) + \bar\nu_e(p_{\nu_e}) + \nu_\mu(p_{\nu_\mu}) \,,
\end{align}
proceeds in the same way as the derivation of modified Urca production/absorption rates. In analogy to \cref{eq:Gamma-MU-abs-0}, and with the matrix element from \cref{eq:Msq-assisted-mu-decay}, the rate is given by
\begin{align}
    \Gamma_\text{amd}
        &= \int \! \bigg[ \prod_{j=1}^2 \frac{\diff^3 p_j}{(2\pi)^3} \bigg]
           \frac{\diff^3 p_e}{2 E_e (2\pi)^3}
           \frac{\diff^3 p_{\bar\nu_e}}{2 E_{\bar\nu_e} (2\pi)^3}
           \frac{\diff^3 p_{\nu_\mu}}{2 E_{\nu_\mu} (2\pi)^3}
           \frac{(2\pi)^4}{2E_\mu} \delta^{(4)}(p_f - p_i) \,
           \mathscr{L}_\text{blocking} \sum_\text{spins} \left| \mathcal{M}_\text{amd}\right|^2\,,
    \label{eq:Gamma-amd-0}
\end{align}
Note that the matrix element from \cref{eq:Msq-assisted-mu-decay} uses a Lorentz-invariant normalization for all external particles, including the nucleons, hence the extra factors of $m_p$ in the denominator. To shorten the notation, we define $p_3 \equiv p_e$, $p_4 \equiv p_{\nu_e}$, and $p_5 \equiv p_{\nu_\mu}$. The Pauli blocking factor in \cref{eq:Gamma-amd-0} is
\begin{align}
    \mathscr{L}_\text{blocking} = f_1 (1-f_2) (1-f_3) \,.
\end{align}
The angular integral evaluates to
\begin{align}
    \int\!\bigg[\prod_{j=1}^5 \diff \Omega_j \bigg] \, \delta^{(3)}(\vec{p}_f - \vec{p}_i)
         &= (4\pi)^2 \int\!\bigg[\prod_{j=1}^3 \diff \Omega_j \bigg]
            \delta^{(3)}(\vec{p}_3 + \vec{p}_2 - \vec{p}_1 - \vec{p}_\mu) \notag\\
         &= \frac{(4\pi)^2}{|\vec{p}_3|^2} \int\!\bigg[\prod_{j=1}^3 \diff \Omega_j \bigg]
            \delta(|\vec{p}_3| - |\vec{p}_2 - \vec{p}_1 - \vec{p}_\mu|)
            \delta^{(2(}(\Omega_3 - \Omega_{1-2-\mu}) \notag\\
         &= \frac{(4\pi)^2}{|\vec{p}_3|^2} \int\!\bigg[\prod_{j=1}^2 \diff \Omega_j \bigg]
            \delta(|\vec{p}_3| - \sqrt{|\vec{p}_2|^2 + |\vec{p}_1 + \vec{p}_\mu|^2
                                     - 2 |\vec{p}_2| |\vec{p}_1 + \vec{p}_\mu| \cos\theta_2})
                                                      \notag\\
         &= \frac{2\pi (4\pi)^2}{|\vec{p}_3| |\vec{p}_2|}
            \int\! \diff \Omega_1 \frac{1}{|\vec{p}_1 + \vec{p}_\mu|} \notag\\[0.2cm]
         &\simeq \frac{2\pi (4\pi)^3}{|\vec{p}_3| |\vec{p}_2| |\vec{p}_1|}
\end{align}
In the last step, where we have evaluated the integral $\int \! \diff \Omega_1 \, |\vec{p}_1 + \vec{p}_\mu|^{-1} \simeq 4\pi / \max(|\vec{p}_1|, |\vec{p}_\mu|)$, we have used the fact that the proton Fermi momentum is typically larger than the muon Fermi momentum.
Replacing the momentum integrals by energy integrals then leads to
\begin{align}
    \Gamma_\text{amd}
        &= \frac{(m_p^*)^2}{2 (2\pi)^7}
           \int \! \bigg[ \prod_{j=1}^5 \diff E_j \bigg]
           \frac{E_{\bar\nu_e} E_{\nu_\mu}}{E_\mu}
           \delta(E_{\nu_\mu} + E_{\nu_e} + E_e + E_2 - E_\mu - E_1) \,
           \mathscr{L}_\text{blocking} \sum_\text{spins} \left| \mathcal{M}_\text{amd}\right|^2
                                                 \notag\\
        &= \frac{(m_p^*)^2 \mu_e E_\mu^2}{2 (2\pi)^7 \beta^8}
           \int \! \bigg[ \prod_{j=1}^5 \diff x_j \bigg]
           x_{\bar\nu_e}^2 x_{\nu_\mu}^2 \,
           \delta\Big( \sum_{j=1}^3 x_j - \hat{x}_\nu \Big)
           \bigg[ \prod_{j=1}^3 \frac{1}{1 + e^{x_j}} \bigg]
           \sum_\text{spins} \left| \widetilde{\mathcal{M}}_\text{amd}\right|^2 \,,
    \label{eq:Gamma-amd-1}
\end{align}
where in the second line we have again defined $x_j \equiv \pm \beta (E_j - \mu_j)$ (with a plus sign for incoming particles and a minus sign for outgoing particles), $x_{\nu_\mu} \equiv \beta E_{\nu_\mu}$, $x_{\nu_e} \equiv \beta E_{\nu_e}$, and $\hat{x}_\nu = x_{\nu_e} + x_{\nu_\mu} - \beta (E_\mu - \mu_e)$.  We have moreover pulled the energy-dependent factors out of the squared matrix element from \cref{eq:Msq-assisted-mu-decay}, and in the resulting prefactor have set the electron energy equal to the corresponding Fermi energy.

We first evaluate the integral over the nucleon and charged lepton momenta,
\begin{align}
    J_\text{amd}
        \equiv \int \bigg[ \prod_{j=1}^3 \diff x_j \frac{1}{1 + e^{x_j}} \bigg]
               \delta\Big( \sum_{j=1}^3 x_j - \hat{x}_\nu \Big) \,.
    \label{eq:J-amd-1}
\end{align}
This expression is identical to the one we found in the case of muon production via the direct Urca process, $J_\text{DU}$, in \cref{eq:J-DU-1}, so we can immediately read off the result from \cref{eq:J-DU-2}:
\begin{align}
  J_\text{amd} = \frac{\pi^2 + \hat{x}_\nu^2}{2(1 + e^{\hat{x}_\nu})}
  \label{eq:J-amd-2}
\end{align}
The remaining integral over dimensionless neutrino energies, $x_{\bar\nu_e}$ and $x_{\nu_\mu}$, is then
\begin{align}
    I_\text{amd}
        &\equiv \int_0^\infty \!\diff x_{\bar\nu_e} \diff x_{\nu_\mu}
            x_{\bar\nu_e}^2 x_{\nu_\mu}^2
            \frac{\pi^2 + \hat{x}_\nu^2}{2(1 + e^{\hat{x}_\nu})}
                                                    \notag\\[0.2cm]
        &= - 2 \, (C^2 + \pi^2) \, \Li_6(-e^{-C})
           - 24 \, C \, \Li_7(-e^{-C})
           - 84 \, \Li_8(-e^{-C}) \,.
    \label{eq:x-nu-int-amd}
\end{align}
with the shorthand notation $C \equiv \beta (\mu_e - E_\mu)$.  We conclude that the rate for assisted muon decay is
\begin{align}
    \Gamma_\text{amd}
        &= \frac{16 (m_p^*)^2 \mu_e E_\mu^2 \alpha^2 G_F^2}
                {\pi^5 \beta^8 m_\mu^8} I_\text{amd} \,.
    \label{eq:Gamma-amd-2}
\end{align}

\section{Muon decay width and neutrino spectra}
\label{sec:mu-width}

In this section, we calculate the muon decay width including Pauli blocking effects in presence of a degenerate electron gas. Moreover, we discuss the computation of the resulting neutrino spectra as a function of the electron Fermi energy.

The muon decay width in presence of a dense background of electrons with chemical potential $\mu_e$ is given by
\begin{align}
    \diff \Gamma &= \frac{(2\pi)^7}{2m_\mu} \sum_\text{spins} |\mathcal{M}|^2
                    \diff \Phi_2(q;p_1,p_2) \diff \Phi_2(P_\mu;q,p_e) \diff q^2
                    [1 - f(|\mathbf{p}_e|,\mu_e)]\,.
    \label{eq:dGamma-mu}
\end{align}
Here $q^2$ is the invariant mass of the two outgoing neutrinos, whose individual four-momenta are labeled $p_1$ and $p_2$ (where $q/2=|\mathbf{p}_1|=|\mathbf{p}_2|$), while $P_\mu$ and $p_e$ are the four-momenta of the muon and electron, respectively. The last term on the right-hand side is the Pauli-blocking factor, with the Fermi--Dirac distribution from \cref{eq:FD-distribuion}.  Note that we have sub-divided the usual three-body phase space into the product of two two-body decays (see Eq. (49.13) in Ref.~\cite{ParticleDataGroup:2020ssz}), with
\begin{align}
    \diff\Phi_2 = \frac{|\mathbf{p}|}{4(2\pi)^6 M} \diff\Omega(\theta,\phi)\,.
\end{align}
In the above, $M$ is the center-of-mass energy of the two-body system, while $\diff\Omega = \diff(\cos\theta) \diff\phi$ is the solid angle and $\mathbf{p}$ is the three-momentum of one of the outgoing particles in the two-body center-of-mass frame (neglecting particle masses).

\subsection{Decay Width for Muons Decaying at Rest}

\begin{figure}
    \centering
    \makebox[1.2\textwidth][c]{
        \hspace{-3.5cm}
        \includegraphics{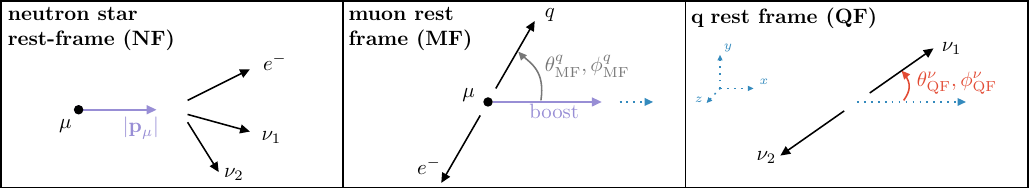}
    }
    \caption{Sketch of the three different frames where kinematic quantities relevant to muon decay are defined.}
    \label{fig:muon-kinematics-frames}
\end{figure}

While we are ultimately interested in the decay width and the neutrino spectra for boosted muons, it is useful to first compute these quantities in the muon rest frame. To avoid confusion we will label frame dependent quantities with neutron star rest frame (NF), muon rest frame (MF) and finally $q$ rest frame (QF), where $q$ is the four-momentum of the two-neutrino system, illustrated in \cref{fig:muon-kinematics-frames}.
Specifying these frames leads to the simplifications
\begin{align}
    P_\mu^\text{MF} &= (m_\mu; \mathbf{0}) \,,
    \qquad&\qquad
    p_1^\text{QF} &= |\mathbf{p}_1|(1; \cth{QF}, \cph{QF}\sth{QF}, \sph{QF}\sth{QF})\,,\\
    p_e^\text{MF} &= E_e(1; -1, 0, 0) \,,
    \qquad&\qquad
    p_2^\text{QF} &= |\mathbf{p}_2|(1; -\cth{QF}, -\cph{QF}\sth{QF}, -\sph{QF}\sth{QF})\,.
\end{align}
Here $p_1^\text{QF}$ and $p_2^\text{QF}$ are the neutrino momenta defined in the QF, and we abbreviate $\cth{QF} \equiv \cos\tht{QF}$, $\sth{QF} \equiv \sin\tht{QF}$. Finally, for the muon at rest the boost is absent. This allows us to choose the $q$-vector along the $x$-axis, corresponding to $\tht{MF}=0$. Note that we are neglecting both electron and neutrino masses which allows the electron energy to be expressed as
\begin{align}
    q^2 &= (P_\mu^\text{MF} - p_e^\text{MF})^2
         = m_\mu^2 - 2 E_e^\text{MF} m_\mu
    \qquad \Longrightarrow \qquad
    E_e^\text{MF} = \frac{m_\mu^2-q^2}{2m_\mu}\,.
    \label{eq:Ee-q-relation}
\end{align}
The boost factors for the Lorentz transformation $\Lambda_\text{QF--MF}$ from the QF to the MF are
\begin{align}
    \gamma = \frac{m_\mu^2+q^2}{2 m_\mu q} \,,
    \qquad\qquad \qquad
    \beta\gamma = \frac{m_\mu^2 - q^2}{2 m_\mu q}\,.
    \label{eq:QF-MF-boost-factors}
\end{align}
Taking this boost to be along the $x$-axis we obtain
\begin{align}
    p_1^\text{MF} &= \Lambda_\text{QF--MF} \, p_1^\text{QF} \,, \notag\\
                   &= \left( \frac{1}{4m_\mu}\left[m_\mu^2 + q^2 +\cth{QF}(m_\mu^2 - q^2)\right];
                            \frac{1}{4m_\mu}\left[m_\mu^2 - q^2 +\cth{QF}(m_\mu^2 + q^2)\right] ,
                            \frac{q \, \cph{QF} \sth{QF}}{2} ,
                            \frac{q \, \sph{QF} \sth{QF}}{2} \right) \,,\\
    p_2^\text{MF} &=\Lambda_\text{QF--MF} \, p_2^\text{QF} \notag\,,\\
                   &= \left( \frac{1}{4m_\mu}\left[m_\mu^2 + q^2 -\cth{QF}(m_\mu^2 - q^2)\right];
                            \frac{1}{4m_\mu}\left[m_\mu^2 - q^2 -\cth{QF}(m_\mu^2 + q^2)\right],
                           -\frac{q \, \cph{QF} \sth{QF}}{2} ,
                           -\frac{q \, \sph{QF} \sth{QF}}{2} \right) \,. 
\end{align}
The Lorentz-invariant matrix element, expressed in terms of muon rest frame quantities, is
\begin{align}
    \sum_\text{spins} |\mathcal{M}|^2
        &= 64 \, G_F^2 (p_e^\text{MF} \cdot p_1^\text{MF}) (p_2^\text{MF}\cdot P_\mu^\text{MF}) \notag\\
        &= 4 G_F^2 (m_\mu^2-q^2) (1-\cth{QF}) \left[m_\mu^2(1+\cth{QF}) + q^2(1-\cth{QF})\right] \,.
\end{align}
Putting everything together and performing the angular integrals over those angles which do not appears in the matrix element or in the Fermi--Dirac distribution yields
\begin{align}
    \diff\Gamma &= \frac{G_F^2}{128\pi^3 m_\mu^3}
                   (m_\mu^2 - q^2)^2 (1 - \cth{QF})
                   \left[m_\mu^2 (1 + \cth{QF}) + q^2 (1 - \cth{QF})\right]
                   [1 - f(|\mathbf{p}_e|,\mu_e)] \diff q^2 \diff(\cth{QF}) \,.
    \label{eq:ctheta_dist}
\end{align}
It can be verified that in the limit $\mu_e\to 0$ the usual muon decay width $\Gamma_\text{SM} = G_F^2 m_\mu^5 / (192 \pi^3)$ is recovered. The above differential width is required to determine the distribution of the angle \tht{QF} for the boosted case and need not be further integrated here. 

For the following section we also need the distribution of the electron energies. This is easily obtained by using \cref{eq:Ee-q-relation} to write $\diff q^2 = 2 m_\mu \diff E_e^\text{MF}$. This leads to
\begin{align}
   \lim_{\mu_e\to0} \frac{\diff\Gamma}{\diff E_e}
       &= \frac{G_F^2 m_\mu^4}{48\pi^3} (3 - 2x_e) x_e^2 \,,
       \qquad \text{with} \quad
       x_e = \frac{2E_e^\text{MF}}{m_\mu} \,.
   \label{eq:Ee-dist}
\end{align}

\subsection{Decay Width for Boosted Muons}

\begin{figure}
    \centering
    \includegraphics[width=\textwidth]{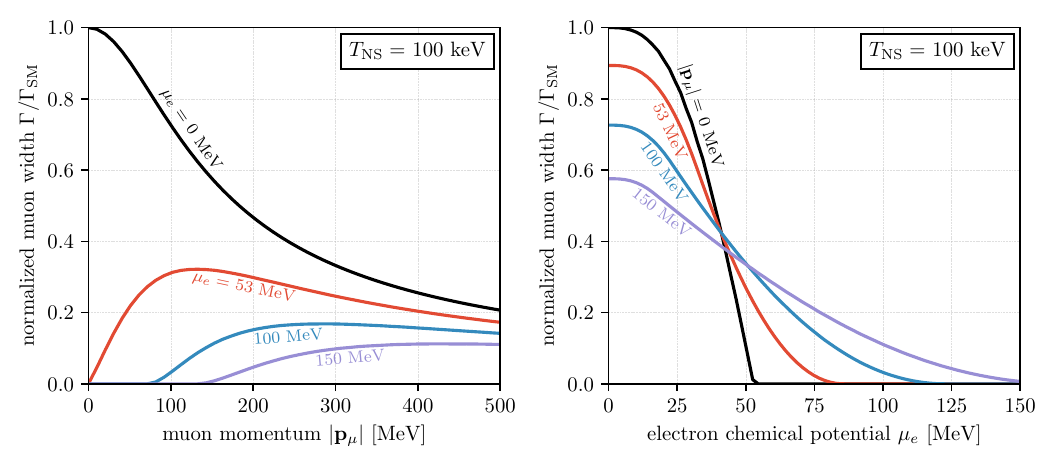}
    \caption{Muon decay width as a function of both the muon momentum (left) and the the electron chemical potential (right) at a typical neutron star temperature $T = \SI{100}{\keV}$.}
    \label{fig:muon-width-results}
\end{figure}

We boost the muon relative to the rest frame of the neutron star's degenerate electron gas and again calculate its width, carefully taking into account Pauli blocking. We begin with the calculation of the muon width in the rest frame of the neutron star. The first quantity that we need is the energy of the electron in the NF. To simplify this expression we choose the boost along the $x$-direction with the appropriate boost factors
\begin{align}
    \gamma_\text{MF--NF} = \frac{E_\mu}{m_\mu} \,,
    \qquad\qquad
    \beta_\text{MF--NF} = \sqrt{1-\gamma_\text{MF--NF}^{-2}} = \frac{|\vec{p}_\mu|}{m_\mu} \,.
\end{align}
We define the four-momenta in the MF as
\begin{align}
    P_\mu^\text{MF} &= (m_\mu; \mathbf{0}) \,,
    \qquad&\qquad
    p_e^\text{MF}   &= E_e^\text{MF} (1; -\cthp{MF}, -\sthp{MF} \cphp{MF}, -\sthp{MF} \sphp{MF}) \,.
\end{align}
After boosting to the NF, we have
\begin{align}
    p_e^\text{NF} &= \left( \frac{E_e^\text{MF} (E_\mu - |\mathbf{p}_\mu| \cthp{MF})}{m_\mu};
                            \frac{E_e^\text{MF} (|\mathbf{p}_\mu| - E_\mu \cthp{MF})}{m_\mu},
                            -E_e^\text{MF} \sthp{MF} \cphp{MF},
                            -E_e^\text{MF} \sthp{MF} \sphp{MF}\right) \,.
    \label{eq:pe-NF}
\end{align}
The required value for the electron momentum in the Pauli-blocking factor is therefore
\begin{align}
    |\mathbf{p}_e^\text{NF}|
        &= \frac{(m_\mu^2 - q^2) (E_\mu - |\mathbf{p}_\mu| \cthp{MF})}{2m_\mu^2} \,.
    \label{eq:electron-energy-NF}
\end{align}
As we cannot choose both the boost of the muon and the direction of the two-neutrino subsystem to be aligned along a single axis we must define boosts in an arbitrary direction and then re-do the angular integrals compared to the muon-decay-at-rest case above. For a boost in the direction of a general unit vector $\hat{\vec{v}}$, the Lorentz transformation matrix is
\begin{align}
    \Lambda(\hat{\vec{v}},\gamma, \beta) &=
        \begin{pmatrix}
           \gamma                      & \gamma \beta \hat{\vec{v}} \\
           \gamma\beta \hat{\vec{v}}^T & I_3 + (\gamma - 1) \hat{\vec{v}}^T \hat{\vec{v}}
        \end{pmatrix} \,.
\end{align}
Choosing specifically $\hat{\vec{v}} = \hat{\mathbf{v}}_\text{QF--MF}$ to define the transformation $\Lambda_\text{QF--MF} \equiv \Lambda(\hat{\vec{v}},\gamma, \beta)$ from the QF to the MF, we transform the neutrino four-vectors from the QF to MF,
\begin{align}
    p_1^\text{QF} &= \frac{q}{2}\left(1; \cth{QF}, \sth{QF} \cph{QF},  \sth{QF} \sph{QF}\right)\,,
    \qquad&\qquad
    p_2^\text{QF} &= \frac{q}{2}\left(1;-\cth{QF},-\sth{QF} \cph{QF}, -\sth{QF} \sph{QF}\right)\,, \\
    p_1^\text{MF} &= \Lambda_\text{QF--MF} p_1^\text{QF}\,,
    \qquad&\qquad
    p_2^\text{MF} &= \Lambda_\text{QF--MF} p_2^\text{QF}\,.
\end{align}
We do not write out the expressions for $p_1^\text{MF}$ and $p_2^\text{MF}$ explicitly here as they are rather lengthy. The matrix element in the MF, on the other hand, is relatively compact,
\begin{align}
   \sum_\text{spins} |\mathcal{M}|^2
       &= 4 G_F^2 (m_\mu^2 - q^2) (1 + \cth{QF} \cthp{MF} + c_{\ph{QF}-\php{MF}} \sph{QF} \sthp{MF}) \notag \\
          &\qquad\times\left[ m_\mu^2 + q^2 - (m_\mu^2 - q^2) (\cth{QF} \cthp{MF} + c_{\ph{QF}-\php{MF}} \sph{QF} \sthp{MF}) \right] \,.
\end{align}
Putting all the pieces together yields the final differential decay width 
\begin{align}
    \diff \Gamma^\text{boosted}
        &= \frac{(2\pi)^7}{2 E_\mu} \sum_\text{spins} |\mathcal{M}|^2
           \left[ \frac{|\mathbf{p}^\text{MF}_e|}{4(2\pi)^6 m_\mu} \diff\Omega(\thtp{MF},\php{MF}) \right]
        \notag \\ 
        &\qquad \qquad \times \left[ \frac{|\mathbf{p}_1^\text{QF}|}{4(2\pi)^6 q} \diff\Omega(\tht{QF},\ph{QF}) \right]
           \big[ 1 - f(|\mathbf{p}^\text{NF}_e|,\mu_e) \big] \diff q^2 \,,
    \label{eq:dGamma-boosted}
\intertext{and its integral,}
       \Gamma^\text{boosted}
        &= \frac{G_F^2}{384 m_\mu^3 \pi^4} \int_0^{m_\mu^2} \! \diff q^2 \int_{-1}^1 \diff (\cthp{MF})
           \left( m_\mu^6 - 3 m_\mu^2 q^4 + 2 q^6 \right) \big[1 - f(|\mathbf{p}^\text{NF}_e|,\mu_e) \big] \,.
    \label{eq:Gamma-boosted}
\end{align}
Between \cref{eq:dGamma-boosted,eq:Gamma-boosted} we have performed all the angular integrals which are independent of the electron momentum $|\mathbf{p}_e^\text{NF}|$. The remaining integrations over $q^2$ and \cthp{MF} do not yield closed formed solutions and must be done numerically. Results for $\Gamma^\text{boosted}$ are shown in \cref{fig:muon-width-results}.

\subsection{Neutrino spectra}

\begin{figure}
    \centering
    \includegraphics[width =0.6\textwidth]{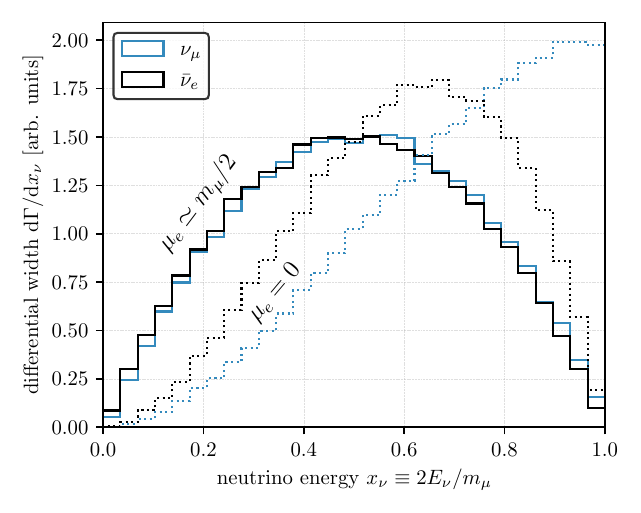}
    \caption{Neutrino energy spectra from muon decay at rest with (solid) and without (dashed) non-zero electron chemical potential. Blue and black lines represent the $\nu_\mu$ and $\bar{\nu}_e$ flavors, respectively. Note that in this plot, we have not included gravitational redshift or neutrino oscillations.}
    \label{fig:neutrino-decay-spectra}
\end{figure}

Based on the above determination of the muon width, the probability of a muon decaying can be determined at each random-walk step in our simulations. If a muon is determined to decay we need to draw the energies of the $\bar{\nu}_e$ and $\nu_\mu$ neutrinos based on the decaying muon's momentum and the Fermi-energy of the electrons at the given radius of the neutron star.  We use the following procedure:
\begin{description}[before={\renewcommand\makelabel[1]{\bfseries ##1.}}]

\item[{\scshape 1}]{\scshape Draw $\thtp{MF}$, $\php{MF}$ and $E_e^\text{MF}$.} We begin by drawing the angles $\thtp{MF}$, $\php{MF}$ between the two-neutrino system and the muon boost vector from uniform distributions, $\thtp{MF} \in [0,\pi]$ and $\php{MF} \in [0,2\pi]$. We randomly choose the electron energy in the muon rest frame ($E_e^\text{MF}$) using the distribution from \cref{eq:Ee-dist}. 

\item[{\scshape 2}]{\scshape Determine $E_e^\text{NF}$.} Once chosen in the muon rest frame, the electron energy is boosted to the NF using \cref{eq:pe-NF} and compared to the Fermi energy of the neutron star electrons at that radius, $p_{Fe}(r)$. We then require $E_e^\text{NF} > p_{Fe}(r)$, i.e., here we neglect temperature effects. If this criterion is not fulfilled we go back to step 1 and choose new $\thtp{MF}$, $\php{MF}$, and $E_e^\text{MF}$ until a configuration is found that allows for an electron energy above the Fermi surface of the electron gas.  

\item[{\scshape 3}]{\scshape Draw \tht{QF}.} The next step requires randomly drawing the angle of the outgoing neutrinos with respect to the center-of-mass of the two-neutrino system. We use the distribution from \cref{eq:ctheta_dist} to draw \tht{QF} randomly (with $q^2$ already fixed by the previous steps according to \cref{eq:electron-energy-NF}). Note that we drop the Pauli-blocking term $1 - f(|\mathbf{p}^\text{NF}_e|,\mu_e)$ in \cref{eq:ctheta_dist} when drawing from this distribution for \tht{QF} as we have already assured that the chosen electron energy lies above the Fermi surface. Having chosen \tht{QF}, the neutrino 4-momenta $p_1^\text{QF}$ and $p_2^\text{QF}$ are unique determined. (The remaining angle \ph{QF} is unimportant given the initial choice of muon boost in the $x$-direction.)

\item[{\scshape 4}]{\scshape Determine $p_1^\text{NF}$ and $p_2^\text{NF}$:} With the neutrino 4-momenta determined in the QF, we finally apply the transformation $p_i^\text{NF} = \Lambda_\text{MF--NF}\Lambda_\text{QF--MF}p_i^\text{QF}$,
for both neutrino four-vectors, yielding the desired outgoing neutrino energies in the rest frame of the neutron star. 
\end{description}
The resulting neutrino energy distributions are shown in \cref{fig:neutrino-decay-spectra} for different values of the electron chemical potential. For clarity, we show spectra from muon decay at rest and before taking into account gravitational redshift and neutrino oscillations. The key takeaway is that larger electron chemical potentials significantly soften the neutrino spectra, in addition to smearing out the spectral-shape differences between the different neutrino flavors.

\section{Neutron star Equations of State}
\label{sec:eos}

\begin{figure*}
    \centering
    \includegraphics[width=0.495\textwidth]{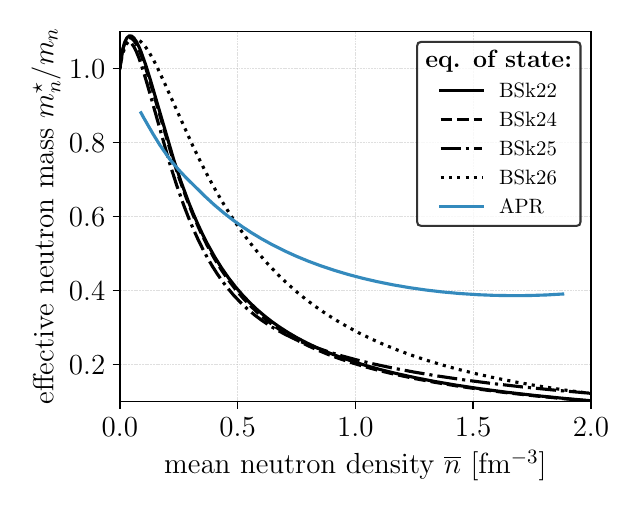}
    \includegraphics[width=0.495\textwidth]{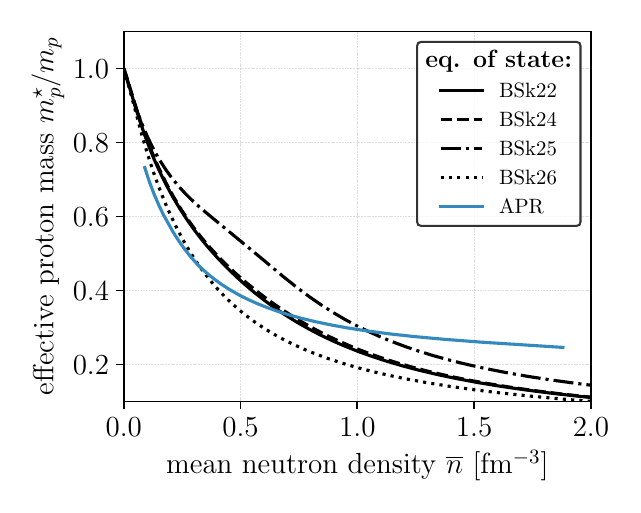}
    \caption{Effective neutron mass (left) and effective proton mass (right) as a function of the mean neutron density for the four Skyrme-type functionals, BSk22 through BSk26, as well as the default APR equation of state used in \nscool. For the DD-MEX and NL3$\omega\rho$ EOS we use an APR-like effective masses as an approximation, see \cref{sec:rel-mean-field-eos}. }
    \label{fig:effect-nucleon-masses}
\end{figure*}

In this section we provide additional information about our implementation of the various equations of state (EOS) used in the body of the paper. Beyond the Akmal--Pandharipande--Ravenhall (APR) equation of state already implemented in \nscool we have added Skyrme-type equations of state (BSk2X) as well as both the DD-MEX and NL3$\omega\rho$ EOS for the neutron star core to \nscool.

\subsection{Skyrme-Type Equations of State}

Our starting point for the BSk2X EOS are the fitting functions for the macroscopic properties of the neutron star matter provided by Ref.~\cite{Pearson:2018tkr}, building upon earlier work~\cite{Potekhin:2013qqa}. Beyond the relationship between mean nucleon density, pressure, and energy density, \nscool also requires the effective nucleon masses as well as the relative abundances of protons, electrons and muons. Once this information is given, the Tolman–Oppenheimer–Volkoff (TOV) equation is solved numerically and a profile is generated. Note that we modify only the EOS of the neutron star core, matching the Skyrme-type core models to the default crust model implemented in \nscool. The crust--core boundary is taken be at the densities indicated in Table~14 of Ref.~\cite{Pearson:2018tkr}.

The effective nucleon masses which parameterize the effects of many-body interactions of the respective nucleons, sensitively control a host of the neutron star's macroscopic properties, see for example Ref.~\cite{Brack:1985vp}. The deviation of the effective masses $m^\star_{n,p}$ from the free nucleon masses $m_{n,p}$ are given by Eq.~(A10) in Ref.~\cite{Chamel:2009yx}, namely
\begin{align}
    \frac{m_q^\star}{m_q} &= \bigg[1 + \frac{m_q}{2} \Big( \sum_i X_i \Big) \bar{n}\bigg]^{-1} \,,
\end{align}
with
\begin{align}
    \sum_i X_i &= t_1 \left[\left(1 + \frac{1}{2} x_1 \right)
                          - \left(\frac{1}{2} + x_1\right) Y_q\right]
                + t_2 \left[\left(1 + \frac{1}{2} x_2 \right)
                          + \left(\frac{1}{2} + x_2\right) Y_q\right] \notag\\
               &+ t_4 \left[\left(1+\frac{1}{2} x_4\right)
                          - \left(\frac{1}{2} + x_4\right) Y_q \right] \bar{n}^\beta
                + t_5 \left[\left(1 + \frac{1}{2} x_5\right)
                          + \left(\frac{1}{2} + x_5\right) Y_q\right] \bar{n}^\gamma \,.
\end{align}
Here $q = \{n,p\}$ labels either the neutron or proton effective mass, $Y_q$ is the relative abundance, and $\bar{n}$ is the mean neutron density of the neutron star matter under consideration. The $t_i$ and $x_i$ are fit parameters. Our implementation of this expression has been cross-checked for the parameter values given in Ref.~\cite{Constantinou:2014hha}, and a comparison has been made with \nscool's default APR equation of state \cite{Akmal:1998cf}. For the case at hand we show the effective masses for the up-to-date Skyrme parameters of Ref.~\cite{Goriely:2013xba} in \cref{fig:effect-nucleon-masses}, with the relevant parameters entering summarized in \cref{tab:eos-parameters}.

\begin{table}
    \centering
    \renewcommand{\arraystretch}{1.4}
    \begin{tabular}{ccccc}
        \toprule
                                     & \textbf{BSk22} & \textbf{BSk24} & \textbf{BSk25} & \textbf{BSk26} \\
        \hline
        $t_1$ [MeV fm$^5$]           &  404.461 & 395.766 & 431.093 & 439.536 \\
        $t_2$ [MeV fm$^5$]           & $\simeq 0$ & $\simeq 0$ & $\simeq 0$ & $\simeq 0$  \\ 
        $t_4$ [MeV fm$^{5+3\beta}$]  & $-100$ & $-100$ & $-200$ & $-100$ \\ 
        $t_5$ [MeV fm$^{5+3\gamma}$] & $-150$ & $-150$ & $-150$ & $-120$ \\
        $x_1$                        & 0.0627540 &  0.0563535 & 0.111366 & $-0.404961$ \\
        $x_2 t_2$ [MeV fm$^5$]       & $-1396.13$ & $-1389.61$ & $-1387.47$ & $-1147.70$ \\
        $x_4$                        & 2 & 2 & 2 & -3 \\
        $x_5$                        & $-11$ & $-11$ & $-11$ & $-11$  \\
        $\beta$                      & $1/2$ & $1/2$ & $1/2$ & $1/6$ \\
        $\gamma$                     & $1/12$ & $1/12$ & $1/12$ & $1/12$ \\\toprule
    \end{tabular}
    \caption{Relevant parameters of the BSk-type Skyrme models from Ref.~\protect\cite{Goriely:2013xba}.}
    \label{tab:eos-parameters}
\end{table}

\begin{figure*}
    \centering
    \includegraphics[width=0.6\textwidth]{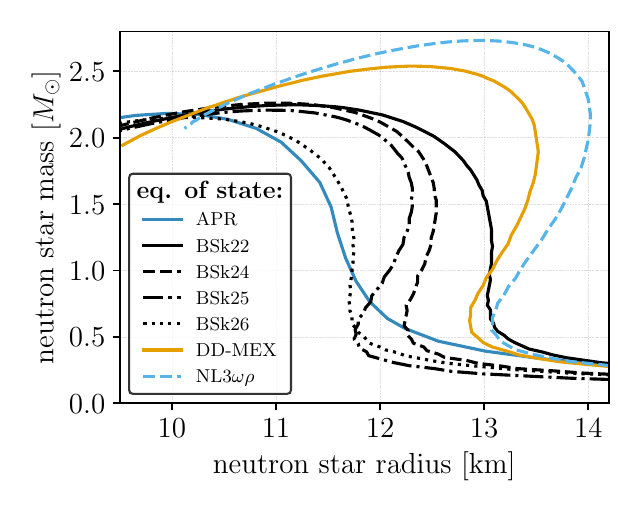}
    \caption{Mass versus radius curves for the different neutron star equations of state used in this work.}
    \label{fig:mass-radius-relation}
\end{figure*}

The final validation for the Skyrme-type neutron star core EOS is the mass--radius relations obtained through integrating the TOV equation. The results of which are shown in \cref{fig:mass-radius-relation}. Small differences from the curves in Fig.~(28) of Ref.~\cite{Pearson:2018tkr} arise from gluing the two different crust and core EOS together as discussed above. Regarding the impact of the EOS on the neutrino fluxes from muon decay, differences between different EOS chiefly arise from the different core radii.

\subsection{Relativistic mean field equations of state}
\label{sec:rel-mean-field-eos}

For comparison to the Skyrme-type models we also show results using two mean field equations of state, NL3$\omega\rho$ \cite{Horowitz:2000xj, Horowitz:2001yn, Grill:2014aea, Pais:2016xiu} and DD-MEX \cite{Lalazissis:2005, Rather:2021azv}. For both DD-MEX and NL3$\omega\rho$ EOS we utilize the public CompOSE library and code \cite{Typel:2013rza, Oertel2017, CompOSECoreTeam:2022ddl} to build neutron star EOS and profiles for use in \nscool. While for the DD-MEX we have also cross-checked these results using additional material supplied by the authors of Ref.~\cite{Rather:2021azv}. Note that the effective masses are required to run \texttt{NSCool}, which are missing when using the EOS data from CompOSE. To run \texttt{NSCool} for these equations of state we use dummy effective masses that resemble the APR EOS. These masses typically enter only linearly in the particle production rates, leading to additional order-one uncertainties in these rates.

\section{Backgrounds}
\label{sec:backgrounds}

Here, we outline how we have estimated the backgrounds shown in \cref{fig:spectrum-comparison}.
\begin{itemize}
  \item \textbf{Solar neutrinos.}  We take the solar neutrino fluxes from ref.~\cite{Bahcall:2004pz}, in particular from the solar model dubbed BS05(AGS,OP) in this reference.  Even though in a realistic detector it will be possible to suppress the solar neutrino background by a factor of a few using directional information, in \cref{fig:spectrum-comparison} we conservatively plot the unsuppressed solar neutrino flux.  As the Sun only produces neutrinos, but no antineutrinos, this background is relevant only in DUNE, but not in HyperKamiokande and JUNO.

  \item \textbf{Reactor neutrinos.}  We take this background from ref.~\cite{Beacom:2003nk}, where it was calculated for SuperKamiokande, and rescale it to the larger fiducial mass of HyperKamiokande. $\bar\nu_e$ from reactor neutrinos are not a relevant background in DUNE, where the main detection channel $\nu_e + \iso{Ar}{40} \to e^- + \iso{K}{40}^*$ is sensitive only to neutrinos, but not anti-neutrinos.

  \item \textbf{Atmospheric neutrinos.} We use the fluxes of atmospheric neutrinos computed in ref.~\cite{Battistoni:2005pd} using the FLUKA Monte Carlo code~\cite{Ferrari:2005zk, Boehlen:2014}. To the best of our knowledge, this is the only available calculation that covers the energy range between \SI{10}{MeV} and \SI{100}{MeV}.
  
  \item \textbf{Diffuse supernova neutrinos.} Our estimate for the diffuse supernova neutrino background in the $\nu_e$ and $\bar\nu_e$ channels is taken from ref.~\cite{Moller:2018kpn}.
\end{itemize}

\bibliographystyle{JHEP}
\bibliography{refs}

\providecommand{\href}[2]{#2}\begingroup\raggedright\begin{thebibliography}{100}

\bibitem{Oertel:2016bki}
M.~Oertel, M.~Hempel, T.~Kl\"ahn, and S.~Typel, {\it {Equations of state for
  supernovae and compact stars}},  {\em Rev. Mod. Phys.} {\bf 89} (2017), no.~1
  015007, [\href{http://www.arxiv.org/abs/1610.03361}{{\tt 1610.03361}}].

\bibitem{Yakovlev:2000jp}
D.~Yakovlev, A.~Kaminker, O.~Y. Gnedin, and P.~Haensel, {\it {Neutrino emission
  from neutron stars}},  {\em Phys. Rept.} {\bf 354} (2001) 1,
  [\href{http://www.arxiv.org/abs/astro-ph/0012122}{{\tt astro-ph/0012122}}].

\bibitem{Cohen:1970}
J.~M. {Cohen}, W.~D. {Langer}, L.~C. {Rosen}, and A.~G.~W. {Cameron}, {\it
  {Neutron Star Models based on an Improved Equation of State}},  {\em \apss}
  {\bf 6} (Feb., 1970) 228--239.

\bibitem{Garani:2018kkd}
R.~Garani, Y.~Genolini, and T.~Hambye, {\it {New Analysis of Neutron Star
  Constraints on Asymmetric Dark Matter}},  {\em JCAP} {\bf 05} (2019) 035,
  [\href{http://www.arxiv.org/abs/1812.08773}{{\tt 1812.08773}}].

\bibitem{Bell:2019pyc}
N.~F. Bell, G.~Busoni, and S.~Robles, {\it {Capture of Leptophilic Dark Matter
  in Neutron Stars}},  {\em JCAP} {\bf 06} (2019) 054,
  [\href{http://www.arxiv.org/abs/1904.09803}{{\tt 1904.09803}}].

\bibitem{Garani:2019fpa}
R.~Garani and J.~Heeck, {\it {Dark matter interactions with muons in neutron
  stars}},  {\em Phys. Rev. D} {\bf 100} (2019), no.~3 035039,
  [\href{http://www.arxiv.org/abs/1906.10145}{{\tt 1906.10145}}].

\bibitem{Dror:2019uea}
J.~A. Dror, R.~Laha, and T.~Opferkuch, {\it {Probing muonic forces with neutron
  star binaries}},  {\em Phys. Rev. D} {\bf 102} (2020), no.~2 023005,
  [\href{http://www.arxiv.org/abs/1909.12845}{{\tt 1909.12845}}].

\bibitem{Hamaguchi:2022wpz}
K.~Hamaguchi, N.~Nagata, and M.~E. Ramirez-Quezada, {\it {Neutron Star Heating
  in Dark Matter Models for the Muon $g-2$ Discrepancy}},
  \href{http://www.arxiv.org/abs/2204.02413}{{\tt 2204.02413}}.

\bibitem{Akmal:1998cf}
A.~Akmal, V.~Pandharipande, and D.~Ravenhall, {\it {The Equation of state of
  nucleon matter and neutron star structure}},  {\em Phys. Rev. C} {\bf 58}
  (1998) 1804--1828, [\href{http://www.arxiv.org/abs/nucl-th/9804027}{{\tt
  nucl-th/9804027}}].

\bibitem{Taninah:2019cku}
A.~Taninah, S.~E. Agbemava, A.~V. Afanasjev, and P.~Ring, {\it {Parametric
  correlations in energy density functionals}},  {\em Phys. Lett. B} {\bf 800}
  (2020) 135065, [\href{http://www.arxiv.org/abs/1910.13007}{{\tt
  1910.13007}}].

\bibitem{Friedman:1986}
J.~L. {Friedman}, J.~R. {Ipser}, and L.~{Parker}, {\it {Rapidly Rotating
  Neutron Star Models}},  {\em \apj} {\bf 304} (May, 1986) 115.

\bibitem{Komatsu:1989zz}
H.~Komatsu, Y.~Eriguchi, and I.~Hachisu, {\it {Rapidly rotating general
  relativistic stars. I - Numerical method and its application to uniformly
  rotating polytropes}},  {\em Mon. Not. Roy. Astron. Soc.} {\bf 237} (1989)
  355--379.

\bibitem{Cook:1993qr}
G.~B. Cook, S.~L. Shapiro, and S.~A. Teukolsky, {\it {Rapidly rotating neutron
  stars in general relativity: Realistic equations of state}},  {\em Astrophys.
  J.} {\bf 424} (1994) 823.

\bibitem{Stergioulas:1994ea}
N.~Stergioulas and J.~L. Friedman, {\it {Comparing models of rapidly rotating
  relativistic stars constructed by two numerical methods}},  {\em Astrophys.
  J.} {\bf 444} (1995) 306,
  [\href{http://www.arxiv.org/abs/astro-ph/9411032}{{\tt astro-ph/9411032}}].

\bibitem{Franzon:2016iai}
B.~Franzon, V.~Dexheimer, and S.~Schramm, {\it {Internal composition of
  proto-neutron stars under strong magnetic fields}},  {\em Phys. Rev. D} {\bf
  94} (2016), no.~4 044018, [\href{http://www.arxiv.org/abs/1606.04843}{{\tt
  1606.04843}}].

\bibitem{Silva:2020oww}
H.~O. Silva, G.~Pappas, N.~Yunes, and K.~Yagi, {\it {Surface of
  rapidly-rotating neutron stars: Implications to neutron star parameter
  estimation}},  {\em Phys. Rev. D} {\bf 103} (2021), no.~6 063038,
  [\href{http://www.arxiv.org/abs/2008.05565}{{\tt 2008.05565}}].

\bibitem{Konstantinou:2022}
A.~{Konstantinou} and S.~M. {Morsink}, {\it {Universal Relations for the
  Increase in the Mass and Radius of a Rotating Neutron Star}},  {\em \apj}
  {\bf 934} (Aug., 2022) 139, [\href{http://www.arxiv.org/abs/2206.12515}{{\tt
  2206.12515}}].

\bibitem{Watanabe:2020vas}
C.~Watanabe, K.~Yanase, and N.~Yoshinaga, {\it {Searching optimum equations of
  state of neutron star matter in strong magnetic fields with rotation}},  {\em
  PTEP} {\bf 2020} (2020), no.~10 103E04.

\bibitem{Hessels:2006ze}
J.~W.~T. Hessels, S.~M. Ransom, I.~H. Stairs, P.~C.~C. Freire, V.~M. Kaspi, and
  F.~Camilo, {\it {A radio pulsar spinning at 716-Hz}},  {\em Science} {\bf
  311} (2006) 1901--1904,
  [\href{http://www.arxiv.org/abs/astro-ph/0601337}{{\tt astro-ph/0601337}}].

\bibitem{Manchester:2004bp}
R.~N. Manchester, G.~B. Hobbs, A.~Teoh, and M.~Hobbs, {\it {The Australia
  Telescope National Facility pulsar catalogue}},  {\em Astron. J.} {\bf 129}
  (2005) 1993, [\href{http://www.arxiv.org/abs/astro-ph/0412641}{{\tt
  astro-ph/0412641}}]. \url{https://www.atnf.csiro.au/research/pulsar/psrcat/}.

\bibitem{Tauris:2011ck}
T.~M. Tauris, N.~Langer, and M.~Kramer, {\it {Formation of millisecond pulsars
  with CO white dwarf companions - I. PSR J1614-2230: Evidence for a neutron
  star born massive}},  {\em Mon. Not. Roy. Astron. Soc.} {\bf 416} (2011)
  2130, [\href{http://www.arxiv.org/abs/1103.4996}{{\tt 1103.4996}}].

\bibitem{Liu:2011ss}
W.-M. Liu and W.-C. Chen, {\it {On the progenitors of millisecond pulsars by
  the recycling evolutionary channel}},  {\em Mon. Not. Roy. Astron. Soc.} {\bf
  416} (2011) 2285, [\href{http://www.arxiv.org/abs/1106.1567}{{\tt
  1106.1567}}].

\bibitem{Tauris:2012jp}
T.~M. Tauris, N.~Langer, and M.~Kramer, {\it {Formation of millisecond pulsars
  with CO white dwarf companions - II. Accretion, spin-up, true ages and
  comparison to MSPs with He white dwarf companions}},  {\em Mon. Not. Roy.
  Astron. Soc.} {\bf 426} (2012) 1601--1627,
  [\href{http://www.arxiv.org/abs/1206.1862}{{\tt 1206.1862}}].

\bibitem{Li:2021zfx}
Z.~Li, X.~Chen, H.-L. Chen, and Z.~Han, {\it {The Maximum Accreted Mass of
  Recycled Pulsars}},  {\em Astrophys. J.} {\bf 922} (2021), no.~2 158,
  [\href{http://www.arxiv.org/abs/2108.02554}{{\tt 2108.02554}}].

\bibitem{Bandyopadhyay:1997kh}
D.~Bandyopadhyay, S.~Chakrabarty, and S.~Pal, {\it {The Quantizing magnetic
  field and quark--hadron phase transition in a neutron star}},  {\em Phys.
  Rev. Lett.} {\bf 79} (1997) 2176--2179,
  [\href{http://www.arxiv.org/abs/astro-ph/9703066}{{\tt astro-ph/9703066}}].

\bibitem{Chakrabarty:1997ef}
S.~Chakrabarty, D.~Bandyopadhyay, and S.~Pal, {\it {Dense nuclear matter in a
  strong magnetic field}},  {\em Phys. Rev. Lett.} {\bf 78} (1997) 2898--2901,
  [\href{http://www.arxiv.org/abs/astro-ph/9703034}{{\tt astro-ph/9703034}}].

\bibitem{Broderick:2000pe}
A.~Broderick, M.~Prakash, and J.~M. Lattimer, {\it {The Equation of state of
  neutron star matter in strong magnetic fields}},  {\em Astrophys. J.} {\bf
  537} (2000) 351, [\href{http://www.arxiv.org/abs/astro-ph/0001537}{{\tt
  astro-ph/0001537}}].

\bibitem{Suh:2000ni}
I.-S. Suh and G.~J. Mathews, {\it {Cold ideal equation of state for strongly
  magnetized neutron star matter: Effects on muon production and pion
  condensation}},  {\em Astrophys. J.} {\bf 546} (2001) 1126--1136,
  [\href{http://www.arxiv.org/abs/astro-ph/9912301}{{\tt astro-ph/9912301}}].

\bibitem{Mao:2001cv}
G.-J. Mao, V.~N. Kondratyev, A.~Iwamoto, Z.~Li, X.~Wu, W.~Greiner, and I.~N.
  Mikhailov, {\it {Neutron star composition in strong magnetic fields}},  {\em
  Chin. Phys. Lett.} {\bf 20} (2003) 1238,
  [\href{http://www.arxiv.org/abs/astro-ph/0111374}{{\tt astro-ph/0111374}}].

\bibitem{Broderick:2001qw}
A.~E. Broderick, M.~Prakash, and J.~M. Lattimer, {\it {Effects of strong
  magnetic fields in strange baryonic matter}},  {\em Phys. Lett. B} {\bf 531}
  (2002) 167--174, [\href{http://www.arxiv.org/abs/astro-ph/0111516}{{\tt
  astro-ph/0111516}}].

\bibitem{Wei:2005aga}
F.~X. Wei, G.~J. Mao, C.~M. Ko, L.~S. Kisslinger, H.~Stoecker, and W.~Greiner,
  {\it {Effect of isovector-scalar meson on neutron star matter in strong
  magnetic fields}},  {\em J. Phys. G} {\bf 32} (2006) 47,
  [\href{http://www.arxiv.org/abs/nucl-th/0508065}{{\tt nucl-th/0508065}}].
  Nucleon effective masses in B-fields. Proton mass is strongly affected,
  neutron mass isn't.

\bibitem{Noronha:2007wg}
J.~L. Noronha and I.~A. Shovkovy, {\it {Color-flavor locked superconductor in a
  magnetic field}},  {\em Phys. Rev. D} {\bf 76} (2007) 105030,
  [\href{http://www.arxiv.org/abs/0708.0307}{{\tt 0708.0307}}]. [Erratum:
  Phys.Rev.D 86, 049901 (2012)].

\bibitem{Rabhi:2009ih}
A.~Rabhi, H.~Pais, P.~K. Panda, and C.~Providencia, {\it {Quark-hadron phase
  transition in a neutron star under strong magnetic fields}},  {\em J. Phys.
  G} {\bf 36} (2009) 115204, [\href{http://www.arxiv.org/abs/0909.1114}{{\tt
  0909.1114}}].

\bibitem{Ferrer:2010wz}
E.~J. Ferrer, V.~de~la Incera, J.~P. Keith, I.~Portillo, and P.~L. Springsteen,
  {\it {Equation of State of a Dense and Magnetized Fermion System}},  {\em
  Phys. Rev. C} {\bf 82} (2010) 065802,
  [\href{http://www.arxiv.org/abs/1009.3521}{{\tt 1009.3521}}].

\bibitem{Strickland:2012vu}
M.~Strickland, V.~Dexheimer, and D.~P. Menezes, {\it {Bulk Properties of a
  Fermi Gas in a Magnetic Field}},  {\em Phys. Rev. D} {\bf 86} (2012) 125032,
  [\href{http://www.arxiv.org/abs/1209.3276}{{\tt 1209.3276}}].

\bibitem{Sinha:2013dfa}
M.~Sinha, X.-G. Huang, and A.~Sedrakian, {\it {Strange quark matter in strong
  magnetic fields within a confining model}},  {\em Phys. Rev. D} {\bf 88}
  (2013), no.~2 025008, [\href{http://www.arxiv.org/abs/1306.3300}{{\tt
  1306.3300}}].

\bibitem{Casali:2013jka}
R.~H. Casali, L.~B. Castro, and D.~P. Menezes, {\it {Hadronic and hybrid stars
  subject to density dependent magnetic fields}},  {\em Phys. Rev. C} {\bf 89}
  (2014), no.~1 015805, [\href{http://www.arxiv.org/abs/1307.2651}{{\tt
  1307.2651}}].

\bibitem{Lopes:2014vva}
L.~L. Lopes and D.~Menezes, {\it {On Magnetized Neutron Stars}},  {\em JCAP}
  {\bf 08} (2015) 002, [\href{http://www.arxiv.org/abs/1411.7209}{{\tt
  1411.7209}}].

\bibitem{Dexheimer:2016yqu}
V.~Dexheimer, B.~Franzon, R.~O. Gomes, R.~L.~S. Farias, S.~S. Avancini, and
  S.~Schramm, {\it {What is the magnetic field distribution for the equation of
  state of magnetized neutron stars?}},  {\em Phys. Lett. B} {\bf 773} (2017)
  487--491, [\href{http://www.arxiv.org/abs/1612.05795}{{\tt 1612.05795}}].

\bibitem{Coelho:2016lcf}
E.~L. Coelho, M.~Chiapparini, and R.~P. Negreiros, {\it {Cooling of neutron
  stars and emissivity of neutrinos by the direct Urca process under influence
  of a strong magnetic field}},  {\em J. Phys. Conf. Ser.} {\bf 706} (2016),
  no.~5 052011. Contains plots of the Urca rate with and without B field, and
  of the NS composition with and without B field.

\bibitem{Gomes:2017zkc}
R.~O. Gomes, B.~Franzon, V.~Dexheimer, and S.~Schramm, {\it {Many-body forces
  in magnetic neutron stars}},  {\em Astrophys. J.} {\bf 850} (2017), no.~1 20,
  [\href{http://www.arxiv.org/abs/1709.01017}{{\tt 1709.01017}}].

\bibitem{Negreiros:2018cjk}
R.~Negreiros, C.~Bernal, V.~Dexheimer, and O.~Troconis, {\it {Many Aspects of
  Magnetic Fields in Neutron Stars}},  {\em Universe} {\bf 4} (2018), no.~3 43.

\bibitem{Gomes:2019paw}
R.~O. Gomes, H.~Pais, V.~Dexheimer, C.~Provid\^encia, and S.~Schramm, {\it
  {Limiting magnetic field for minimal deformation of a magnetized neutron
  star}},  {\em Astron. Astrophys.} {\bf 627} (2019) A61,
  [\href{http://www.arxiv.org/abs/1902.08146}{{\tt 1902.08146}}].

\bibitem{Thapa:2020ohp}
V.~B. Thapa, M.~Sinha, J.~J. Li, and A.~Sedrakian, {\it {Equation of State of
  Strongly Magnetized Matter with Hyperons and $\Delta$-Resonances}},  {\em
  Particles} {\bf 3} (2020), no.~4 660--675,
  [\href{http://www.arxiv.org/abs/2010.00981}{{\tt 2010.00981}}].

\bibitem{Rather:2021azv}
I.~A. Rather, U.~Rahaman, V.~Dexheimer, A.~A. Usmani, and S.~K. Patra, {\it
  {Heavy Magnetic Neutron Stars}},  {\em Astrophys. J.} {\bf 917} (2021), no.~1
  46, [\href{http://www.arxiv.org/abs/2104.05950}{{\tt 2104.05950}}].

\bibitem{Chatterjee:2021wsr}
D.~Chatterjee, J.~Novak, and M.~Oertel, {\it {Structure of ultra-magnetised
  neutron stars}},  {\em Eur. Phys. J. A} {\bf 57} (2021), no.~8 249,
  [\href{http://www.arxiv.org/abs/2108.13733}{{\tt 2108.13733}}].

\bibitem{Rather:2022bmm}
I.~A. Rather, A.~A. Rather, V.~Dexheimer, I.~Lopes, A.~A. Usmani, and S.~K.
  Patra, {\it {Magnetic-field Induced Deformation in Hybrid Stars}},  {\em
  Astrophys. J.} {\bf 943} (2023), no.~1 52,
  [\href{http://www.arxiv.org/abs/2209.06016}{{\tt 2209.06016}}].

\bibitem{Price:2006fi}
D.~Price and S.~Rosswog, {\it {Producing ultra-strong magnetic fields in
  neutron star mergers}},  {\em Science} {\bf 312} (2006) 719,
  [\href{http://www.arxiv.org/abs/astro-ph/0603845}{{\tt astro-ph/0603845}}].

\bibitem{Goldreich:1992}
P.~{Goldreich} and A.~{Reisenegger}, {\it {Magnetic Field Decay in Isolated
  Neutron Stars}},  {\em \apj} {\bf 395} (Aug., 1992) 250.

\bibitem{Jones:2005km}
P.~B. Jones, {\it {Type II superconductivity and magnetic flux transport in
  neutrons stars}},  {\em Mon. Not. Roy. Astron. Soc.} {\bf 365} (2006)
  339--344, [\href{http://www.arxiv.org/abs/astro-ph/0510396}{{\tt
  astro-ph/0510396}}].

\bibitem{Elfritz:2015vom}
J.~G. Elfritz, J.~A. Pons, N.~Rea, K.~Glampedakis, and D.~Vigan\`o, {\it
  {Simulated magnetic field expulsion in neutron star cores}},  {\em Mon. Not.
  Roy. Astron. Soc.} {\bf 456} (2016), no.~4 4461--4474,
  [\href{http://www.arxiv.org/abs/1512.07151}{{\tt 1512.07151}}].

\bibitem{Bransgrove:2017jzs}
A.~Bransgrove, Y.~Levin, and A.~Beloborodov, {\it {Magnetic field evolution of
  neutron stars \textendash{} I. Basic formalism, numerical techniques and
  first results}},  {\em Mon. Not. Roy. Astron. Soc.} {\bf 473} (2018), no.~2
  2771--2790, [\href{http://www.arxiv.org/abs/1709.09167}{{\tt 1709.09167}}].

\bibitem{Igoshev:2021ewx}
A.~P. Igoshev, S.~B. Popov, and R.~Hollerbach, {\it {Evolution of Neutron Star
  Magnetic Fields}},  {\em Universe} {\bf 7} (2021), no.~9 351,
  [\href{http://www.arxiv.org/abs/2109.05584}{{\tt 2109.05584}}].

\bibitem{Maggiore:2007ulw}
M.~Maggiore, {\em {Gravitational Waves. Vol. 1: Theory and Experiments}}.
\newblock Oxford Master Series in Physics. Oxford University Press, 2007.

\bibitem{Gamow:1970}
G.~Gamow and S.~Ulam, {\em My World Line; an Informal Autobiography}.
\newblock Viking Press, 1970.

\bibitem{Chiu:1964zza}
H.-Y. Chiu and E.~Salpeter, {\it {Surface X-Ray Emission from Neutron Stars}},
  {\em Phys. Rev. Lett.} {\bf 12} (1964) 413--415.

\bibitem{Shapiro:1983du}
S.~L. Shapiro and S.~A. Teukolsky, {\em {Black holes, white dwarfs, and neutron
  stars: The physics of compact objects}}.
\newblock {John Wiley \& Sons, Ltd}, 1983.

\bibitem{Haensel:1995}
P.~{Haensel}, {\it {URCA Processes in Dense Matter and Neutron Star Cooling}},
  {\em Space Science Reviews} {\bf 74} (Nov., 1995) 427--436.

\bibitem{Yakovlev:1995}
D.~G. {Yakovlev} and K.~P. {Levenfish}, {\it {Modified URCA process in neutron
  star cores}},  {\em Astronomy and Astrophysics} {\bf 297} (May, 1995) 717.

\bibitem{Page:2006ud}
D.~Page and S.~Reddy, {\it {Dense Matter in Compact Stars: Theoretical
  Developments and Observational Constraints}},  {\em Ann. Rev. Nucl. Part.
  Sci.} {\bf 56} (2006) 327--374,
  [\href{http://www.arxiv.org/abs/astro-ph/0608360}{{\tt astro-ph/0608360}}].

\bibitem{Page:2005fq}
D.~Page, U.~Geppert, and F.~Weber, {\it {The Cooling of compact stars}},  {\em
  Nucl. Phys. A} {\bf 777} (2006) 497--530,
  [\href{http://www.arxiv.org/abs/astro-ph/0508056}{{\tt astro-ph/0508056}}].

\bibitem{1998nspt.conf..183P}
D.~{Page}, {\it {Thermal Evolution of Isolated Neutron Stars}},  in {\em
  Neutron Stars and Pulsars: Thirty Years after the Discovery} (N.~{Shibazaki},
  ed.), p.~183, Jan., 1998.

\bibitem{Shternin:2007ee}
P.~Shternin and D.~Yakovlev, {\it {Electron-muon heat conduction in neutron
  star cores via the exchange of transverse plasmons}},  {\em Phys. Rev. D}
  {\bf 75} (2007) 103004, [\href{http://www.arxiv.org/abs/0705.1963}{{\tt
  0705.1963}}].

\bibitem{Beacom:2010kk}
J.~F. Beacom, {\it {The Diffuse Supernova Neutrino Background}},  {\em Ann.
  Rev. Nucl. Part. Sci.} {\bf 60} (2010) 439--462,
  [\href{http://www.arxiv.org/abs/1004.3311}{{\tt 1004.3311}}].

\bibitem{Sartore:2010}
N.~{Sartore}, E.~{Ripamonti}, A.~{Treves}, and R.~{Turolla}, {\it {Galactic
  neutron stars. I. Space and velocity distributions in the disk and in the
  halo}},  {\em \aap} {\bf 510} (Feb., 2010) A23,
  [\href{http://www.arxiv.org/abs/0908.3182}{{\tt 0908.3182}}].

\bibitem{Horowitz:2000xj}
C.~J. Horowitz and J.~Piekarewicz, {\it {Neutron star structure and the neutron
  radius of Pb-208}},  {\em Phys. Rev. Lett.} {\bf 86} (2001) 5647,
  [\href{http://www.arxiv.org/abs/astro-ph/0010227}{{\tt astro-ph/0010227}}].

\bibitem{Lalazissis:2005}
G.~A. {Lalazissis}, T.~{Nik{\v{s}}i{\'c}}, D.~{Vretenar}, and P.~{Ring}, {\it
  {New relativistic mean-field interaction with density-dependent meson-nucleon
  couplings}},  {\em \prc} {\bf 71} (Feb., 2005) 024312.

\bibitem{Friman:1979ecl}
B.~L. Friman and O.~V. Maxwell, {\it {Neutron Star Neutrino Emissivities}},
  {\em Astrophys. J.} {\bf 232} (1979) 541--557.

\bibitem{Gusakov:2004mj}
M.~E. Gusakov, D.~G. Yakovlev, P.~Haensel, and O.~Y. Gnedin, {\it {Direct Urca
  process in a neutron star mantle}},  {\em Astron. Astrophys.} {\bf 421}
  (2004) 1143--1148, [\href{http://www.arxiv.org/abs/astro-ph/0404165}{{\tt
  astro-ph/0404165}}].

\bibitem{github}
J.~Kopp and T.~Opferkuch, ``{GitHub repository:
  \protect\url{https://github.com/koppj/leaky-neutron-stars}}.'' 2022.

\bibitem{Tolman:1939jz}
R.~C. Tolman, {\it {Static solutions of Einstein's field equations for spheres
  of fluid}},  {\em Phys. Rev.} {\bf 55} (1939) 364--373.

\bibitem{Oppenheimer:1939ne}
J.~R. Oppenheimer and G.~M. Volkoff, {\it {On massive neutron cores}},  {\em
  Phys. Rev.} {\bf 55} (1939) 374--381.

\bibitem{Glendenning:1997wn}
N.~Glendenning, {\em Compact Stars: Nuclear Physics, Particle Physics and
  General Relativity}.
\newblock Astronomy and Astrophysics Library. Springer New York, 2012.

\bibitem{Zdunik:2016vza}
J.~L. Zdunik, M.~Fortin, and P.~Haensel, {\it {Neutron star properties and the
  equation of state for the core}},  {\em Astron. Astrophys.} {\bf 599} (2017)
  A119, [\href{http://www.arxiv.org/abs/1611.01357}{{\tt 1611.01357}}].

\bibitem{Gnedin:1995lgf}
O.~Gnedin and D.~Yakovlev, {\it {Thermal conductivity of electrons and muons in
  neutron star cores}},  {\em Nucl. Phys. A} {\bf 582} (1995) 697--716.

\bibitem{Esteban:2018azc}
I.~Esteban, M.~Gonzalez-Garcia, A.~Hernandez-Cabezudo, M.~Maltoni, and
  T.~Schwetz, {\it {Global analysis of three-flavour neutrino oscillations:
  synergies and tensions in the determination of $\theta_{23}$, $\delta_{CP}$,
  and the mass ordering}},  {\em JHEP} {\bf 01} (2019) 106,
  [\href{http://www.arxiv.org/abs/1811.05487}{{\tt 1811.05487}}]. see
  \url{http://www.nu-fit.org}; we use numbers from the NuFit~4.1 fit.

\bibitem{Pearson:2018tkr}
J.~M. Pearson, N.~Chamel, A.~Y. Potekhin, A.~F. Fantina, C.~Ducoin, A.~K.
  Dutta, and S.~Goriely, {\it {Unified equations of state for cold
  non-accreting neutron stars with Brussels\textendash{}Montreal functionals
  \textendash{} I. Role of symmetry energy}},  {\em Mon. Not. Roy. Astron.
  Soc.} {\bf 481} (2018), no.~3 2994--3026,
  [\href{http://www.arxiv.org/abs/1903.04981}{{\tt 1903.04981}}]. [Erratum:
  Mon.Not.Roy.Astron.Soc. 486, 768 (2019)].

\bibitem{Xiang:2018}
M.~{Xiang}, J.~{Shi}, X.~{Liu}, H.~{Yuan}, B.~{Chen}, Y.~{Huang}, C.~{Wang},
  Y.~{Wu}, Z.~{Tian}, Z.~{Huo}, H.~{Zhang}, and M.~{Zhang}, {\it {Stellar Mass
  Distribution and Star Formation History of the Galactic Disk Revealed by
  Mono-age Stellar Populations from LAMOST}},  {\em \apjs} {\bf 237} (Aug.,
  2018) 33, [\href{http://www.arxiv.org/abs/1807.04592}{{\tt 1807.04592}}].

\bibitem{Alsing:2017bbc}
J.~Alsing, H.~O. Silva, and E.~Berti, {\it {Evidence for a maximum mass cut-off
  in the neutron star mass distribution and constraints on the equation of
  state}},  {\em Mon. Not. Roy. Astron. Soc.} {\bf 478} (2018), no.~1
  1377--1391, [\href{http://www.arxiv.org/abs/1709.07889}{{\tt 1709.07889}}].

\bibitem{Bahcall:2004pz}
J.~N. Bahcall, A.~M. Serenelli, and S.~Basu, {\it {New solar opacities,
  abundances, helioseismology, and neutrino fluxes}},  {\em Astrophys. J.
  Lett.} {\bf 621} (2005) L85--L88,
  [\href{http://www.arxiv.org/abs/astro-ph/0412440}{{\tt astro-ph/0412440}}].

\bibitem{Battistoni:2005pd}
G.~Battistoni, A.~Ferrari, T.~Montaruli, and P.~Sala, {\it {The atmospheric
  neutrino flux below 100-MeV: The FLUKA results}},  {\em Astropart. Phys.}
  {\bf 23} (2005) 526--534.

\bibitem{Moller:2018kpn}
K.~M\o{}ller, A.~M. Suliga, I.~Tamborra, and P.~B. Denton, {\it {Measuring the
  supernova unknowns at the next-generation neutrino telescopes through the
  diffuse neutrino background}},  {\em JCAP} {\bf 05} (2018) 066,
  [\href{http://www.arxiv.org/abs/1804.03157}{{\tt 1804.03157}}].

\bibitem{Beacom:2003nk}
J.~F. Beacom and M.~R. Vagins, {\it {GADZOOKS! Anti-neutrino spectroscopy with
  large water Cherenkov detectors}},  {\em Phys. Rev. Lett.} {\bf 93} (2004)
  171101, [\href{http://www.arxiv.org/abs/hep-ph/0309300}{{\tt
  hep-ph/0309300}}].

\bibitem{Super-Kamiokande:2021jaq}
{\bf Super-Kamiokande} {\bf Collaboration}, K.~Abe {\em et~al.}, {\it {Diffuse
  supernova neutrino background search at Super-Kamiokande}},  {\em Phys. Rev.
  D} {\bf 104} (2021), no.~12 122002,
  [\href{http://www.arxiv.org/abs/2109.11174}{{\tt 2109.11174}}].

\bibitem{Ferrari:2005zk}
A.~Ferrari, P.~R. Sala, A.~Fasso, and J.~Ranft, {\it {FLUKA: A multi-particle
  transport code (Program version 2005)}},  2005.
\newblock available from \url{https://cds.cern.ch/record/898301}.

\bibitem{Boehlen:2014}
T.~Böhlen, F.~Cerutti, M.~Chin, A.~Fassò, A.~Ferrari, P.~Ortega, A.~Mairani,
  P.~Sala, G.~Smirnov, and V.~Vlachoudis, {\it The fluka code: Developments and
  challenges for high energy and medical applications},  {\em Nuclear Data
  Sheets} {\bf 120} (2014) 211 -- 214.

\bibitem{Cocco:2004ac}
A.~Cocco, A.~Ereditato, G.~Fiorillo, G.~Mangano, and V.~Pettorino, {\it
  {Supernova relic neutrinos in liquid argon detectors}},  {\em JCAP} {\bf 12}
  (2004) 002, [\href{http://www.arxiv.org/abs/hep-ph/0408031}{{\tt
  hep-ph/0408031}}].

\bibitem{DUNE:2020ypp}
{\bf DUNE} {\bf Collaboration}, B.~Abi {\em et~al.}, {\it {Deep Underground
  Neutrino Experiment (DUNE), Far Detector Technical Design Report, Volume II:
  DUNE Physics}},  \href{http://www.arxiv.org/abs/2002.03005}{{\tt
  2002.03005}}.

\bibitem{Capozzi:2018dat}
F.~Capozzi, S.~W. Li, G.~Zhu, and J.~F. Beacom, {\it {DUNE as the
  Next-Generation Solar Neutrino Experiment}},  {\em Phys. Rev. Lett.} {\bf
  123} (2019), no.~13 131803, [\href{http://www.arxiv.org/abs/1808.08232}{{\tt
  1808.08232}}].

\bibitem{Zhu:2018rwc}
G.~Zhu, S.~W. Li, and J.~F. Beacom, {\it {Developing the MeV potential of DUNE:
  Detailed considerations of muon-induced spallation and other backgrounds}},
  {\em Phys. Rev. C} {\bf 99} (2019), no.~5 055810,
  [\href{http://www.arxiv.org/abs/1811.07912}{{\tt 1811.07912}}].

\bibitem{Hyper-Kamiokande:2018ofw}
{\bf Hyper-Kamiokande} {\bf Collaboration}, K.~Abe {\em et~al.}, {\it
  {Hyper-Kamiokande Design Report}},
  \href{http://www.arxiv.org/abs/1805.04163}{{\tt 1805.04163}}.

\bibitem{JUNO:2021vlw}
{\bf JUNO} {\bf Collaboration}, A.~Abusleme {\em et~al.}, {\it {JUNO physics
  and detector}},  {\em Prog. Part. Nucl. Phys.} {\bf 123} (2022) 103927,
  [\href{http://www.arxiv.org/abs/2104.02565}{{\tt 2104.02565}}].

\bibitem{JUNO:2023vyz}
{\bf JUNO} {\bf Collaboration}, A.~Abusleme {\em et~al.}, {\it {JUNO
  sensitivity to the annihilation of MeV dark matter in the galactic halo}},
  \href{http://www.arxiv.org/abs/2306.09567}{{\tt 2306.09567}}.

\bibitem{Lattimer:1991ib}
J.~Lattimer, M.~Prakash, C.~Pethick, and P.~Haensel, {\it {Direct URCA process
  in neutron stars}},  {\em Phys. Rev. Lett.} {\bf 66} (1991) 2701--2704.

\bibitem{Schwenk:2002fq}
A.~Schwenk, B.~Friman, and G.~E. Brown, {\it {Renormalization group approach to
  neutron matter: Quasiparticle interactions, superfluid gaps and the equation
  of state}},  {\em Nucl. Phys. A} {\bf 713} (2003) 191--216,
  [\href{http://www.arxiv.org/abs/nucl-th/0207004}{{\tt nucl-th/0207004}}].

\bibitem{Page:2009fu}
D.~Page, J.~M. Lattimer, M.~Prakash, and A.~W. Steiner, {\it {Neutrino Emission
  from Cooper Pairs and Minimal Cooling of Neutron Stars}},  {\em Astrophys.
  J.} {\bf 707} (2009) 1131--1140,
  [\href{http://www.arxiv.org/abs/0906.1621}{{\tt 0906.1621}}].

\bibitem{proton-superfluidity-ref}
T.~Takatsuka, {\it {Proton Superfluidity in Neutron-Star Matter}},  {\em
  Progress of Theoretical Physics} {\bf 50} (11, 1973) 1754--1755,
  [\href{http://www.arxiv.org/abs/https://academic.oup.com/ptp/article-pdf/50/5/1754/5209126/50-5-1754.pdf}{{\tt
  https://academic.oup.com/ptp/article-pdf/50/5/1754/5209126/50-5-1754.pdf}}].

\bibitem{Kublbeck:1990xc}
J.~Kublbeck, M.~Bohm, and A.~Denner, {\it {Feyn Arts: Computer Algebraic
  Generation of Feynman Graphs and Amplitudes}},  {\em Comput. Phys. Commun.}
  {\bf 60} (1990) 165--180.

\bibitem{Hahn:2000kx}
T.~Hahn, {\it {Generating Feynman diagrams and amplitudes with FeynArts 3}},
  {\em Comput. Phys. Commun.} {\bf 140} (2001) 418--431,
  [\href{http://www.arxiv.org/abs/hep-ph/0012260}{{\tt hep-ph/0012260}}].

\bibitem{Mertig:1990an}
R.~Mertig, M.~Bohm, and A.~Denner, {\it {FEYN CALC: Computer algebraic
  calculation of Feynman amplitudes}},  {\em Comput. Phys. Commun.} {\bf 64}
  (1991) 345--359.

\bibitem{Shtabovenko:2016sxi}
V.~Shtabovenko, R.~Mertig, and F.~Orellana, {\it {New Developments in FeynCalc
  9.0}},  {\em Comput. Phys. Commun.} {\bf 207} (2016) 432--444,
  [\href{http://www.arxiv.org/abs/1601.01167}{{\tt 1601.01167}}].

\bibitem{Shtabovenko:2020gxv}
V.~Shtabovenko, R.~Mertig, and F.~Orellana, {\it {FeynCalc 9.3: New features
  and improvements}},  {\em Comput. Phys. Commun.} {\bf 256} (2020) 107478,
  [\href{http://www.arxiv.org/abs/2001.04407}{{\tt 2001.04407}}].

\bibitem{ParticleDataGroup:2020ssz}
{\bf Particle Data Group} {\bf Collaboration}, P.~A. Zyla {\em et~al.}, {\it
  {Review of Particle Physics}},  {\em PTEP} {\bf 2020} (2020), no.~8 083C01.

\bibitem{Potekhin:2013qqa}
A.~Y. Potekhin, A.~F. Fantina, N.~Chamel, J.~M. Pearson, and S.~Goriely, {\it
  {Analytical representations of unified equations of state for neutron-star
  matter}},  {\em Astron. Astrophys.} {\bf 560} (2013) A48,
  [\href{http://www.arxiv.org/abs/1310.0049}{{\tt 1310.0049}}].

\bibitem{Brack:1985vp}
M.~Brack, C.~Guet, and H.~B. Hakansson, {\it {Selfconsistent semiclassical
  description of average nuclear properties. A Link between microscopic and
  macroscopic models}},  {\em Phys. Rept.} {\bf 123} (1985) 275--364.

\bibitem{Chamel:2009yx}
N.~Chamel, S.~Goriely, and J.~M. Pearson, {\it {Further explorations of
  Skyrme-Hartree-Fock-Bogoliubov mass formulas. XI. Stabilizing neutron stars
  against a ferromagnetic collapse}},  {\em Phys. Rev. C} {\bf 80} (2009)
  065804, [\href{http://www.arxiv.org/abs/0911.3346}{{\tt 0911.3346}}].

\bibitem{Constantinou:2014hha}
C.~Constantinou, B.~Muccioli, M.~Prakash, and J.~M. Lattimer, {\it {Thermal
  properties of supernova matter: The bulk homogeneous phase}},  {\em Phys.
  Rev. C} {\bf 89} (2014), no.~6 065802,
  [\href{http://www.arxiv.org/abs/1402.6348}{{\tt 1402.6348}}].

\bibitem{Goriely:2013xba}
S.~Goriely, N.~Chamel, and J.~M. Pearson, {\it {Further explorations of
  Skyrme-Hartree-Fock-Bogoliubov mass formulas. 13. The 2012 atomic mass
  evaluation and the symmetry coefficient}},  {\em Phys. Rev. C} {\bf 88}
  (2013), no.~2 024308.

\bibitem{Horowitz:2001yn}
C.~J. Horowitz and J.~Piekarewicz, {\it {Density dependence of charge symmetry
  breaking}},  {\em Phys. Rev. C} {\bf 63} (2001) 011303.

\bibitem{Grill:2014aea}
F.~Grill, H.~Pais, C.~Provid\^encia, I.~Vida\~na, and S.~S. Avancini, {\it
  {Equation of state and thickness of the inner crust of neutron stars}},  {\em
  Phys. Rev. C} {\bf 90} (2014), no.~4 045803,
  [\href{http://www.arxiv.org/abs/1404.2753}{{\tt 1404.2753}}].

\bibitem{Pais:2016xiu}
H.~Pais and C.~Provid\^encia, {\it {Vlasov formalism for extended relativistic
  mean field models: The crust-core transition and the stellar matter equation
  of state}},  {\em Phys. Rev. C} {\bf 94} (2016), no.~1 015808,
  [\href{http://www.arxiv.org/abs/1607.05899}{{\tt 1607.05899}}].

\bibitem{Typel:2013rza}
S.~Typel, M.~Oertel, and T.~Kl\"ahn, {\it {CompOSE CompStar online supernova
  equations of state harmonising the concert of nuclear physics and
  astrophysics compose.obspm.fr}},  {\em Phys. Part. Nucl.} {\bf 46} (2015),
  no.~4 633--664, [\href{http://www.arxiv.org/abs/1307.5715}{{\tt 1307.5715}}].

\bibitem{Oertel2017}
M.~Oertel, M.~Hempel, T.~Kl\"ahn, and S.~Typel, {\it Equations of state for
  supernovae and compact stars},  {\em Rev. Mod. Phys.} {\bf 89} (Mar, 2017)
  015007.

\bibitem{CompOSECoreTeam:2022ddl}
{\bf CompOSE Core Team} {\bf Collaboration}, S.~Typel {\em et~al.}, {\it
  {CompOSE Reference Manual}},  {\em Eur. Phys. J. A} {\bf 58} (2022), no.~11
  221, [\href{http://www.arxiv.org/abs/2203.03209}{{\tt 2203.03209}}].

\end{thebibliography}\endgroup

\end{document}